\documentclass[hypertex,twoside,letterpaper,10pt]{report}

\usepackage{setspace}
\usepackage{anysize}
\usepackage{graphicx}
\usepackage{bm}
\usepackage{amsmath}
\usepackage{amssymb}
\usepackage{fancybox}
\usepackage[colorlinks=true,citecolor=blue]{hyperref}

\newcommand{\fr}[2]{{\textstyle \frac{#1}{#2}}}
\newcommand{\tin}[1]{{\mbox{\tiny $ #1 $}}}
\newcommand{\mui}{\mu_{\mbox{\tiny I}}}
\newcommand{\llang}{\langle\!\langle}
\newcommand{\rrang}{\rangle\!\rangle}
\newcommand{\xone}{\tilde\xi_1}
\newcommand{\xtwo}{\tilde\xi_2}

\marginsize{2.5cm}{2.5cm}{2.5cm}{2.5cm}
\begin{document}
\pagenumbering{roman}

\begin{titlepage}

\begin{flushleft}
  \textsc{~~~~~~~~~~~~~~~~~~~~~~Pontificia Universidad Cat\'{o}lica de Chile}  \\
  \textsc{~~~~~~~~~~~~~~~~~~~~~~Facultad de F\'{\i}sica}
\end{flushleft}
\noindent
\begin{picture}(400, 1)
   \put(0,30){\includegraphics[scale=.5]{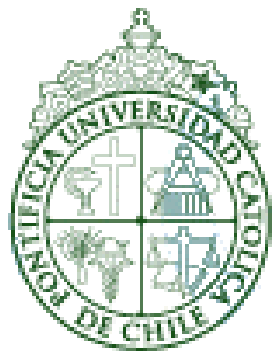}}
\end{picture}

\begin{center}
     {\large \textbf{CHIRAL DYNAMICS AND PION PROPERTIES AT FINITE
TEMPERATURE AND ISOSPIN CHEMICAL POTENTIAL}}\\
\ \\
by\\
 {\bf CRISTI\'AN LUIS VILLAVICENCIO REYES}
\end{center}

\begin{center}
\begin{minipage}[c]{.5\textwidth}
  A thesis submitted to Facultad de F\'{\i}sica, Pontificia Universidad
Cat\'olica
de Chile, in partial fulfillment of the requirements for the degree  of Doctor
of Philosophy.
\end{minipage}
\end{center}

\vspace{.5cm}

\begin{flushright}
  \begin{tabular}{rcl}
     \textsc{Advisor} & : & Dr. Marcelo Loewe  \\
     \textsc{Examining Committee} 
     				& : &   Dr. Jorge Alfaro\\
                                 &  &   Dr. Marco A. D\'{\i}az\\
				  &  &  Dr. Claudio Dib\\
                                 &  &  Dr. Andreas Reisenegger \\
\end{tabular}
\end{flushright}

\vspace{.5cm}
\vfill
\begin{center}
  Agosto, 2004                \\
  \textsc{Santiago -- Chile}                  
\end{center}

\end{titlepage}

\newpage

\chapter*{Acknowledgments}

I would like to acknowledge to the Facultad de F\'{\i}sica of the  Pontificia
Universidad Cat\'olica de Chile, for accepting me in the graduate program and
for all the facilities during my studies. I also wish to acknowledge the
professors of the Facultad de F\'{\i}sica for all the courses I had the
opportunity to attend and for the knowledge I acquired during my
doctoral period. I am grateful to the students and staff of the Facultad de
F\'{\i}sica for the support and help given to me.

My sincere acknowledgment to CONICYT, for the financial support during the
last three years of the development of my thesis with a {\it Beca de
apoyo para la rea\-lizacion de tesis doctoral} and with a {\it Beca de t\'ermino
de tesis
doctoral}. I also like to acknowledge DIPUC, for the financial support given at
the begining of my studies in the doctoral program.

I would like to thank my parents, for their support during all my studies, 
in particular for their comprehension and support of my ``strange" choice of
studying physics.

Finally, I would like give my special acknowledge to my advisor, Prof. Marcelo
Loewe, for his enthusiastic support during all these years, his dedication and
specially his patience to understand what I was doing, to try to explain me why
something could or could not be, and for giving me all the tips necessary to
make me understand what I was doing.

\tableofcontents

\noindent
\chapter*{Abstract}
\addcontentsline{toc}{section}{Abstract}

\noindent
The thermal and density corrections, in terms of the isospin chemical
potential $\mui $, to the mass of the pions, the decay constant and different
condensates are studied in the framework of the $SU(2)$ low energy effective
chiral lagrangian at finite temperature in the two phases: The first phase
$|\mui|<m$ and the second phase $|\mui|>m$, being $m$ the tree-level pion
mass. As a function of temperature for $\mui  =0$, the mass remains quite
stable, starting to grow for very high values of $T$,  confirming previous
results. However, there are interesting corrections to the mass when both
effects (temperature and chemical potential) are simultaneously present. At zero
temperature the $\pi ^{\pm}$ should condense when $\mui = \pm m_{\pi }$. At
finite $T$, the condensed pion acquires a thermal mass in such a way that a
mixture, like in a superfluid, of a condensed and normal phase appears.

\chapter*{Preface}
\addcontentsline{toc}{section}{Preface}

There is a reasonable agreement that Quantum Chromodynamics (QCD) is
the right theory that explains strong interactions. The QCD degrees of
freedom are quark fields and gluon gauge fields, which possess color charge
and, up till now, have not been detected as isolated free particles.
The theory cannot explain the confinement of the quarks and gluons inside the
hadrons, and it is not possible to obtain, in general, analytical solutions,
except for processes at high energies or ``hard processes'' (also equivalent to
small distances) where, due to the asymptotic freedom, perturbation theory is
possible \cite{Gross:1973id,Gross:1973ju,Gross:1974cs,Politzer:1973fx}.
Non-perturbative methods like the Operator Product Expansion (OPE) and QCD Sum
Rules \cite{Shifman:1978bx} or the Nambu Jona-Lasinio (NJL) effective model
\cite{Nambu:1961tp} include corrections to the perturbative analysis, with
quark-gluons degrees of freedom, in terms of condensates, responsible for the
hadronic resonances at low energy.
Numerical simulations of QCD on the Lattice
\cite{'tHooft:1977hy,'tHooft:1981ht}, on the other side, turn out to be a real
non-perturbative approach to the theory, starting from first principles.

Chiral Effective field theories, like the Linear Sigma Model
\cite{Gell-Mann:1960np}, the Non-Linear Sigma Model \cite{Weinberg:1968de} and
Chiral Perturbation Theory ($\chi$PT)
\cite{Weinberg:1978kz,Gasser:1983yg,Gasser:1984ux}, have been a strong and
useful tool for understanding QCD at low energies or ``soft processes'', since
they involve hadronic degrees of freedom preserving the symmetries of QCD and
allow us to establish a connection with the concepts of Current Algebra
(the precursor of QCD).
Among all hadrons, pions play a special role in the dynamics of hadronic matter
because they are the lightest hadrons and therefore are produced at a higher
rate in high-energy reactions. The discussion of pion dynamics is a precious
window to explore the non-perturbative structure of QCD.

 In the last 20 years, there has been a special interest in hot and (or) dense
processes in scenarios like the early universe, relativistic heavy ion
collisions, and inside neutron stars.
High energy processes in a thermal bath have been studied in different frames,
searching for new phenomena and states of matter like Chiral Symmetry
Restoration (see for example \cite{Nowak:1996aj}), the Quark-Gluon Plasma
\cite{Muller:1983ed,Kogut:2004su,Kodama:2004zb} and confinement of quarks and
gluons into hadrons, Color Glass Condensate
\cite{Iancu:2000hn,Ferreiro:2001qy,McLerran:2004fg}, etc. In all these cases we
deal with different phase transitions. A main problem in this context is to find
the appropriate order parameters, in terms of observable quantities associated
to the behavior of a quasi-particle in a medium.

The dependence of the pion properties on finite temperature  has been studied
in a variety of frameworks, such as thermal QCD-Sum Rules
\cite{Dominguez:1996kf}, Chiral Perturbation Theory at one loop
\cite{Gasser:1987ah} and two loops \cite{Schenk:1993ru,Toublan:1997rr}, the
Linear Sigma Model \cite{Larsen:1985ei,Contreras:1989gi,Dominguez:1993kr}, the
Mean Field Approximation \cite{Pisarski:1981mq,Barducci:1991rh}, the Virial
Expansion \cite{Leutwyler:1990uq,Schenk:1991xe}, etc.
There seems to be a reasonable agreement that the pion mass $m_{\pi}(T)$ is
essentially independent of $T$, except possibly near the critical temperature
$T_{c}$, where $m_{\pi}(T)$ increases with $T$, and that the pion decay
constant $f_\pi(T)$ vanishes at the critical temperature.

The introduction of in-medium processes via isospin chemical potential has been
studied at zero temperature in $\chi$PT
\cite{Son:2000by,Kogut:2001id,Splittorff:2001fy} in both phases
($|\mu_I|\lessgtr m_\pi$) at tree level.

 In-medium properties at finite density have been discussed for a variety of
phenomena, as for example, the chiral condensates
\cite{Peng:2003jh,Peng:2003nq}, the anomalous decays of pions and etas
\cite{Costa:2003mz}, etc.
As the density increases, both the quark condensates and the decay rates
diminish.

Interesting results concerning the structure of the QCD phase diagram,
including temperature effects,  have been achieved for the case when we have
simultaneously baryon chemical potential and isospin chemical potential.
Two completely different approaches have confirmed a qualitative change in the
phase diagram as soon as the isospin chemical potential starts to grow
\cite{Toublan:2003tt,Barducci:2003un}.
These  problems concerning the structure of the phase transition diagram, have
also been handled in the frame of two color QCD in four dimensions (a QCD-like
model) \cite{Splittorff:2002xn} as well as in three dimensions
\cite{Dunne:2003ji}.
The  problem with baryonic chemical potential has been considered in the frame
of $\chi$PT \cite{Alvarez-Estrada:1995mh} and also using the finite pion number
chemical potential \cite{Ayala:2002qy}.

Different properties of QCD (or QCD-inspired models) under these kind of
circumstances have also been analyzed in the lattice approach.
For QCD with three and two colors, extensive work has been carried out
\cite{Kogut:2001if,Kogut:2002cm,Kogut:2002tm,Kogut:2002zg,Kogut:2003ju} in
connection with the behavior of different order parameters in several phase
transitions in the $\mu /T$ plane as, for example, the transition to a diquark
phase, the chiral condensate, etc.

Isospin asymmetry does exist in nature in the case of neutron
stars. Since 1971, the idea of a pion condensate in the core of
neutron stars has been considered in connection with the cooling
process of a neutron star (see for example \cite{Pethick:2004}).
This idea is a motivation to study the behavior of pions at
extreme isospin densities, searching for phase transitions and
considering the possibility of having such scenario, isospin
asymmetric regions, in RHIC and ALICE experiments. Eventually,
these considerations are relevant in early stages of the universe.

The main results of this thesis concerns the discussion of thermal radiative
corrections, including finite isospin chemical potential role on both pion
phases, enlarging in this way the existing literature on the isospin
chemical potential effects in the pion phase transitions
\cite{Splittorff:2002xn,Kogut:2004su}. The results of this thesis with respect
to the thermal masses and decay constants in the first phase can be found
in \cite{Loewe:2002tw} and the behavior of the thermal masses in the second
phase in \cite{Loewe:2004mu}.
The non-tivial evolution of pion masses and decay constants may have also
important phenomenological consequences for the diagnosis of the pion gas, in
the central rapidity region of relativistic heavy ion collisions, when looking
for signals as dilepton or photo-production.

This thesis is organized as follows: In chapter \ref{intro}, I will introduce
Chiral Perturbation theory, the renormalization procedure involved, and the
finite temperature and density formalism.
In chapter \ref{prop}, the derivation of the effective Lagrangian, the
propagators necessary for radiative corrections, and the expansion criteria to
solve the problem of the non-diagonal propagators in the second phase will be
presented.
In chapter \ref{phase1}, the self-energy corrections, decay constants, and
condensates in the first phase will be derived.
Chapter \ref{musimm} and \ref{muggm} will be devoted to the calculation of the
self-energy corrections and condensates in the regions $\mui\gtrsim m$ and
$\mui\gg m$.
Finally, in chapter \ref{conc}, I will summarize the results and conclusions,
presenting also an outlook.

\newpage
\pagenumbering{arabic}

\chapter{Introduction}\label{intro}

This chapter is  a brief introduction about the theories and tools that will be
used to compute radiative corrections within a thermal bath for the case of a
pure pion gas; i.e., $\chi$PT and Thermo-Field Dynamics.   

\section{Chiral Perturbation Theory}

$\chi$PT is a successful effective field theory that makes use of the chiral
symmetry, which is an intrinsic symmetry of massless fermions. 
Since light quarks ($u$, $d$) possess masses of a few MeV (the $s$ quark can
also be handled in this frame), it is reasonable to think of them as massless,
allowing a separation of the quark fields in right- and left-handed components.
Then, it is possible to construct an effective Lagrangian with hadronic degrees
of freedom that preserves chiral symmetry. 
The idea of the construction of an effective theory in this context was first
proposed by Weinberg \cite{Weinberg:1978kz}.

The chiral symmetry breaking is usually parametrized by a non-vanishing pion
decay constant $f_\pi$, using Partial Conservation of Axial Current (PCAC)
\cite{Gell-Mann:1960np,Nambu:1960xd}.  
According to the Goldstone theorem \cite{Goldstone:1962es}, a consequence of
the spontaneous symmetry breaking will be the appearance of massless bosons. 
In the case of $\chi$PT, where the $SU(N)_R\times SU(N)_L$ chiral symmetry
grou reduces to a $SU(N)_V$, $N^2-1$ massless bosons appear, according to the
number of generators of the group $SU(N)$ that do not annihilate the vacuum. 
In the case of $SU(2)$, for the two lightest flavors $u$ and $d$, the massless
goldstone bosons are associated with the three pion pseudo-scalar fields.
In fact pions are pseudo-goldstone bosons, due to the intrinsic mass of the
quarks. 

In the same way as it happens in the theory of superconductivity, we expect
that quarks should form Cooper pairs of bound quark-antiquark states, due to
strong attractive interactions, specially if the quarks are massless. 
The vacuum state with a quark pair condensate is characterized by the order
parameter $\langle \bar qq\rangle \neq 0$, which also breaks the chiral
symmetry, allowing the quarks to acquire effective masses as they move through
the vacuum. 
Then the pions acquire a mass due to the explicit symmetry breaking induced by
the small quark masses and also due to the existence of this quark condensate.

One of the advantages of $\chi$PT, for  example in the case of two flavors, is
that we can forget about the isosinglet scalar field $\sigma$ because it is
heavier than the pseudo-scalar fields ($m_\sigma\sim (400-1200)MeV$ in
comparison with $m_{\pi^\pm}\approx 140MeV$).
Then, it is possible to concentrate on pion interactions only, in a scenario
like a pion gas, especially when dealing with finite-temperature corrections. 
In effective models, like the linear $\sigma$-model, it is possible to discuss
the properties of the $\sigma$-meson, and its influence in the low-energy
dynamics \cite{Contreras:1989gi,Dominguez:1993kr}.
Here, I will only concentrate on the pion dynamics, dominant when finite
temperature and density effects are taken into account.

A new comprehensive review of $\chi$PT can be found in \cite{Scherer:2002tk}.

\subsection{Chiral Lagrangian}

Let us proceed in the frame of the $SU(2)$ chiral perturbation theory. 
The most general chiral invariant expression for a QCD-extended Lagrangian
under the presence of external hermitian-matrix auxiliary fields, has the form
\begin{equation}
{\cal L}_{ QCD}(s,p,v,a)   =  {\cal L}_{ QCD}^0
 +{\cal L}_{QCD}^A + \bar q \gamma^\mu (v_\mu +\gamma_5 a_\mu)q
 -\bar q(s-i\gamma_5p)q,
\end{equation}
where ${\cal L}_{ QCD}^0$ is the usual QCD Lagrangian with  zero mass,
${\cal L}_A$ is the anomalous part\footnote{In principle, it includes also the
anomalies of the fermion determinant but it will not be considered here,
because it does not contribute to the corrections involved in this work} and 
$s$, $p$, $v$, and $a$ are scalar, pseudo-scalar vector and  axial external
fields defined as 
\begin{equation}
\begin{array}{cc}
v_\mu = v_\mu^a(x)\tau^a/2,  & \qquad s = s^0(x) +s^a(x)\tau^a,\\ 
a_\mu = a_\mu^a(x)\tau^a/2,  & \qquad p = p^0(x) +p^a(x)\tau^a
\end{array}
\end{equation}
where $s_\mu^a$, $p_\mu^a$, $v_\mu^a$, $a_\mu^a$ are real functions and
$\tau^a$ are the Pauli matrices. 
Consider the independent local $SU(2)_R\times SU(2)_L$ transformations of the
left- and right-handed quarks in flavor space (chiral transformations)
\begin{equation}
\begin{array}{lr}
( 1 +\gamma_5)q \rightarrow g_R( 1+\gamma_5)q ,
& \qquad
\big( 1 -\gamma_5\big)q\rightarrow g_L( 1-\gamma_5)q,
\end{array}
\end{equation}
which induce gauge transformations of the external fields
\begin{eqnarray}
 v'_\mu +a'_\mu &=& g_R( v_\mu +a_\mu)g_R^\dag +ig_R\partial_\mu
  g^\dag_R,\\
 v'_\mu -a'_\mu &=& g_L( v_\mu -a_\mu)g_L^\dag +ig_L\partial_\mu
  g^\dag_L,\\
 s'+ip' &=& g_R( s +ip)g_L^\dag.
\label{transform}
\end{eqnarray}
Then, it is possible to construct a generalized chiral-invariant Lagrangian
containing only pion degrees of freedom, where the external fields transform in
the same way as in Eq.(\ref{transform}). 
The effective low-energy Lagrangian will be expressed as an expansion in powers
of momentum $P$ on a certain scale $\Lambda_\chi$, 
\begin{equation}
{\cal L}_{\chi}={\cal L}_2+{\cal L}_4+{\cal L}_6+\dots +{\cal L}_{\chi}^A,
\end{equation}
where ${\cal L}_n \sim{\cal O}(P^n)$ and ${\cal L}_{\chi}^A$ is the
Wess-Zumino-Witten Lagrangian (the anomaly contribution).

According to Weinberg power counting \cite{Weinberg:1978kz}, $\partial, a, v,
s, p\sim{\cal O}(P)$. This means that the external momentum must be smaller
than a certain scale $\Lambda_\chi$ in order to proceed with an expansion in a
series of powers of momentum\footnote{We will see afterwards that
$\Lambda_\chi=4\pi f\sim 1.2GeV$, where $f$ is the pion decay constant in the
chiral limit}.

We will start with the ${\cal O}(P^2)$ chiral Lagrangian
\begin{eqnarray}
    {\cal L}_2  &=& \frac{f^2}{4}Tr\left[(D_\mu U)^\dag D^\mu U+U^\dag \chi
+ \chi^\dag U\right]
\end{eqnarray}
with
\begin{eqnarray}
 D_\mu U &=& \partial_\mu U-i[v_\mu, U]-i\{ a_\mu, U\},\\
 \chi &=& 2B(s+ip),\\
 U &=& \bar U^\tin{1/2}(e^{i\pi^a\tau^a/f})\bar U^\tin{1/2}.
 \label{eq:fields}
\end{eqnarray}
At this point, $f$ and $B$ are arbitrary constants with dimension of mass, and
$\bar U$ is the vacuum expectation value of the unitary-matrix field $U$.

The most general ${\cal O}(P^4)$ chiral Lagrangian has the form 
\begin{eqnarray}
{\cal L}_4 &=&  \fr{1}{4}l_1\left( Tr\left[(D_\mu U)^\dag D^\mu
           U\right]\right)^2  \nonumber\\
 && +  \fr{1}{4}l_2Tr\left[(D_\mu U)^\dag D_\nu
           U\right]Tr\left[(D^\mu U)^\dag D^\nu U\right]\nonumber\\ 
 &&+  \fr{1}{16}(l_3+l_4)\left(Tr\left[\chi U^\dag
           +U\chi^\dag\right]\right)^2\nonumber\\ 
 &&+  \fr{1}{8}l_4Tr\left[(D_\mu U)^\dag D^\mu U\right]Tr\left[\chi
           U^\dag + U\chi^\dag    \right]\nonumber\\ 
 &&+  l_5Tr\left[\big(L_{\mu\nu}U +UR_{\mu\nu}\big)\big(U^\dag
           L^{\mu\nu}+R^{\mu\nu}U^\dag\big)\right]\nonumber\\
 &&  + l_6Tr\left[iL_{\mu\nu}D^\mu U(D^\nu U)^\dag +iR_{\mu\nu}(D^\mu
           U)^\dag D^\nu U\right]\\ 
 && -\fr{1}{16}l_7\left(Tr\left[\chi
           U^\dag-U\chi^\dag\right]\right)^2\nonumber\\ 
 &&+  \fr{1}{4}(\tilde{h}_1+\tilde{h}_3)Tr[\chi^\dag\chi]\nonumber\\
 &&
-\fr{1}{8}(\tilde{h}_1-\tilde{h}_3)\left\{Tr\left[\chi^2+{\chi^\dag}^2\right]
           -Tr\left[\chi\right]^2
           -Tr\left[\chi^\dag\right]^2\right\}\nonumber\\  
&&
-2\tilde{h}_2Tr\left[L_{\mu\nu}L^{\mu\nu}+R_{\mu\nu}R^{\mu\nu}\right],\nonumber
\end{eqnarray} 
with
\begin{eqnarray}
    L_{\mu\nu}  = \partial_\mu l_\nu -\partial_\nu l_\mu +i[l_\mu,l_\nu],
                    &&\quad l_\mu       = v_\mu -a_\mu\\
    R_{\mu\nu}  = \partial_\mu r_\nu -\partial_\nu r_\mu +i[r_\mu,r_\nu],
                    &&\quad r_\mu       = v_\mu +a_\mu.
\end{eqnarray}
The $l_i$ are the  original Gasser and Leutwyler coupling constants ~
\cite{Gasser:1983yg,Gasser:1984ux} and the $\tilde{h}_i$ are couplings to pure
external fields, and their values depend on the model.
For calculations at the one-loop level, it is enough to keep terms up to ${\cal 
 O}(P^4)$, as we will see soon.

We can see that our extended QCD-Lagrangian  contains the external fields $s$,
$p$, $v$, $a$, coupled to the scalar, pseudo-scalar, vector and axial-vector
currents, respectively.  
Now, it is easy to derive these currents and other quantities in the low-energy
theory. 
For example, let us consider the axial current. Using the high-energy QCD
Lagrangian, it is given by
\begin{equation}
A^a_\mu = \fr{1}{2}\bar q\gamma_\mu\gamma_5\tau^aq = \frac
{\delta}{\delta a_\mu^a}S_{QCD}[s,p,v,a],
\end{equation}
where $S_{QCD}[s,p,v,a]$ is the QCD action.  
In the low-energy description of QCD, we can obtain this current following the
same procedure:
\begin{equation}
A^a_\mu = \frac{\delta S_{\chi}}{\delta a_\mu^a} =
  A_{(1)\mu}^a +A_{(3)\mu}^a +\dots +A_{(A)\mu}^a,
\end{equation}
where the index $n$ in $A_{(n)\mu}^a$ denote ${\cal O}(P^{n})$.

If we want to calculate condensates, for example the chiral condensate, (also
known as quark condensate) $\langle \bar qq\rangle$, the procedure is the same:
\begin{equation}
 \langle 0|\bar qq|0\rangle 
 =\langle 0|J_s|0\rangle = -\langle
  0|\frac{\delta}{\delta s^0}S_{QCD}[s,p,v,a]|0\rangle.
\end{equation}
The chiral condensate in the low-energy regime can be identified with
\begin{equation}
\langle\bar qq\rangle  = -\langle 0|\frac{\delta S_{\chi}}{\delta
  s^0}|0\rangle = \langle\bar qq\rangle_{(1)} +\langle\bar
  qq\rangle_{(3)} +\cdots +\langle\bar qq\rangle_{(A)}
\end{equation}
and it is proportional to $B$ in the low energy effective theory.

\subsection{Power counting, $B$, $f$ and 
$\Lambda_\chi$}\label{intro.power-counting}

It is obvious that the effective Lagrangian is difficult to handle because it
is expressed in terms of an infinite series in powers of the fields involved;
for such reason we need to truncate this series.
Consider the case of massless QCD without external fields. 
Proceeding as we did in the previous section, the Lagrangian, the axial current
and the chiral condensate are (see App. \ref{expansion}) 
\begin{eqnarray}
{\cal L}_2(0,0,0,0) &=& \frac{f^2}{4}Tr\left[\partial U\partial U^\dag\right],\\
A_{(1)\mu}^a(0,0,0,0) &=& \frac{f^2}{4}Tr\big[ i\partial_\mu
  U\{\fr{1}{2}\tau^a,U^\dag\} -i\{\fr{1}{2}\tau^a,U\}\partial_\mu
  U^\dag\big],\\
\langle\bar qq\rangle_{(1)}(0,0,0,0) &=& -\langle
  0|\fr{1}{2}Bf^2Tr\left[U+U^\dag\right]|0\rangle .
\end{eqnarray}
Note that this Lagrangian is independent of the vacuum state value.
The lower contribution to these quantities (tree-level), expanding in pion
fields, are
\begin{eqnarray}
{\cal L}_{2,2} &=& \fr{1}{2}(\partial\bm{\pi})^2,\\
 A_{(1,1)\mu}^a &=& -f\partial_\mu\pi^a,\\
 \langle\bar qq\rangle_{(1,0)} &=& -2Bf^2,\label{qq-tree-level}
\end{eqnarray}
respectively, where the sub-indices $(n,m)$ mean ${\cal O}(P^{n}\pi^m)$. 

 According to the PCAC relation,\footnote{Strictly speaking, the concept PCAC
is related to the divergence of the axial current, but actually is known by
the relation in eq. (\ref{PCAC-tree-level}).} 
\begin{equation}
\langle 0|A_\mu^a(x)|\pi^b(p)\rangle =
ip_\mu f_\pi \delta^{ab}e^{-ipx},
\end{equation}
where $f_\pi$ is the pion-decay constant. 
In the case of $A_{(1,1)\mu}^a$, 
\begin{equation}
\langle 0|A_{(1,1)\mu}^a(x)|\pi^b(p)\rangle = -f\partial_\mu \langle
0|\pi^a(x)|\pi^b(p)\rangle = ip_\mu f\delta^{ab}e^{-ipx}.
\label{PCAC-tree-level}
\end{equation}

From eqs. (\ref{qq-tree-level}) and (\ref{PCAC-tree-level}) we conclude
that $f$ is associated to $f_\pi$  and $Bf^2$ is related to the quark
condensate. In fact $f$ and $-2Bf^2$ are the pion decay constant and the quark
condensate, respectively, in the chiral limit.\footnote{ If we take into account
higher terms in the expansion, it can be proved that their contribution is
negligible in the chiral limit.}

\medskip

Consider now the case of QCD with massive quarks without external fields:
${\cal L}_{QCD}(M,0,0,0)$, where $M$ is the quark mass-matrix (in the SU(2)
case, $M=diag (m_u,m_d)$). 
The free effective Lagrangian in terms of the $U$ fields is 
\begin{equation}
{\cal L}_2=\frac{f^2}{4}Tr\big[\partial U\partial U^\dag
   +2BM(U+U^\dag)\big].
\end{equation}
The effective potential is then
\begin{equation}
V_{eff}=-\frac{f^2}{4}Tr[2BM(\bar U+\bar U^\dag)].
\end{equation}
By construction, the vacuum expectation value of the fields $U$ is defined with
the fields $\bm{\pi^a}=0$. 
A general expression for $\bar U$ is 
\begin{equation} 
\bar U=\cos\varphi +i\bm{\tau}\cdot\bm{n}\sin\varphi,
\end{equation}
where $\bm{n}$ is a unitary vector and $\varphi$, for now, is an arbitrary
angle to be derived. 
It is easy  to see that the value of $\bar U$ that minimizes the effective
potential is $\bar U =1$.

Now, expanding the Lagrangian in terms of the pion fields, we get the free
Lagrangian:
\begin{equation}
{\cal L}_{2,2} = \fr{1}{2}\big[ (\partial\bm{\pi})^2
  -(m_u+m_d)B\bm{\pi}^2\big].
\end{equation}
The constant $(m_u+m_d)B$ is then associated to the pion mass $m_\pi^2$. 
We will refer to the tree-level pion mass as $m^2=(m_u+m_d)B$.  
 
 If we compare with eqq. (\ref{qq-tree-level}) and (\ref{PCAC-tree-level}), we
get the well-known Gell-Mann Oakes Renner (G-MOR) relation at the tree level
\cite{Gell-Mann:1968rz}
\begin{equation}
-\fr{1}{2}(m_u+m_d)\langle\bar qq\rangle = m_\pi^2f_\pi^2.
\end{equation}
This relation opened a new way to our understanding of strong hadron dynamics
and chiral symmetry, since it relates the the pion masses and decay constants
with parameters of the fundamental theory: the quark masses and the quark
condensate.
 
The next terms of the Lagrangian are powers of the form
$f^2(\bm{\pi}^2/f^2)^{m}$. 
If we impose that the vacuum expectation value of $\langle \pi^2/f^2\rangle
\sim p^2/\Lambda_\chi^2$, we can neglect higher terms in the Lagrangian. 
In fact, if we use dimensional regularization (see App. \ref{dim_reg})
\begin{eqnarray}
&& \langle \pi^2/f^2\rangle 
 \rightarrow \frac{1}{f^2}\langle 0|\pi^a(x)\pi^b(x)|0\rangle \nonumber\\
 &&\hspace{1.5cm} = \frac{\Lambda^{4-d}}{f^2} \int\frac{d^dk}{(2\pi)^4}
  \frac{i\delta_{ab}}{k^2-m^2+i\epsilon}
 = \delta_{ab}\frac{m^2}{(4\pi)^2f^2}\left[ \ln\frac{m^2}{\Lambda^2}
   -\lambda\right]
\end{eqnarray}
where $\lambda$ is the $\overline{MS}$ divergent term and $\Lambda$ is a scale
factor associated to the dimensional regularization procedure. 
Forgetting for a moment the divergent term (keeping in mind, however, that the
theory is renormalizable in an effective sense), certainly the term $m^2\ln
m^2/\Lambda^2$ is in the range of ${\cal O}(P^2)$ for a reasonable value of
$\Lambda$. 
We can set then the chiral-scale factor as $\Lambda_\chi = 4\pi f$. Since
$f_\pi \sim 93MeV$, the theory is valid up to $\Lambda_\chi \sim 1.2GeV$.

Now we can truncate the series according to the number of loop corrections. 
For example in the calculation of 1-loop self-energy corrections, we need to
consider corrections up to ${\cal O}(\pi^4)$ in the fields. 
If they include the ${\cal L}_{2,4}$ which is of ${\cal O}(P^6)$, then we must
take into account the next contributions of the ${\cal L}_{2n}$ up to ${\cal
O}(P^6)$ 
\begin{eqnarray}
{\cal L}_{tree} &=& {\cal L}_{2,0} +{\cal L}_{2,1} +{\cal L}_{2,2},\\
{\cal L}_{1-loop} &=& \sum_{m=0}^{4}{\cal L}_{2,m} +{\cal L}_{4,0}
 +{\cal L}_{4,1} +{\cal L}_{4,2}\label{L 1-loop},\\ 
{\cal L}_{2-loop} &=& \sum_{m=0}^6{\cal L}_{2,m} +\sum_{m=0}^4{\cal L}_{4,m}
 +{\cal L}_{6,0} +{\cal L}_{6,1} +{\cal L}_{6,2},\\
etc. && \nonumber
\end{eqnarray}

I would like to remark at this moment that, if we go to higher orders in the
expansion, we will loose predictive power.
In fact, we need experimental values of some observables to fix the different
coupling constants, at each order in the expansion. 
For higher orders we have more and more couplings, since the theory is only
renormalizable in an effective sense. 
This point will be discussed more in detail in the next sub-section.

\subsection{Renormalization}\label{renormalization}

The chiral Lagrangian is not renormalizable in the usual sense, since it
involves an infinite number of couplings. 
Nevertheless, by construction, the different coupling constants involve the
infinities that appear in dimensional regularization:
\begin{equation}
l_i(\Lambda)=\frac{\gamma_i}{32\pi^2}\left[ \bar l_i
  +\ln\frac{m^2}{\Lambda^2} -\lambda\right],
\end{equation}
where the different $\bar l_i$ and $\gamma_i$ are tabulated in App.
\ref{l-constants}

The renormalization procedure for self-energy corrections works as follows:
The corrected propagator  is
\begin{eqnarray}
\bm{D}_{c} &=& \bm{D}-i\bm{D}\bm{\Sigma}\bm{D}+\cdots \quad
  = \bm{D}(1+i\bm{\Sigma}\bm{D})^{-1},\nonumber\\
\bm{D}_{c}^{-1} &=& \bm{D}^{-1}+i\bm{\Sigma},
\end{eqnarray}
where $\bm{D}$ is the free propagator matrix defined as 
\begin{equation}
 \bm{D}_{ij}^{-1}=i\int d^4xe^{ipx}\frac{\delta^2
 S_{2,2}[\pi]}{\delta \pi_{\tin{i}}\delta\pi_{\tin{j}}}(x),
 \label{D^-1}
\end{equation}
and $\bm{\Sigma}$ is the self-energy  matrix. 
Usually $\bm{D}$ and $\bm{\Sigma}$ are diagonal (or anti-diagonal); but we will
see that this will not be the case in the second phase where $|\mui| >m$.

Following the usual renormalization procedure, we re-scale the fields
$\pi^i=\sqrt{Z_i}\pi_R^i$ with $Z_i=1+\delta_{Z_i}$. 
$\bm{\Sigma}$ and $\delta_{Z_i}$ are  of the order of the perturbative
parameter $P^2/\Lambda_\chi^2\equiv \delta_\chi$. Since $[\bm{\Sigma}]=2$ in
mass units and $\delta_{Z_i}$ is dimensionless we see that $\bm{\Sigma} \sim
{\cal O}(P^2\delta_\chi)$ and $\delta_{Z_i} \sim {\cal O}(\delta_\chi)$. 
The renormalized propagator is
\begin{eqnarray}
i D_{R}^{-1}(p)_\tin{ij}
&=& Z_{ij}D_c^{-1}(p)_\tin{ij}\nonumber\\
  &=& Z_{ij}iD^{-1}(p)_\tin{ij}
  -\Sigma(p)_\tin{ij}+{\cal O}(P^2\delta_\chi^2)\nonumber\\
  &=& iD^{-1}(p)_\tin{ij}
  -\Sigma_{R}(p)_\tin{ij}
\end{eqnarray}
with $ Z_{ij}\equiv\sqrt{Z_iZ_j}$. The value of $Z$ is chosen in such a way
that the corrected propagator does not have corrections proportional to $p^2$,
or, in other words, the coefficient that multiply the term $p^2$ in the
corrected propagator must be $1$.

As an example, let us consider a free propagator and the self energy correction
of the form
\begin{eqnarray}
iD^{-1}(p) &=& (p_0-b)^2-\bm{p}^2-a^2,\\
\Sigma (p) &=& \Sigma^\tin{0}+\Sigma^\tin{1}p_0+\Sigma^\tin{2}p^2.
\label{SigmaGeneral}
\end{eqnarray}
This expression for $\Sigma$, as we will see, is valid in the first phase.

Choosing $Z=1+\Sigma^\tin{2}$, the renormalized self-energy will be
\begin{eqnarray}
\Sigma_R(p) &=& \big[\Sigma^\tin{0}+(a^2-b^2)\Sigma^\tin{2}\big]
 +\big[\Sigma^\tin{1}+2b\Sigma^\tin{2}\big]p_0\nonumber\\
  &\equiv& \Sigma_R^\tin{0}+\Sigma_R^\tin{1}p_0.
\end{eqnarray}

\subsubsection{Expansion in terms of mass corrections}

Unfortunately, in general, the self-energy $\Sigma$  will not have the form of
eq.(\ref{SigmaGeneral}), but could be a complicated function of the external
momenta, as indeed will be the case for the second phase in the region of high
isospin chemical potential values. 
As we want to compute mass corrections, it is possible to expand the
self-energy in terms of these corrections. 
In the rest frame, where $\bm{p}=0$, the energy will be $p_0=m_R=m_t+\delta m$,
where $m_R$ is the renormalized mass, $m_t$ is the tree level mass and $\delta
m$ is the correction due to the self-energy terms. 
Then
\begin{eqnarray}
\Sigma (m_R) &=& \Sigma (m_t)
    +\Sigma ^\prime(m_t)\delta m
    +\fr{1}{2}\Sigma ^{\prime\prime}(m_t)(\delta m)^2
    +\dots\nonumber\\
    &=& \Sigma^\tin{0}[m_t]+\Sigma^{1}[m_t]m_R
    +\Sigma^\tin{2}[m_t]m_R^2+{\cal O}(\delta m)^3,
\end{eqnarray}
where the argument inside the brackets indicates the mass around which the
self-energy expansion was computed.

The masses are defined as the poles of the determinant of the propagator matrix
at zero 3-momentum, i.e. they will correspond to the solutions of
\begin{equation}
 \left|\bm{D^{-1}}(p)\right| _{\vec p=0}=0.
\label{det}
\end{equation}

As we said before, if  the self-energy has a complicated form, we expand the
renormalized mass in the rest frame in powers of corrections $\delta m$ to the
tree-level mass. 
Since the renormalized masses will be extracted as solutions of $\left|
\tilde{\bm{ D}}^{-1}_R(m_R)\right|\equiv 0$, we only need to compute the
$\delta m$ corrections.

\subsubsection{PCAC and condensates}

In the case of the PCAC relation, where the axial vector current is saturated 
with one pion $\langle 0|A_\mu^a|\pi^b\rangle$, the physical pion is the
renormalized pion $\pi_R^a$. 
The relation will then be
\begin{equation}
\langle 0|A_\mu^a(\pi)|\pi_R^b\rangle = 
\langle 0|A_\mu^a(\sqrt{Z}\pi_R)|\pi_R^b\rangle .
\end{equation}

As the tree-level part of the axial-vector current is proportional to a single
pion, the square root of the renormalization constant $\sqrt{Z_a}$ can be
expanded in terms of $\delta_{Z_a}$, which is of order $\delta_\chi$:
\begin{equation}
\sqrt{Z_a} = \sqrt{1+\delta_{Z_a}} 
= 1+\fr{1}{2}\delta_{Z_a} +{\cal O}(\delta_\chi^2).
\end{equation}
The different $Z_a$ values are set by the self-energy correction conditions,
explained before.

For the case of condensates, the renormalization procedure is not necessary,
since all the divergences coming from radiative corrections cancel with the
$l_i$ constants.

\section{Finite temperature and chemical potential}

The properties of a system in thermal equilibrium can be determined from the
grand partition function
\begin{equation}
{\cal Z} = Tre^{-\beta H'}, \quad H' = H-\sum_i\mu_iQ_i,
\end{equation}
where $H$ is the Hamiltonian of the theory, $\beta = 1/T$ is the inverse of the
temperature\footnote{In our unit system, the Boltzmann constant $k_B$ is taken
as $k_B=1$}, being the Lagrange multiplier of the energy, and $\mu_i$ are the
chemical potentials where $\beta\mu_i$ are the Lagrange multipliers of the
different conserved quantities. With this, the thermal average of an observable
is
\begin{equation}
\langle {\cal O}\rangle = \frac{1}{\cal
  Z}Tr\big[e^{-\beta H'}{\cal O}\big].
\end{equation} 
In field theory, if we consider the fields $\phi_i$, representing the different
particles of the theory, the partition function is then given as a functional
integral
\begin{equation}
{\cal Z}
= \int \{D\phi\}\langle\{\phi(x)\}|e^{-\beta H'}|\{\phi(x)\}\rangle 
={\cal N} \int \{D\phi\} e^{iS_\beta'[\phi_i]}
\end{equation}
integrated along an imaginary time path \cite{Matsubara:1955ws} with
\begin{equation}
S_\beta' = \int_{x_0}^{x_0-i\beta}dt\big[L +\sum_i\mu_iQ_i],
\label{extended Lagrangian}
\end{equation}
satisfying the Kubo-Martin-Schwinger (KMS) boundary conditions
\cite{Kubo:1957mj,Martin:1959jp}
\begin{equation}\label{boundaryconditions}
\begin{array}{rl}
\phi(x_0-i\beta,\bm{x})=\phi(x_0,\bm{x}) & \textrm{for bosons},\\
\phi(x_0-i\beta,\bm{x})=-\phi(x_0,\bm{x}) & \textrm{for fermions}.
\end{array}
\end{equation}

\subsection{Thermo-Field Dynamics}

\begin{figure}
\begin{center}
\includegraphics{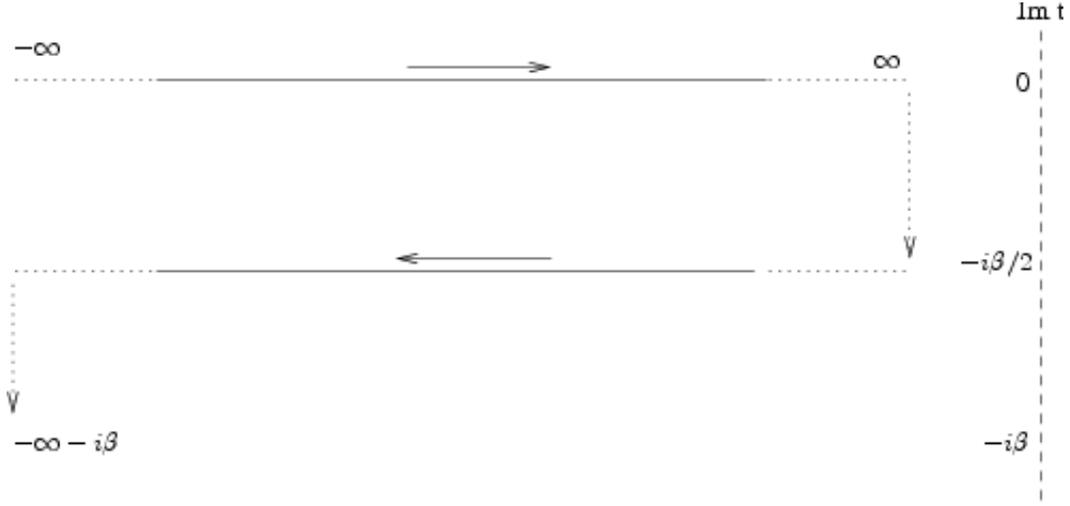}
\end{center}
\caption{\footnotesize Integration path in TFD. The fields that ``live" in the
imaginary time $x_0-i\beta /2$ are called Thermal Ghosts.}
\label{path}
\end{figure}

To construct the usual generating functional, we can set $x_0 = -\tau$ and
follow a path of integration in the complex time-plane as in Figure \ref{path}.

If we take the limit $\tau\rightarrow\infty$, the fields will be split into
two regions; the ``real'' fields that live on the real axis of time, and the so
called thermal ghosts. 
This means that the Fock-space is duplicated \cite{Umezawa:1982nv}. 
This formalism is named Thermo-Field Dynamics (TFD). 
The formal algebraic formulation of TFD was proposed by \cite{Ojima:1981ma}. 
The path integral formulation for the fermion fields, and 
the real-time Feynman rules for gauge theories with fermions, at finite
temperature and density, were developed by \cite{Kobes:1984vb} and
\cite{Niemi:1983nf,Matsumoto:1983by}.

One of the consequences of the appearance of thermal ghosts is that we will get
four two-point propagators instead of one for each field, i.e., if we denote by
$\phi_1=\phi(x)$ the real field and by $\phi_2=\phi(x_0-i\beta/2,\bm{p})$ the
ghost fields, we have
\begin{equation}
D_{T_{ij}}(x,y;T)=\llang 0|T\phi_i(x)\phi_j(y)|0\rrang,
\end{equation}
where  by $|0\rrang$  we understand the thermal vacuum state or ``populated''
vacuum state, and the propagator with the real fields $D_{Th_{11}}$ is the
so-called Dolan-Jackiw propagator (DJp) \cite{Dolan:1973qd}.

The prescriptions for loop calculations are:
\begin{itemize}
\item The thermal insertions appear only through loop calculations.
\item Thermal ghosts must be included in calculations at two or more
  loops
\end{itemize}

Since in this work we will deal only with one-loop corrections, we need only to
consider the DJp. 
I will not discuss the  construction of the DJp, which isbased on the KMS
boundary conditions indicated in eq.\ref{boundaryconditions} and the time-order
of the fields along the path described in Fig.\ref{path}. 
A general formula for the DJp for bosonic fields is 
\begin{eqnarray}
D_{DJ}(p) &=& \int \frac{dk_0}{2\pi i}\lim_{\eta\to 0}
\left[\frac{D(k_0+i\eta,\bm{p})
    -D(k_0-i\eta,\bm{p})}{k_0-p_0-i\epsilon}\right]\nonumber\\
&& \quad +n_B(p_0)\big[D(p_0+i\epsilon,\bm{p})
  -D(p_0-i\epsilon,\bm{p})\big],
\label{DJp formula}
\end{eqnarray}
where $-iD$ is the free propagator in momentum space defined in equation
(\ref{D^-1}) and $n_B$ is the Bose-Einstein distribution defined as 
\begin{equation}
n_B(x) = \frac{1}{e^{\beta x}-1}.
\end{equation}

For a free propagator of the form
\begin{equation}
iD^{-1}(p) = p^2 -2ap_0 -b,
\end{equation}
the corresponding DJp, following eq.\ref{DJp formula} is
\begin{equation}
D_{DJ}(p)= \frac{i}{p^2 -2ap_0 -b+i\epsilon} +2\pi n_B(|p_0|)\delta(p^2 -2ap_0
-b).
\end{equation}

To distinguish the propagator at zero temperature and the pure thermal part of
the DJp, we will refer to them as the free-propagator and the thermal
insertion, respectively.

\subsection{Covariant formalism and finite chemical potential}
Consider the case of a fermionic field within a dense medium, asymmetric with
respect to baryon number. 
If the baryon number density is $n=\psi^\dag\psi$, the introduction of a
chemical potential into the Lagrangian density as a Lagrange multiplier of this
conserved quantity gives
\begin{eqnarray}
{\cal L} +\mu n &=& \bar\psi(i\gamma\partial-m)\psi +\mu\psi^\dag\psi
= \bar\psi(i\gamma\partial +\gamma_0\mu-m)\psi\nonumber\\
 &=& \bar\psi(i\gamma^\mu[\partial_\mu -i\mu\delta_{\mu 0}]-m)\psi.
\end{eqnarray}
This means that the chemical potential acts like a gauge field
\cite{Weldon:1982aq,Actor:1985xp}. 
The corresponding propagator is
\begin{equation}
S_F(p)=\frac{i(\gamma p +\gamma_0\mu +m)}{(p_0+\mu)^2-\bm{p}^2-m^2+i\epsilon}.
\end{equation}

The presence of a thermal bath breaks Lorentz invariance, since it plays the
role of a privileged reference frame. 
Nevertheless, formally we can ``restore'' the Lorentz symmetry in such a way
that it is possible to give a covariant form to the equations, through the
introduction of the fourt-velocity $u_\mu$  between the observer and the
thermal bath \cite{Weldon:1982aq}. 
In the rest frame (i.e. where the thermal bath is at rest with respect the
observer), we have 
\begin{equation}
u=(1,0,0,0).
\end{equation}
Although  this work is performed in the rest frame, we use the symbol $u_\mu =
\delta_{\mu 0}$ recalling that in principle it is possible to use a covariant
formalism.

\subsection{QCD at finite isospin chemical potential}

As was indicated in the previous section, the chemical potential in
eq.(\ref{extended Lagrangian}) is added to the Lagrangian as a Lagrange
multiplier of a conserved charge.  
In this thesis we will consider the isospin chemical potential related to the
isospin number, which indicates the difference of the baryon number between
the $u$ and the $d$ fields. The isospin density number operator is defined as
\begin{equation}
n_I \equiv \fr{1}{2}(u^\dag u -d^\dag d) = q^\dag\frac{\tau_3}{2}q 
\end{equation}
where $q= (u,d)$ in the case of two flavors. 
Then the QCD-Lagrangian with massive quarks acquires the form
\begin{eqnarray}
{\cal L}_{QCD}(M,0,0,0) +\mui n_I &=& {\cal L}^0_{QCD} +{\cal
L}_{QCD}^A -\bar qMq +\mui \bar q\gamma\cdot u\frac{\tau_3}{2}q \nonumber\\  
  &=& {\cal L}_{QCD}(M,0,\fr{1}{2}\mui\tau_3u,0).
\end{eqnarray}

\chapter{Tree-level masses, thermal propagators and vertices at finite isospin
chemical potential.}\label{prop}

For radiative calculations, we need the different propagators derived
from the free Lagrangian.
To obtain the effective Lagrangian at the one-loop level, first we have
to calculate the vacuum expectation value of the field $U$, named $\bar U$.
Taking the ${\cal O}(P^2)$ Lagrangian and setting $\pi^a=0$ we have that the
effective potential is
\begin{equation}
V_{eff} = -\frac{f^2}{4}Tr\left[ \mui^2\left[\fr{1}{2}\tau_3,\bar
U\right]\left[\fr{1}{2}\tau_3,\bar U^\dag\right]+2BM\left(\bar U+\bar
U^\dag\right)\right].
\end{equation}
Taking the general expression for $\bar U$
\begin{equation}
\bar U = \cos\varphi + i\bm{\tau}\cdot\bm{n}\sin\varphi,
\end{equation}
with
\begin{equation}
\bm{n} = \cos\phi\sin\theta\bm{e_1} +\sin\phi\sin\theta\bm{e_2}
+\cos\theta\bm{e_3},
\end{equation}
the previous expression becomes
\begin{equation}
V_{eff} = -f^2\left( \fr{1}{2}\mui^2\sin^2\varphi\sin\theta -m^2\cos\varphi
\right).
\end{equation}
Minimizing the effective potential, we find that the vacuum expectation value
of the $U$ fields depends on $\mui$ \cite{Kogut:2000ek,Kogut:2001id} and is
given by
\begin{equation}
\bar U = \left\{
\begin{array}{cl}
1 & \quad \mbox{if} \quad \mui^2 < m^2\\
\frac{m^2}{\mui^2} +i[\tau_1\cos\phi +\tau_2\sin\phi]\sqrt{1-\fr{m^4}{\mui^4}} &
\quad\mbox{if}\quad \mui^2 > m^2
\end{array}\right.
\end{equation}

Now, we can expand the chiral Lagrangian in terms of the pion degrees of freedom
up to the necessary powers for one-loop radiative corrections. The two phases,
$|\mui|<m$ and $|\mui|>m$, will be referred to as the first phase and the second
phase, respectively. The expansion of the chiral Lagrangian is explained in App.
\ref{expansion.lagrangian}.

\section{First phase}

For the case of the first phase, the relevant Lagrangian for one-loop
corrections is
${\cal L}_{1-loop} = {\cal L}_{2,2} +{\cal L}_{2,4} +{\cal L}_{4,2}$ as was
indicated in Eq. (\ref{L 1-loop}),
where the ${\cal L}_{i,j}$ terms are
\begin{eqnarray}
{\cal L}_{2,2} &=& \frac{1}{2}\left[\left(\partial\pi_0\right)^2
 -m^2\pi_0^2\right]
+\left|\partial_I\pi\right|^2-m^2\left|\pi\right|^2\label{L_22-phase1}\\
{\cal L}_{2,4} &=& \frac{1}{4!}\frac{m^2}{f^2}\pi_0^4
 +\frac{1}{6f^2}\left[-4\left|\partial_I\pi\right|^2\left|\pi\right|^2
 +\left(\partial\left|\pi\right|^2\right)^2
 +m^2\left(\left|\pi\right|^2\right)^2\right]\nonumber\\
 && +\frac{1}{6f^2}\left[ -2\left|\partial_I\pi\right|^2\pi_0^2
 -2\left(\partial\pi_0\right)^2\left|\pi\right|^2
 +\partial\pi_0^2\cdot\partial\left|\pi\right|^2
 +m^2\pi_0^2\left|\pi\right|^2\right]\label{L_24 phase1}\\
{\cal L}_{4,2} &=& \frac{m^2}{f^2}\left[2l_4\left|\partial_I\pi\right|^2
 -2m^2(l_3+l_4)\left|\pi\right|^2
+l_4\left(\partial\pi_0\right)^2
-m^2(l_3+l_4-\epsilon_{ud}^2l_7)\pi_0^2\right],\nonumber\\
\label{L_42 phase1}
\end{eqnarray}
using the standard definition of the pion fields
\begin{equation}
\pi_0=\pi_3, \qquad \pi_\pm = \fr{1}{\sqrt{2}}(\pi_1\mp i\pi_2).
\end{equation}
In the previous
expressions
\begin{equation}
 |\pi|^2 \equiv \pi^+\pi^- , \qquad
|\partial_I\pi|^2=(\partial_I^+ \pi^+)(\partial_I^- \pi^-),
\end{equation}
and the derivative $\partial_{I\pm}$ that appears in the Lagrangian is defined
as
\begin{equation}
\partial_{I}^\pm\equiv
\partial\mp i\mu_Iu.
\end{equation}
This definition of the covariant derivative is
natural, since we know
\cite{Actor:1985xp}  that the chemical potential is introduced as
the zero component of an external ``gauge" field.
The term $\epsilon_{ud}$ in Eq. \ref{L_42 phase1} is the quark masses ratio
$(m_u-m_d)/(m_u+m_d)$.

To obtain the DJp, first consider the equations of motion in momentum space,
extracted from the free Lagrangian ${\cal L}_{2,2}$ in Eq. (\ref{L_22-phase1}).
\begin{equation}
i\bm{D}^{-1}
   = i\left(
  \begin{array}{ccc}
   D^{-1}_\tin{00} & 0                & 0                \\
   0               & 0                &  D^{-1}_\tin{+-} \\
   0               &  D^{-1}_\tin{-+} & 0
  \end{array}\right)
\end{equation}

The
non-zero elements of the free propagator matrix
are\footnote{Note that
  $D^{-1}(p)_\tin{ij}$ is different from $1/D(p)_\tin{ij}$
  in this case because  $D_\tin{+-}$ and $D_\tin{-+}$ are anti-diagonal.}
\begin{eqnarray}
iD^{-1}(p)_\tin{00} &=& p^2 -m^2\\
iD^{-1}(p)_\tin{+-} &=& (p -\mui  u)^2  -m^2 \\
iD^{-1}(p)_\tin{-+} &=& iD^{-1}(-p)_\tin{+-}.
\end{eqnarray}
It is easy to derive the tree-level masses, because the propagator matrix has
no crossed terms.
Then, the equations of motion act independently on each pion
field.
The previous equations are the equations of motion in momentum space
$iD^{-1}(p)_{ij}$ respect to the field $\pi_j$.
From these equations, it is easy to associate the different tree-level masses,
obtaining
\begin{equation}
m_0 = m, \qquad\qquad m_\pm = m \mp \mui,
\label{tree level masses - phase1}
\end{equation}
where the the different pion masses are defined as their rest energy.

Since our calculation will be at the one-loop level, we do not need, as was
said previously, the full formalism of thermo field dynamics including thermal
ghosts.
Then, by the general formula in Eq. (\ref{DJp formula}), the DJp are

\begin{eqnarray}
D_{DJ}(p)_\tin{00} &=& \frac{i}{p^2 -m^2 +i\epsilon} +2\pi
  n_B(|p_0|)\delta (p^2 -m^2)\\
D_{DJ}(p)_\tin{+-} &=& \frac{i}{(p +\mui u)^2 -m^2 +i\epsilon}
  +2\pi n_B(|p_0|)\delta\big( (p +\mui u)^2 -m^2\big)\\
D_{DJ}(p)_\tin{-+} &=& D_\tin{+-}(-p;T).
\end{eqnarray}

\begin{figure}
\centering
\includegraphics[scale=1.2]{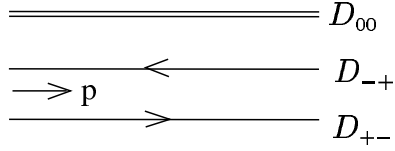}
\caption{\footnotesize  Propagators in the first phase (and in second
phase
for $|\mui|\gtrsim m$).
The double line denotes the neutral propagator.
The single line denotes the charged propagator with an arrow pointing from
\mbox{$+$ to $-$.}}\label{fig props}
\end{figure}

Diagrammatically, the neutral pion propagator will be denoted as a double line.
The $\pm$ propagator will be denoted as a single line with an arrow pointing in
direction from $+$ to $-$, as is indicated in Fig. \ref{fig props}.

The vertices extracted from ${\cal L}_{2,4}$ and ${\cal L}_{4,2}$, shown in
eqs. (\ref{L_24 phase1}) and (\ref{L_42 phase1}), are
\begin{eqnarray}
 V_{2,4}^\tin{0000} &=&
  i\frac{m^2}{f^2}\\
V_{2,4}^\tin{00\pm} &=&
 \frac{i}{3f^2}\Big\{ 2pq
   +2p_\tin{+}p_\tin{-}
    -\big[p+q\big]
   \big[p_\tin{+}+p_\tin{-}\big]
  -\mui\big[p_\tin{+}^0-p_\tin{-}^0\big]
    +m^2-2\mui^2\Big\}\nonumber\\
    &&\\
V_{2,4}^{\tin{\pm\pm}} &=&  \frac{i}{3f^2}\Big\{
 -2p_\tin{+}q_\tin{+}
    -2p_\tin{-}q_\tin{-}
 +\big[p_\tin{+}+q_\tin{+}\big]
  \big[p_\tin{-}+q_\tin{-}\big]\nonumber\\
 &&\hspace{1cm} +4\mui\big[p_\tin{+}^0+q_\tin{+}^0
  -p_\tin{-}^0-q_\tin{-}^0\big]
  +2m^2-8\mui^2\Big\} \\
V_{4,2}^{\tin{00}} &=&
    -2i\frac{m^2}{f^2}
        \big[2pql_4 +m^2(l_3 +l_4 -{\epsilon_{ud}}^2l_7)\big]\\
V_{4,2}^{\tin{-+}} &=&
  2i\frac{m^2}{f^2}\Big\{\big[-p_\tin{+}p_\tin{-}
    +\mui (p_\tin{+}^0-p_\tin{-}^0)+\mui^2\big]l_4
    -m^2(l_3+l_4)\Big\},
\end{eqnarray}
where $p$ and $q$ denote the external momentum of the $\pi^0$ legs and $p_\pm$
and $q_\pm$ denote the external momenta of the $\pi^\pm$ legs.
In all cases, the momentum emerges from the vertex.

\begin{figure}
\centering
\includegraphics[scale=1.1]{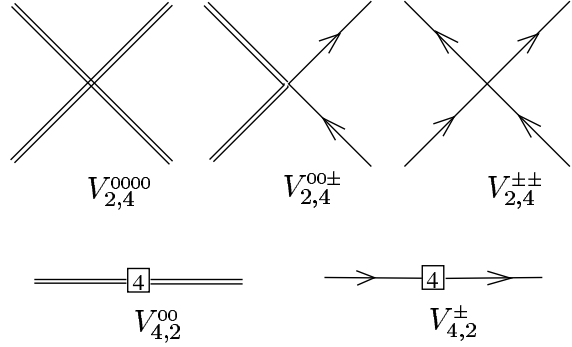}
\caption{\footnotesize  Vertices in the first phase (and in second phase for
$|\mui|\gtrsim m$).
The double line denotes the neutral pion leg. The single line represents the
charged pion leg with an arrow pointing outside from the vertex for the $\pi^+$
and into the vertex for the $\pi^-$ (or $\tilde\pi^+$ and $\tilde\pi^-$ in
second phase for $|\mui|\gtrsim m$)}\label{fig verts}
\end{figure}

Diagrammatically, using the convention for the propagators, a double line will
denote  a neutral pion leg and a single line a charged pions, where an arrow
pointing away from the vertex denotes a $\pi^+$ leg, and an arrow pointing
into  the vertex denotes a $\pi^-$, as can be seen in Fig. \ref{fig verts}.

\section{Second phase}

Let us denote the vacuum expectation value of the $U$ field as
\begin{equation}
\bar U = c+i\tilde\tau_1s,
\end{equation}
with
\begin{equation}
c\equiv\frac{m^2}{\mui^2},\qquad s\equiv\sqrt{1-\frac{m^4}{\mui^4}},
\end{equation}
where, for any vector, the
$\tilde{\mbox{v}}_i$ components refer to the components of a rotated vector by
the azimuthal angle $\phi$:
\begin{equation}
 \tilde{\mbox{v}}_1 =\mbox{v}_1\cos\phi +\mbox{v}_2\sin\phi,
 \qquad
 \tilde{\mbox{v}}_2 = -\mbox{v}_1\sin\phi +\mbox{v}_2\cos\phi,
 \qquad
 \tilde{\mbox{v}}_3 = \mbox{v}_3.
 \label{rotated-vectors}
\end{equation}
The vacuum expectation value of the $U$ fields breaks SU(2)$\rightarrow$U(1).

The expanded Lagrangian, keeping all the details, is given by
\begin{eqnarray}
{\cal L}_{2,2}
 &=&
  \fr{1}{2}(\partial\bm{\pi})^2
   -\fr{1}{2}\mu_\tin{I}^2(s^2\tilde\pi_{\tin{1}}^2+\pi_{\tin{3}}^2)
  -|\mu_\tin{I}|c(\tilde\pi_{\tin{1}}\partial_{\tin{0}}\tilde\pi_{\tin{2}}
   -\tilde\pi_{\tin{2}}\partial_{\tin{0}}\tilde\pi_{\tin{1}})\\
{\cal L}_{2,3}
 &=&
  \frac{1}{f}|\mu_\tin{I}|s\big[(\tilde\pi_{\tin{1}}^2+\pi_{\tin{3}}^2)
   \partial_{\tin{0}}\tilde\pi_{\tin{2}}
   -\fr{1}{2}|\mu_\tin{I}|c\tilde\pi_{\tin{1}}\bm{\pi}^2\big]\\
{\cal L}_{2,4}
 &=&
  \frac{1}{6f^2}\Big\{\big[-(\partial\bm{\pi})^2
   +\mu_\tin{I}^2(s^2\tilde\pi_{\tin{1}}^2
    +\pi_{\tin{3}}^2
    -\fr{3}{4}c^2\bm{\pi}^2)\nonumber\\
   &&
\hspace{3cm}
+2|\mui|c(\tilde\pi_{\tin{1}}\partial_{\tin{0}}\tilde\pi_{\tin{2}}
    -\tilde\pi_{\tin{2}}\partial_{\tin{0}}\tilde\pi_{\tin{1}})\big]
\bm{\pi}^2+(\bm{\pi}\cdot\partial\bm{\pi})^2\Big\}\\
{\cal L}_{4,1}
 &=&
\frac{1}{f}\mu_\tin{I}^4cs[4s^2(l_1+l_2)-2c^2l_3-s^2l_4]\tilde\pi_{\tin{1}}
\\
{\cal L}_{4,2}
 &=&
  \frac{1}{f^2}\mu_\tin{I}^2\Big\{2s^2\big[(\partial\bm{\pi})^2
    +2(\partial_{\tin{0}}\tilde\pi_{\tin{2}})^2\big]l_1
    +2s^2\big[(\partial_{\tin{0}}\bm{\pi})^2
  +(\partial\tilde\pi_{\tin{2}})^2
    +(\partial_{\tin{0}}\tilde\pi_{\tin{2}})^2\big]l_2\nonumber\\
  && \qquad
-2s^2\big[8|\mu_\tin{I}|c\tilde\pi_{\tin{1}}\partial_0\tilde\pi_{\tin{2}}
  +\mu_\tin{I}^2(s^2\bm{\pi}^2
     -3c^2\tilde\pi_{\tin{1}}^2
     -\tilde\pi_{\tin{2}}^2)\big](l_1+l_2)\nonumber\\
  && \qquad +\mu_\tin{I}^2c^2\big[s^2\tilde\pi_{\tin{1}}^2-c^2\bm{\pi}^2\big]l_3
   +\big[c^2(\partial\bm{\pi})^2
        +2|\mu_\tin{I}|c(1-3c^2)
     \tilde\pi_{\tin{1}}\partial_{\tin{0}}\tilde\pi_{\tin{2}}\\
  && \quad\qquad   -\mui^2c^2\{(s^2+1)\bm{\pi}^2
      -(c^2-s^2)\tilde\pi_{\tin{1}}^2
      -\tilde\pi_{\tin{2}}^2\}\big]l_4
   +\mu_\tin{I}^2c^2\epsilon_{\tin{ud}}^2\pi_3^2l_7\Big\}.\nonumber
\end{eqnarray}
Here, $m^2=\mui^2c$ for further considerations, and I will
consider a negative chemical potential
($\mu_\tin{I}=-|\mu_\tin{I}|$) as is the case in neutron stars,
where a condensate of $\pi_-$ quasi-particles might exist
\cite{Pethick:2004}. The same happens for the $\pi_+$, for the
opposite sign of $\mu_\tin{I}$. Note that, in the second phase,
there vertices with three legs will appear from (${\cal
L}_{2,3}$), and with one leg from (${\cal L}_{4,1}$). The latter
is responsible for the counter-terms of the tadpoles, but  they
will not be considered in the approximation we will use next.

\begin{figure}
\centering
\includegraphics[scale=1.2]{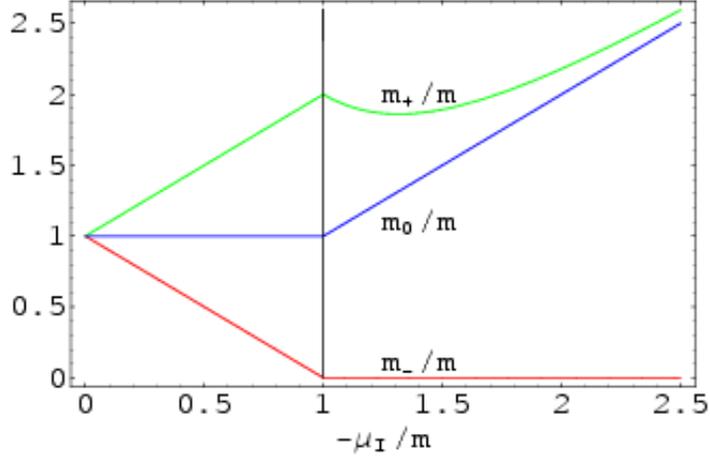}
\caption{\footnotesize Pion masses as a function of the isospin chemical
potential at
tree-level}
\label{figure.mtree}
\end{figure}

The inverse of the free propagator for pions in momentum space,
i.e., the kernel of equations of motion extracted from ${\cal L}_{2,2}$ is
given by the matrix
\begin{equation}
  i\bm{D}^{-1}
  =i\left(
   \begin{array}{ccc}
     D^{-1}_\tin{11} & D^{-1}_\tin{12} & 0             \\
     D^{-1}_\tin{21} & D^{-1}_\tin{22} & 0             \\
     0               & 0               & D^{-1}_\tin{33}
   \end{array}\right)
  =\left(
   \begin{array}{ccc}
     p^2-\mu_\tin{I}^2s^2 & 2i|\mu_\tin{I}|cp_0 & 0          \\
     -2i|\mu_\tin{I}|cp_0 & p^2           & 0          \\
     0              & 0             & p^2-\mu_\tin{I}^2
   \end{array}
  \right)
\end{equation}
To extract the tree-level masses, we calculate the poles of the determinant of
$D$ at zero momentum as is prescribed in \cite{Kogut:2001id}
\begin{equation}
\left|i\bm{D}^{-1}(p)\right|_{\bm{p}=0} =
\big[p_0^2-\mui^2\big]\big[p_0^2-\mui^2(1+3c^2)\big]p_0^2 =0.
\end{equation}
We can identify the masses with respect to the results in Eq.
(\ref{tree level masses - phase1}), which have to match for $\mui=m$. Then, the
tree-level masses for the second phase are
\begin{equation}
m_0 = |\mui|, \qquad m_+=|\mui|\sqrt{1+3c^2}, \qquad m_-=0.
\end{equation}
In Figure \ref{figure.mtree}, we can see the behavior of the masses at
tree-level
in both phases as a function of a negative chemical potential (see
also \cite{Son:2000by} for the treatment at tree-level).

We can identify the $\pi_3=\pi_0$ field, as in
\cite{Kogut:2001id}, through  its mass at $|\mui|=m$ and because it is
diagonal in the propagator matrix.
Therefore, there will be no
difficulties in  handling this propagator,
\begin{equation}
D_{DJ}(p)_\tin{00} =\frac{i}{p^2 -\mui^2 +i\epsilon} +2\pi
n_B(p_0)\delta(p^2-\mui^2).
\label{D00}
\end{equation}

 Unfortunately, this is not the case for the charged pions, which are mixed in a
non-trivial way.\footnote{Although we can set $\phi=0$, I will continue using
the notation with a tilde to recall that the $\tilde\pi$ fields are not the
physical fields}

The free propagator for the charged pions in momentum space is
\begin{equation}
  \tilde{\bm{D}}
    =\frac{i}{[p^4-p^2\mui^2s^2-4\mui^2c^2p_0^2]}
\left(
   \begin{array}{cc}
      p^2            & -2i|\mui|cp_0 \\
      2i|\mui|cp_0  & p^2-\mui^2s^2\\
   \end{array}\right)
\end{equation}
with
$\tilde{\bm{D}}_{ij}\equiv\bm{D}_{ij\neq 3}$.
The Dolan-Jackiw propagators can be constructed with the general
formula indicated in Eq. (\ref{DJp formula}).

 The main
difficulty with this matrix propagator is that it is very
cumbersome to use it in the different loop corrections. This fact
motivates to proceed in a systematic way, through an expansion
in a new appropriate smallness parameter, namely $s$ when
$|\mui|\gtrsim m$ or $c$ when $|\mui|\gg m$.

It is not difficult to realize that the different vertices that
appear in our Lagrangian will correspond to different  powers of
$c$ or $s$, depending on the case. Although it could appear as a
trivial correction, we will keep only the zeroth order in all
calculations. As we will see, this procedure is not trivial at all
and provides us with interesting information about the behavior of
the pion masses as function of temperature and isospin chemical
potential.

If we scale all the parameters with $|\mu_\tin{I}|$ in all
structures needed for one-loop corrections, we have that:

\begin{eqnarray}
\textrm{Propagators} &\rightarrow&
D(p;T,\mu_\tin{I};m)=\frac{1}{\mu_\tin{I}^2}\bar{\bm{D}}(\bar
p;\bar T,c)\\
&&\nonumber\\
l_i ~ \textrm{constants} &\rightarrow&
l_i(\Lambda)=\frac{\gamma_i}{32\pi^2}\left[\bar l_i-\lambda
  -\ln (c\bar\Lambda^2)\right]\\
&&\nonumber\\
\textrm{Vertices} &\rightarrow&
V_{n,m}(p;\mu_\tin{I};f,m,\Lambda)
    =\frac{|\mu_\tin{I}|^n}{f^{m+n-4}}v_{n,m}(\bar p;\bar\Lambda,c)\\
&&\nonumber\\
\textrm{Integrals} &\rightarrow&
\int \frac{d^dk}{(2\pi)^d}\Lambda^{4-d}
    =\mu_\tin{I}^4\int\frac{d^d\bar k}{(2\pi)^d}\bar\Lambda^{4-d}\\
&&\nonumber\\
\textrm{Corrections} &\rightarrow&
\Sigma (p_0;T,\mu_\tin{I};f,m,\Lambda)
=\frac{\mui^4}{(4\pi f)^2}\sigma(\bar p_0;\bar T,c;\bar\Lambda)
\end{eqnarray}
 where
from now on the bar on a parameter means that it is scaled with
$|\mu_\tin{I}|$.  It is possible then to expand the propagator in
powers of $s^n$ for$|\mu_\tin{I}|\gtrsim m$  and $c^n$ when
$|\mu_\tin{I}|\gg m$. Doing the same with the vertices, the resulting radiative
corrections will be expressed in powers of $s$ and $c$.
The problem with divergences is then also fixed since they appear in powers of
$c$ and $s$.

As we said before, we will keep only the $n=0$ terms in the previous expansions.
This approximation is non-trivial, since it allows us to explore the behavior of
the renormalized masses precisely in the vicinity of the transition point and
also in the region of high values of the chemical potential.
Further corrections could also be calculated, taking into account higher order
vertices and propagators.

We will avoid the region where $c\sim s$ (i.e. arround
$c=s=\frac{1}{\sqrt{2}}$) and consider only the region where $c\rightarrow 0$
or $s\rightarrow 0$.
This excluded area becomes smaller when $n$ (the order
of the expansion) starts to grow.
By demanding that
$c^{n+1},s^{n+1}\lesssim \frac{1}{2}(\frac{1}{\sqrt{2}})^{n+1}$,
we achieve this condition. Note that if $n=0$, we would exclude
the region of the chemical potential where $\sqrt{\frac{8}{7}}m^2
\lesssim \mui^2 \lesssim \sqrt{8}m^2$.
We remark again that by
going to higher orders in our expansion, the excluded region
becomes smaller, so this is the best bound.

\subsection{Propagators and vertices  at order $s^0$}

As was said before, the neutral pion propagator is diagonal with respect to the
charged ones.
Its propagator will be the same in both limits at
any order.

The propagator for the charged pions at ${\cal O}(s^0)$ is given
by
\begin{equation}
  \tilde{\bm{D}}
    =\frac{i}{[p^4-4\mu_\tin{I}^2p_0^2]}
\left(
   \begin{array}{cc}
      p^2             & -2i|\mu_\tin{I}|p_0 \\
      2i|\mu_\tin{I}|p_0  & p^2\\
   \end{array}\right)+\frac{1}{\mu_\tin{I}^2}{\cal O}(s^2)
\end{equation}
It is more convenient to work with the combination of
fields
\begin{equation}
\tilde\pi_{\pm}  =\fr{1}{\sqrt{2}}(\tilde\pi_{\tin{1}}\mp
i\tilde\pi_{\tin{2}}),
\end{equation}
where the tilde tilde convention was described in equation
(\ref{rotated-vectors})
We remark that the $\tilde\pi_{\pm}$ fields do not correspond to
the physical charged pion fields but to a combination of them.
This fact is a consequence of the non-trivial vacuum structure in
this phase. This combination is also not trivial due to derivative
terms and the inverse D'Alembertian operator.
I would like to mention that, in the first phase ($\mui < m$), the $\pi_\pm$
correspond effectively to the charged pions and the propagator
becomes anti-diagonal.
 Then, the inverse propagator for charged pions are
\begin{equation}
 i\tilde{\bm{D}}^{-1}
   = i\left(
  \begin{array}{cc}
   D^{-1}_\tin{++} &  D^{-1}_\tin{+-} \\
                   &                  \\
   D^{-1}_\tin{-+} &  D^{-1}_\tin{--}
  \end{array}\right)
  = \left(
   \begin{array}{cc}
    -\fr{1}{2}\mu_\tin{I}^2s^2
    & p^2+2|\mu_\tin{I}|cp_0-\fr{1}{2}\mu_\tin{I}^2s^2 \\
    &                                       \\
    p^2-2|\mu_\tin{I}|cp_0-\fr{1}{2}\mu_\tin{I}^2s^2
    & -\fr{1}{2}\mu_\tin{I}^2s^2
   \end{array}\right)
\end{equation}
and the free propagator matrix is
\begin{equation}
  \tilde{\bm{D}} = i\left(
   \begin{array}{cc}
      0                         & \frac{1}{p^2-2|\mu_\tin{I}|p_0} \\
      \frac{1}{p^2+2|\mu_\tin{I}|p_0}  &                         0\\
   \end{array}\right)+\frac{1}{\mu_\tin{I}^2}{\cal O}(s).
\end{equation}
The Dolan-Jackiw propagators for charged pions, including isospin
chemical potential (enough for one-loop calculations), are
\begin{eqnarray}
  D_{DJ}(p)_{\tin{+-}} &=&
  \frac{i}{p^2-2|\mu_\tin{I}|p_0+i\epsilon} +2\pi n_B(|p_0|)\delta
(p^2-2|\mu_\tin{I}|p_0) +\frac{1}{\mu_\tin{I}^2}{\cal O}(s^2),\\
 D_{DJ}(p)_\tin{-+} &=& D_{DJ}(-p)_\tin{+-},
\end{eqnarray}
and the other propagators, $\bar D_\tin{++}$ and $\bar D_\tin{--}$  are of
order $s^2$, so we will neglect them.
Diagrammatically, the propagators in this limit are the same as in the first
phase (see Fig. \ref{fig props}).
Note that, in the case where $|\mui|=m$, the fields $\tilde\pi^\pm$ can be
identified as the physical fields $\pi^\pm$.

The relevant vertices at order $s^0$ are
\begin{eqnarray}
 V_{2,4}^\tin{0000} &=&
  i\frac{\mui^2}{f^2}\\
V_{2,4}^\tin{00+-} &=&
 \frac{i}{3f^2}\Big\{ 2pq
   +2p_\tin{+}p_\tin{-}
    -\big[p+q\big]
   \big[p_\tin{+}+p_\tin{-}\big]
  +|\mui|\big[p_\tin{+}^0-p_\tin{-}^0\big]
   -\mui^2\Big\}\\
V_{2,4}^{\tin{++--}} &=&  \frac{i}{3f^2}\Big\{
 -2p_\tin{+}q_\tin{+}
    -2p_\tin{-}q_\tin{-}
 +\big[p_\tin{+}+q_\tin{+}\big]
  \big[p_\tin{-}+q_\tin{-}\big]\nonumber\\
  &&\hspace{2cm}
 -4|\mui|\big[p_\tin{+}^0+q_\tin{+}^0
  -p_\tin{-}^0-q_\tin{-}^0\big]
  -6\mui^2\Big\} \\
V_{4,2}^{\tin{00}} &=&
    -2i\frac{\mui^2}{f^2}
        \Big\{2pql_4 +\mui^2(l_3 +l_4 -{\epsilon_{ud}}^2l_7)\Big\}\\
V_{4,2}^{\tin{-+}} &=&
  -2i\frac{\mui^2}{f^2}\Big\{\big[p_\tin{+}p_\tin{-}
    +|\mui| (p_\tin{+}^0-p_\tin{-}^0)]l_4
    +\mui^2l_3\Big\},
\end{eqnarray}
which, diagrammatically, are the same as the first phase (Fig.
\ref{fig verts}),
where $p$ and $q$ denote the external momenta of $\pi^0$ legs and $p_\pm$
and $q_\pm$ denote the external momenta of the $\pi^\pm$ leg.
In all cases, the momentum emerges from the vertex.
These vertices include higher terms of order $s$ and $s^2$, and other vertices
also appear:
$v_{2,2}^{\tin{\pm\pm}}$,
$v_{2,4}^{\tin{00\pm\pm}}$, $v_{2,4}^{\tin{\mp\pm\pm\pm}}$  of ${\cal O}(s^2)$
and
 $v_{2,3}^\tin{\mp\pm\pm}$,$v_{2,3}^\tin{00\pm}$, $v_{4,1}^\tin{\pm}$  of
 ${\cal O}(s)$. As was said, however, since we are
working at zeroth order, these terms will not be considered in the analysis.

\subsection{Propagators and vertices  at order $c^0$}

Proceeding in the same way as in the $|\mu_I|\gtrsim m$ case, the
propagator of the charged pions $\tilde\pi_{\tin{1}}$ and
$\tilde\pi_{\tin{2}}$ at zero temperature at ${\cal O}(c^0)$ in the region
$|\mui|\gg m$ is
\begin{equation}
 \tilde{\bm{D}}
    =i
\left(
   \begin{array}{cc}
      \frac{1}{p^2-\mu_I^2} & 0 \\
                     0      & \frac{1}{p^2} \\
   \end{array}\right)+\frac{1}{\mu_I^2}{\cal O}(c).
\end{equation}

\begin{figure}
\centering\includegraphics[scale=1.2]{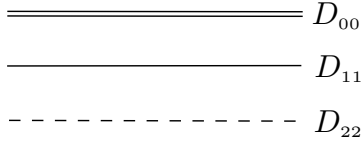}
\caption{ \footnotesize Propagators in the second phase
for \mbox{$|\mui|\gg m$.}}
\label{fig propc}
\end{figure}

A was said before, the $D_{\tin{00}}$ propagator in Eq. (\ref{D00}) remains the
same in all cases. The
Dolan-Jackiw propagators for charged pions are
\begin{eqnarray}
D_{DJ}(p)_\tin{11} &=& D_{DJ}(p)_\tin{00}+\frac{1}{\mu_I^2}{\cal
O}(c^2),\\
D_{DJ}(p)_\tin{22} &=& \frac{i}{p^2+i\epsilon} +2\pi
n_B(|p^0|)\delta (p^2)+\frac{1}{\mu_I^2}{\cal O}(c^2),
\end{eqnarray}
 and the propagators $\bar D_{\tin{12}}$ and $\bar D_{\tin{21}}$ are of order
$c$.
In the case of a propagator with zero mass, it is necessary to introduce a
small fictitious mass as a regulator, i.e. $D_{\tin{22}}^{-1}=\lim_{\eta\to
0}(p^2-\eta^2)$.
Note that, in  the chiral limit ($c=0$), the fields $\pi_{\tin{1}}$
and $\pi_{\tin{0}}$ have the same behavior. The other components
of the propagator matrix are of order $c$.
Diagrammatically, we
will denote the $D_{\tin{11}}$ propagator with a line and
$D_{\tin{22}}$ with a dashed line, as we can see in Figure
\ref{fig propc}. Note that when $|\mui|\rightarrow\infty$, the fields
$\tilde\pi^1$ and $\tilde\pi^2$ can be identified as the physical fields $\pi^+$
and $\pi^-$, respectively.

\begin{figure}
\centering
\includegraphics[scale=1.0]{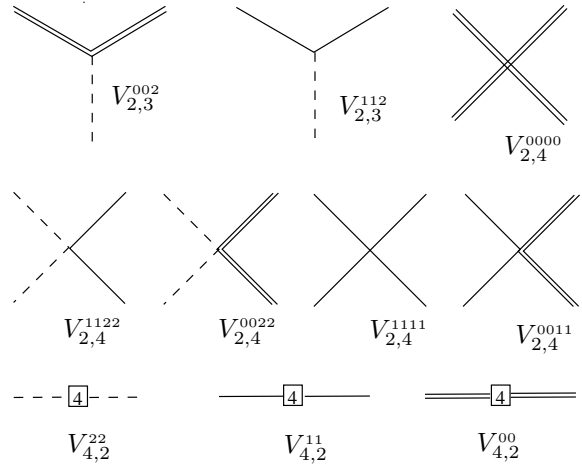}
\caption{\footnotesize  Vertices in the second phase for $|\mui|\gg m$).
The double line denotes the $\pi^0$ leg. The single line represents the
 $\tilde\pi^1$ and the dashed line represent the $\tilde\pi^2$ leg.}
\label{fig vertc}
\end{figure}

The relevant vertices at ${\cal O}(c^0)$ are indicated in Figure
\ref{fig vertc} with the analytical expressions
 \begin{eqnarray}
 V_{2,3}^{\tin{002}} &=& V_{2,3}^{\tin{112}} =
   \frac{2|\mui|}{f} p_\tin{(2)}^0\\
 V_{2,4}^{\tin{0000}} &=&
  i\frac{\mu_I^2}{f^2}\\
 V_{2,4}^{\tin{0011}} &=&
  \frac{i}{3f^2}\Big\{[2p_\tin{(0)}q_\tin{(0)}
    +2p_\tin{(1)}q_\tin{(1)} -\big[p_\tin{(0)} +q_\tin{(0)}\big]
    \big[p_\tin{(1)} +q_\tin{(1)}\big]+ 4\mui^2\Big\}\\
 V_{2,4}^{\tin{0022}} &=&
  \frac{i}{3f^2}\Big\{[2p_\tin{(0)}q_\tin{(0)}
    +2p_\tin{(2)}q_\tin{(2)} -\big[p_\tin{(0)} +q_\tin{(0)}\big]
    \big[p_\tin{(2)} +q_\tin{(2)}\big]+ 2\mui^2\Big\}\\
 V_{2,4}^{\tin{1122}} &=&
  \frac{i}{3f^2}\Big\{[2p_\tin{(1)}q_\tin{(1)}
    +2p_\tin{(2)}q_\tin{(2)} -\big[p_\tin{(1)} +q_\tin{(1)}\big]
    \big[p_\tin{(2)} +q_\tin{(2)}\big]+ 2\mui^2\Big\}\\
V_{4,2}^\tin{00}&=&V_{4,2}^{\tin{11}}=-4i\frac{\mu_I^2}{f^2}\Big\{pql_1
+p^0q^0l_2 +\mui^2(l_1+l_2)\Big\}\\
V_{4,2}^{\tin{22}}&=&-i4\frac{\mui^2}{f^2}\big[pq+2p^0q^0\big](l_1+l_2).
\end{eqnarray}
These vertices include higher corrections of order $c$ and $c^2$.
As in the $|\mui |\gtrsim m$ case, other
vertices of higher order in powers of $c$ also appear: $\bar
v_{2,2}^{\tin{$12$}}$, $v_{2,3}^{\tin{$001$}}$,
 $v_{2,3}^{\tin{$111$}}$, $v_{2,3}^{\tin{$122$}}$,
 $v_{2,4}^{\tin{$0012$}}$, $v_{2,4}^{\tin{$1112$}}$,
 $v_{2,4}^{\tin{$1222$}}$, $v_{4,2}^{\tin{$12$}}$, of ${\cal O}(c)$.

\chapter{Thermal pions in the first phase}\label{phase1}
This chapter will show how to compute radiative corrections to the masses,
condensates and
to the PCAC relation which allows us to extract the pion decay constant.

\section{Masses}

In the previous section, we extracted from the effective Lagrangian the
different vertices and propagators involved in the self-energy corrections.
Using the fact that the anti-diagonal terms of the propagator matrix are the
same, except that they have opposite flow of momentum, 
\begin{equation}
\Sigma(p)_\tin{ij}=\Sigma(-p)_\tin{ji},
\end{equation}
we just need to calculate $\Sigma_\tin{00}$ and $\Sigma_\tin{-+}$.
The loop contributions to these self-energies are shown in Fig. \ref{self-energy
O(s^0)}

\begin{figure}
\begin{minipage}[b]{.5\textwidth}
\centering\fbox{
\includegraphics[scale=1]{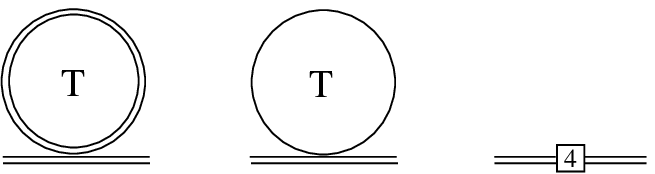}}
\hspace{.5cm}{\bf a.}
\end{minipage}
\begin{minipage}[b]{.5\textwidth}
\centering\fbox{
\includegraphics[scale=1]{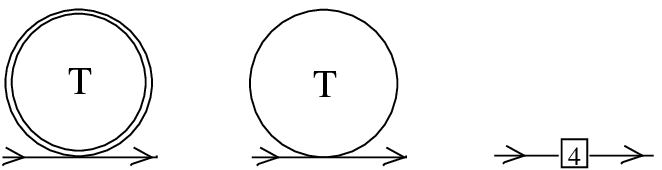}}
\hspace{.5cm}{\bf b.}
\end{minipage}
\caption{\footnotesize Loop contributions to {\bf a.} $\Sigma_\tin{00}$
and  
{\bf b.} $\Sigma_\tin{-+}$}
\label{self-energy O(s^0)}
\end{figure}

The calculation of the different loop corrections is standard. For denominators 
of the form $(k+b u)^2-a^2$ (or delta functions in the thermal insertions), we
just change the variables of integration $k_\mu +bu_\mu\rightarrow k'_\mu$. 
The calculation of the different kinds of integrals can be found in appendices
\ref{dim_reg} and \ref{formulas.thermal_ints}.

Defining $\underline\lambda \equiv \lambda -\ln (m^2/\Lambda^2)$, the resulting
self-energy corrections are
\begin{eqnarray}
\Sigma (p)_\tin{00} 
&=& \alpha\Big\{ p^2 \big[
  \fr{4}{3}\underline\lambda -2\bar l_4 +\fr{8}{3}I\big] -m^2 \big[
  \fr{4}{3}\underline\lambda -2\bar l_4 +\fr{1}{2}\bar l_3
  +32\pi^2\epsilon_{ud}^2l_7 +2I_0 -\fr{4}{3}I\big]\Big\},\nonumber\\
  &&\\
\Sigma (p)_\tin{-+} 
&=& \alpha \Big\{ ( p +\mui u)^2 \big[
  \fr{4}{3}\underline\lambda -2\bar l_4 +\fr{4}{3} ( I_0 +I)\big] +8
( p_0 +\mui)\epsilon(\mui)mJ\nonumber\\
&& \hspace{4cm} -m^2 \big[ \fr{4}{3}\underline\lambda
  -2\bar l_4 +\fr{1}{2}\bar l_3 -\fr{2}{3} ( I_0 -2I)\big]\Big\},\\
\Sigma (p)_\tin{+-} &=& \Sigma (-p)_\tin{-+},
\end{eqnarray}
where  $\alpha = (m/4\pi f)^2$ is the perturbative term that
fixes  the scale of energies in the theory (for energies below
$4\pi f $). The functions $I$, $J$ and $I_n$ are defined as
follows:
\begin{eqnarray}
I   &=& \int_1^\infty dx\sqrt{x^2-1}
        [n_B(mx-|\mui|)+n_B(mx+|\mui|)]\\
J   &=& \int_1^\infty dxx\sqrt{x^2-1}
        [n_B(mx-|\mui|)-n_B(mx+|\mui|)]\\
I_n &=&  \int_1^\infty dxx^{2n}\sqrt{x^2-1}~2n_B(mx).
\end{eqnarray}

In all these radiative corrections, since the loops do not carry external
momentum, it is not necessary to set $\bm{p}=0$. 
The resulting self-energy corrections are proportional to $p^2$, $p_0$ and
constants. 
Proceeding with the renormalization of the corrected propagators, described in
section \ref{renormalization}, we choose the renormalization constants in
such a way that they absorb corrections proportional to $p^2$.
Taking
\begin{eqnarray}
Z_0 &=& 1 +\alpha \big[ \fr{4}{3}\underline\lambda -2\bar l_4
  +\fr{8}{3}I\big], \label{phase1.Z_0}\\
Z_\pm &=& 1 +\alpha \big[ \fr{4}{3}\underline\lambda -2\bar l_4
  +\fr{4}{3} ( I_0 +I)\big],\label{phase1.Z_pm} 
\end{eqnarray}
we have that
\begin{eqnarray}
\Sigma_R(p)_\tin{00} &=& \alpha m^2 \big[ -\fr{1}{2}\bar l_3
  -32\pi^2\epsilon_{ud}^2l_7 -2I_0 +4I\big],\\
\Sigma_R(p)_\tin{-+} &=& \alpha \Big\{ ( p_0 +\mui) \epsilon(\mui)8mJ
+m^2 \big[ -\fr{1}{2}\bar l_3 +2I_0\big]\Big\},\\
\Sigma_R(p)_\tin{+-} &=& \Sigma_R(-p)_\tin{-+}.
\end{eqnarray}

It is important to remark that radiative corrections will leave a
dependence on the chemical potential for the pion mass only for
finite values of temperature. 
In a strict sense, this procedure does not allow us to say anything new about
an eventual chemical potential dependence of the masses at $T=0$ (cold matter),
which is already included in ${\cal L}_2$.

There are no significant difficulties in extracting the masses from the
renormalized propagator matrix, since there are no mixed terms, so the
determinant of the renormalized propagator is
\begin{equation}
\left|\bm{D}_R^{-1}(p)\right|
=D_R^{-1}(p)_\tin{00}~D_R^{-1}(p)_\tin{-+}~D_R^{-1}(p)_\tin{+-}. 
\end{equation}
For finite $T$ and $\mu_I$, we find the following expressions for
the masses :
\begin{eqnarray}
m_{\pi^0}(T,\mui) &=& m\Big\{1 +\alpha\big[-\bar l_3/4
  -16\pi^2\epsilon_{ud}^2l_7 -I_0 +2I\big]\Big\},\label{m_pi0(T,mu)}\\
m_{\pi^+}(T,\mui) &=& m\Big\{1 +\alpha\big[-\bar l_3/4 +I_0
  +4\epsilon(\mui)J\big]\Big\} -\mui,\label{m_pi+(T,mu)}\\
m_{\pi^-}(T,\mui) &=& m\Big\{1 +\alpha\big[-\bar l_3/4 +I_0
  -4\epsilon(\mui)J\big]\Big\} +\mui.\label{m_pi-(T,mu)}
\end{eqnarray}

The term $16\pi^2{\epsilon_{ud}}^2l_7$ that appear in the $m_{\pi^0}$
corrections, which value runs between 0.03 and 0.35 can be neglected.

We can see that, at zero temperature and chemical potential, we obtain the usual
pion mass
\begin{equation}
m_\pi = m[1-\alpha\bar l_3/4]\approx 140~\mbox{MeV}.
\label{phase1.m_pi}
\end{equation}
The different values for the tree-level constants can be found in App.
\ref{f-B-e}.

If the chemical potential of the charged pions vanishes, i.e for
symmetric matter,  at finite $T$ we get the well known result 
\begin{equation}
m_{\pi }(T)=m\Big\{1 +\alpha\big[-\bar l_3/4 +I_0\big]\Big\}
\label{m_pi(T)}
\end{equation}
of chiral perturbation theory \cite{Gasser:1986vb}, see
also \cite{Larsen:1985ei,Contreras:1989gi}.
However, due to radiative corrections to the neutral pion propagator, its mass
will acquire a non-trivial chemical potential dependence for finite values of
temperature.

\begin{figure}
\begin{center}
\includegraphics[scale=1.2]{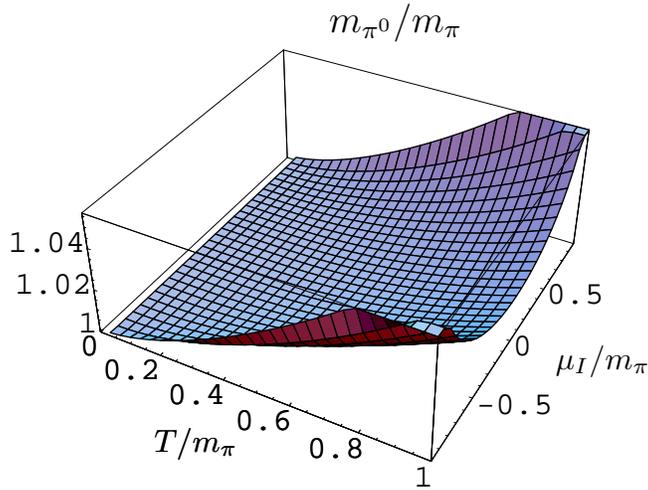}
\end{center}
\caption{\footnotesize  $m_{\pi^0}$ as a function of $T$ and $\mui$ in
units
of $m_\pi$.}
\label{fig.phase1.m03D}
\end{figure}

Figure \ref{fig.phase1.m03D}  shows the dependence of the $\pi^0$ mass as
a function of temperature and isospin chemical potential. It is a growing
function of both variables,  temperature and $|\mui|$.

\begin{figure}
\centering
\includegraphics[scale=.9]{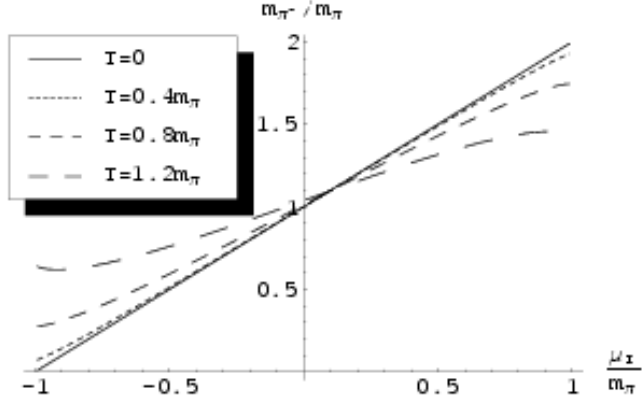}
\caption{\footnotesize  $m_{\pi^-}$ as a function of $\mui$ for fixed
values of
$T$ in units of $m_\pi$.}
\label{fig.phase1.mm}
\end{figure}

From Figure \ref{fig.phase1.mm} we see that, at zero temperature, we
agree with the usual prediction, $m_{\pi^-} = m_{\pi } + \mui$. 
In fact, at zero temperature, the $\pi ^{-}$ should condense when
$\mui = -m_{\pi }$ (the inverse situation occurs for $\pi ^{+}$). 
Now, this situation changes if temperature starts to grow. 
The condensation point no longer occurs at $\mui =-m_\pi$, but it moves to the
left, i.e. to a bigger (in absolute value) chemical potential. 
In the vicinity of  $\mui =m_\pi$ the mass  starts to decrease.
This figure shows the $\pi^-$ mass in the normal phase only, $|\mui|<m_\pi$.
Even though we can consider higher values for the isospin chemical potential in
the first phase, since $m>m_\pi$ (see appendix \ref{f-B-e}), this situation
will be discussed in the next chapter. 

For the case of the $\pi^+$ mass, the behavior is the same, exchanging $\mui$
for $-\mui$.

\section{Decay constant}

In connection with the behavior of $f_{\pi }(T,\mui)$ when $\mui <m_{\pi }$,
we have made use of PCAC, which provides us with a relation between the
renormalized propagator and the pion decay constant.

The axial current is obtained as the functional derivative of the action with
respect to $a_\mu^a$, with $a_\mu=a_\mu^a\tau^a/2$
\begin{equation}
A_\mu^a = \frac{\delta S_\chi}{\delta a_\mu^a}(M,0,\mui u\fr{1}{2}\tau_3,0)
\end{equation}
The expansion of the axial current in terms of powers of the pion fields can be
found in detail in Appendix \ref{expansion.axial}.
Defining $A^\pm_\mu \equiv \frac{1}{\sqrt{2}}\big(A^1_\mu \mp iA^2_\mu\big)$,
the axial current is
\begin{eqnarray}
A_{(1,1)}^0 &=& -f\partial\pi^0\\
A_{(1,1)}^\pm  &=& -f(\partial_{I}\pi)^\pm\nonumber\\
A_{(1,3)}^0 &=& \frac{2}{3f}\big[2|\pi|^2\partial\pi^0
  -\pi^0\partial|\pi |^2\big]\\ 
A_{(1,3)}^\pm    &=& \frac{1}{3f}\big[2(\pi_0^2
  +2|\pi|^2)(\partial_{I}\pi)_\pm -\pi_\pm\partial(\pi_0^2
  +2|\pi|^2)\big]\\  
A_{(3,1)}^0  &=& -\frac{1}{f}2m^2l_4 \partial\pi^0\\
A_{(3,1)}^\pm    &=& -\frac{1}{f}2m^2l_4(\partial_{I}\pi)^\pm 
\end{eqnarray}
\begin{figure}
\centering
\includegraphics[scale=1]{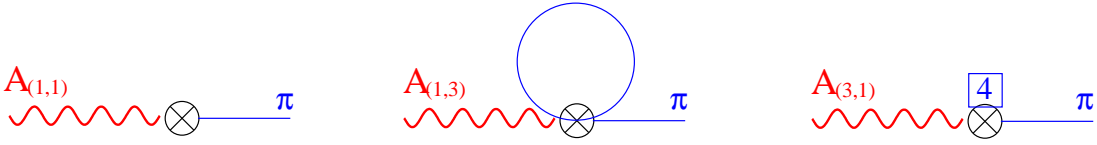}
\caption{\footnotesize Diagramatic contribution to the PCAC relation at one-loop
corrections.}
\label{figure.PCAC}
\end{figure}

Saturating the axial current with a single pion (the reduction formula can be
found in Appendix \ref{LSZ}), we obtain
\begin{eqnarray}
\llang 0|A_\mu^0|\pi_R^0(p)\rrang &=& ip_\mu f\Big\{\sqrt{Z_0}
+\alpha\big[-\fr{2}{3}\underline\lambda +2\bar l_4-\fr{16}{3}I\big]\Big\},\\
\llang 0|A^\pm|\pi_R^\mp(p)\rrang &=&
i(p \pm \mui u)f\Big\{\sqrt{Z_\pm} +\alpha\big[-\fr{2}{3}\underline\lambda
+2\bar l_4-\fr{8}{3}(I_0+I)\big]\Big\}\nonumber\\
&&\mp iu_\mu f\epsilon(\mui)\alpha 8mJ.\label{PCAC.pm-mp}
\end{eqnarray}

\begin{figure}
\centering
{\bf a.}\includegraphics[scale=1.1]{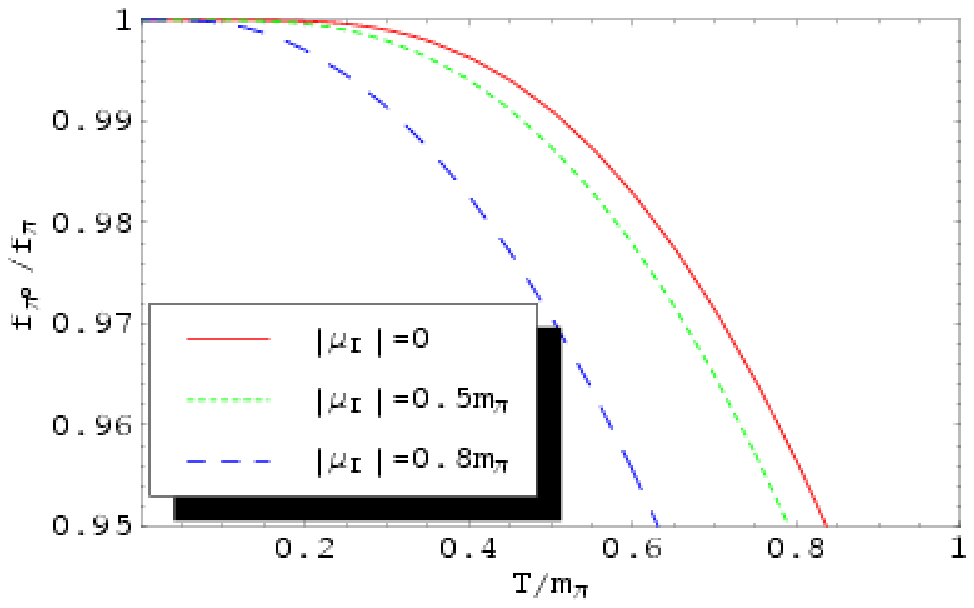}
{\bf b.}\includegraphics[scale=1.1]{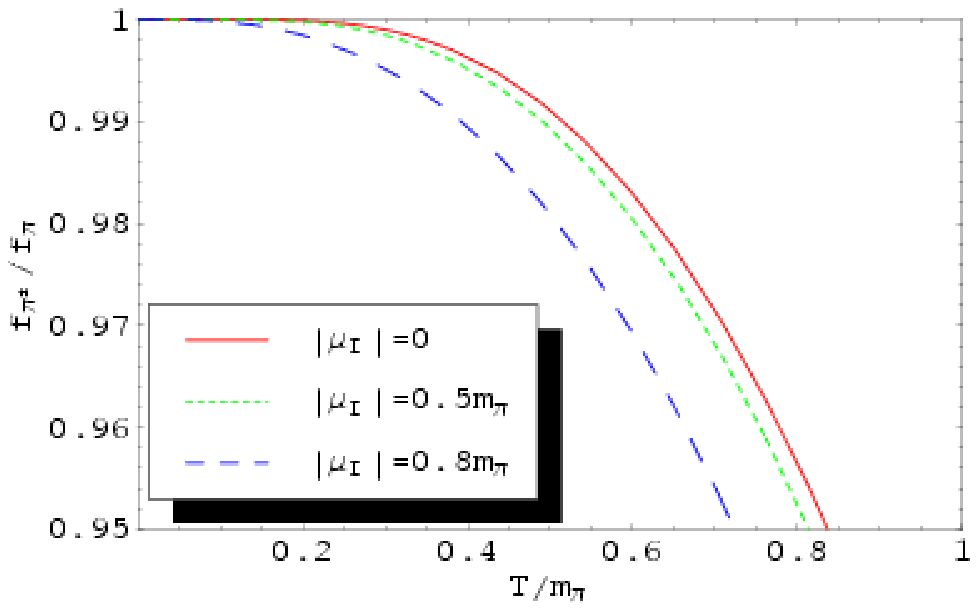}
\caption{\footnotesize {\bf a.} Neutral pion decay constant and {\bf b.} charged
pions decay constant at finite temperature and isospin chemical
 potential}\label{figure.f}
\end{figure}

The different contributions to the PCAC relation in the expansion of the axial
current are shown in Figure \ref{figure.PCAC}.
Using the values of the renormalization constants $Z_a$ in equations
(\ref{phase1.Z_0}) and (\ref{phase1.Z_pm}), the renormalized PCAC relation is
\begin{eqnarray}
\llang 0|A^0|\pi_R^0(p)\rrang &=& ipf\Big\{1
+\alpha\big[\bar l_4 -4 I\big]\Big\},\\
\llang 0|A^\pm|\pi_R^\mp(p)\rrang &=&
i(p \pm \mui u)\Big\{1 +\alpha\big[\bar l_4
  -2(I+I_0)\big]\Big\} \mp iu \epsilon(\mui)8\alpha fmJ.
\end{eqnarray}
Then, we define the effective decay constant as the part proportional to
$p_\mu$, so
\begin{eqnarray}
f_{\pi^0}(T,\mu_I)&\equiv& f\Big\{1
+\alpha\big[\bar l_4 -4 I\big]\Big\},\label{f-pi0(T,mu)}\nonumber\\
f_{\pi^\pm}(T,\mu_I)&\equiv& f\Big\{1 +\alpha\big[\bar l_4
  -2(I+I_0)\big]\Big\}.\label{f_pipm(T,mu)}
\end{eqnarray}
We can see that, at zero temperature and chemical potential, we obtain the usual
pion decay constant
\begin{equation}
f_{\pi}= f\big[1+\alpha\bar l_4\big]\approx 92.5\mbox{MeV}.
\label{phase1.f_pi}
\end{equation}

For an increasing chemical potential, the couplings $f_{\pi }(T,\mui)$ decrease
faster (Fig. \ref{figure.f}).
This effect is enhanced for $f_{\pi _{0}}(T,\mui)$ (Fig. \ref{figure.f}.a)
and is related to the fact that $f_{\pi ^0}(T,\mui)$ only  receives radiative
corrections from charged pion loops; in contrast, $f_{\pi^\pm}(T,\mui)$
receives both loop contributions, from charged pions and for neutral pions. 
For the case of zero chemical potential, $f_\pi(T)$ is again the same one for
the three pions,
\begin{equation}
f_\pi(T)=f\left[1+\alpha\left(\bar l_4-4I_0\right)\right].
\label{f_pi(T)}
\end{equation}

\section{Condensates.}

There are three kinds of condensates relevant for the present analysis: the
chiral condensate $\llang\bar qq\rrang$, responsible for the spontaneous chiral
symmetry breaking; the pion condensate
$\llang\bm{\pi}\rrang$, which gives us information about the pions in the
condensed phase; and the isospin number density $\llang n_I\rrang$, which
tells us about the difference of $u$ and $d$ baryon number inside the
thermal bath.
The expansions of the currents are explained in detail in Appendix
\ref{expansion.currents}. In the case of the first phase, as is expected, the
pion condensate is equal to zero.

\subsection{Chiral condensate.}

The chiral condensate can be defined as the expectation value of the scalar
current in the vacuum, $\llang\bar qq\rrang =\llang 0|J_s|0\rrang$, with the
scalar current for massive quarks at finite isospin chemical potential defined
as
\begin{equation}
J_s =\frac{\delta S_\chi}{\delta s^0}(M,0,\fr{1}{2}\mui u\tau_3,0)
\end{equation}

The components needed for radiative corrections are
\begin{eqnarray}
\llang J_{s(1,0)}\rrang&=&-2Bf^2\\
\llang J_{s(1,2)}\rrang&=&B\llang 0|\pi_0^2+2|\pi|^2|0\rrang\\
\llang J_{s(3,0)}\rrang&=&-4Bm^2(l_3+l_4+\tilde h_1).
\end{eqnarray}
Using the fact that 
\begin{equation}
\llang 0|\pi_a(x)\pi_b(x)|0\rrang=\llang
0|T\pi_a(x)\pi_b(y)|0\rrang_{y=x},
\end{equation}
the corrected chiral condensate turns out to be
\begin{equation}
\llang\bar qq\rrang  (T,\mui)=-2Bf^2\left\{1 
-\alpha\left[\fr{1}{2}\bar l_3-2\bar l_4-32\pi^2\tilde
h_1+2I_0+4I\right]\right\},
\label{phase1.qq(T,mu)}
\end{equation}
where the chiral condensate at zero temperature and chemical potential is
\begin{equation}
\langle\bar qq\rangle= -2Bf^2\big[1-\alpha\big(\fr{1}{2}\bar l_3-2\bar
l_4-32\pi^2\tilde h_1\big)\big],
\label{phase1.qq}
\end{equation}
and the chiral condensate at finite temperature and zero chemical potential is
\begin{equation}
\llang\bar qq\rrang  (T)=-2Bf^2\left\{1 
-\alpha\left[\fr{1}{2}\bar l_3-2\bar l_4-32\pi^2\tilde h_1+6I_0\right]\right\}.
\label{qq(T)}
\end{equation}

The constant $\tilde h_1$, as was said before, is model dependent. If we accept
the G-MOR relation 
\begin{equation}
m_\pi^2f_\pi^2=-\frac{1}{2}(m_u+m_d)\langle\bar qq\rangle,
\end{equation} 
neglecting the term $\epsilon_{ud}^2l_7$ that appears in the $m_{\pi^0}$
corrections, we find from equations (\ref{phase1.m_pi}) and (\ref{phase1.f_pi})
that
\begin{equation}
m_\pi^2f_\pi^2=(m_u+m_d)Bf^2\big[1-\alpha(\fr{1}{2}\bar l_3 -2\bar l_4)\big].
\end{equation}
Comparing with eq. (\ref{phase1.qq}), if we consider the G-MOR relation as
valid, we can set 
$\tilde h_1=0$\footnote{For higher corrections to the G-MOR relation
 see  \cite{Dominguez:1996kf}.}.

\begin{figure}[ht]
\centering
\includegraphics[scale=1]{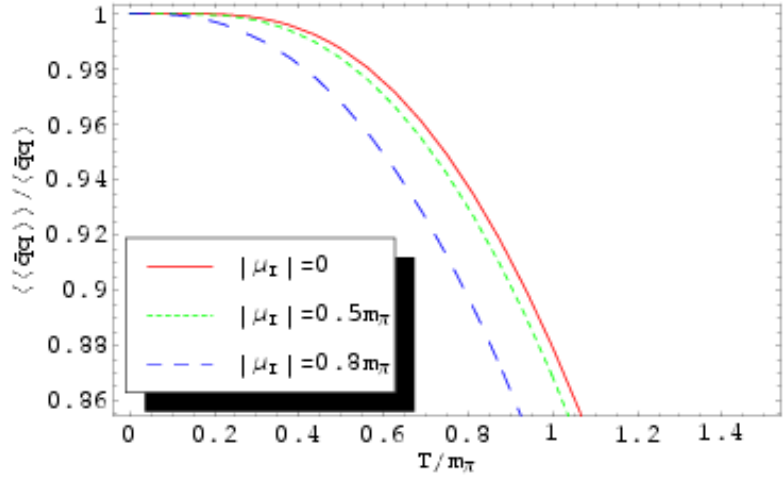}
\caption{\footnotesize Quark condensate as a function of the temperature
and the
isospin chemical potential}
\label{phase1.figure.qq}
\end{figure}

Figure \ref{phase1.figure.qq} shows the chiral condensate at
finite temperature and isospin chemical potential. It has a similar 
behavior as the pion decay constants in Figure \ref{figure.f}.

\subsubsection{G-MOR relation at finite temperature and isospin chemical
potential}

If we set $\tilde h_1=0$, $\epsilon_{ud}^2l_7\sim 0$, from equations
(\ref{m_pi(T)}) and (\ref{f_pi(T)}) we obtain
\begin{equation}
m_\pi(T)^2f_\pi(T)^2=\left[1-\alpha\left(\fr{1}{2}\bar l_3-2\bar l_4
+6I_0\right)\right].
\end{equation}

Compared with equation (\ref{qq(T)}) we can see that the G-MOR relation is still
valid. 

At finite chemical potential, however, this relation in not possible,
since the states are not degenerate. This suggests to extend this relation using
the average value of the results. 

First of all, we have to take into account that the tree-level part of this
average cannot depend on the chemical potential. 
Observing equations
(\ref{m_pi+(T,mu)}) and (\ref{m_pi-(T,mu)}), we can see that the only
possibility is to use the average of the pion masses and the pion decay
constants. 
Defining
\begin{eqnarray}
\overline{m}
_\pi(T,\mui)&\equiv&\fr{1}{3}\left[m_{\pi^+}(T,\mui)
+m_{\pi^-}(T,\mui )+m_{\pi^0}(T,\mui)\right]\nonumber\\
&=& m\left\{1+\alpha\left[-\fr{1}{4}\bar
l_3-\fr{1}{3}32\pi^2\epsilon_{ud}^2l_7+\fr{1}{3}(I_0+2I)\right]\right\},\\
&&\nonumber\\
\overline{f}_\pi(T,\mui)&\equiv&\fr{1}{3}\left[f_{\pi^+}(T,\mui)
+f_{\pi^-}(T,\mui )+f_{\pi^0}(T,\mui)\right]\nonumber\\
&=&
f\left\{1+\alpha\left[\bar l_4-\fr{4}{3}(I_0+2I)\right]\right\},
\end{eqnarray}
and combining the quadratic terms, we find
\begin{equation}
\overline{m}_\pi(T,\mui)^2\overline{f}_\pi(T,\mui)^2
=(m_u+m_d)Bf^2\left\{1+\alpha\left[-\fr{1}{2}\bar l_3 +2\bar l_4
-\fr{1}{3}32\pi^2\epsilon_{ud}^2l_7 -2I_0-4I\right]\right\}.
\end{equation}
This result is exactly $-\fr{1}{2}(m_u+m_d)\llang\bar qq\rrang$ in equation
(\ref{phase1.qq(T,mu)}) but with $\tilde h_1 = \fr{1}{3}\epsilon_{ud}^2l_7$.
We can say that the generalized G-MOR relation is
\begin{equation}
\overline{m}_\pi^2\overline{f}_\pi^2=-\fr{1}{2}(m_u+m_d)\llang\bar qq\rrang.
\end{equation}

\subsection{Isospin-number density.}

The isospin-number density condensate gives us information about the baryon
number difference between $u$ and $d$ quarks in the thermal vacuum or populated
vacuum. The isospin-number density is defined as the 0-Lorentz and 3-isospin
component of the vector current
\begin{equation}
n_I\equiv \fr{1}{2}q^\dag\tau^3q =V_0^3.
\end{equation}

We need to calculate the expectation value of the vector current in the vacuum.
We will see that the only non-vanishing component of the vector current in the
vacuum is the isospin-number density.

Considering the expansion of the vector current explained in appendix
\ref{expansion.vector} and using the fact that 
\begin{equation}
\llang 0|\pi_a|0\rrang =0, \qquad \llang 0|\pi_{a\neq 3}\pi_3|0\rrang =0,
\end{equation}
 we find that the only non-zero component of the vector current in the vacuum,
for the first phase case, is
\begin{equation}
 \llang\bm{V}_{(1,2)\mu}\rrang = \llang
0|-i(\pi_\tin{+}\partial_\mu\pi_\tin{-} -\pi_\tin{-}\partial_\mu\pi_\tin{+})
 +2\mui|\pi|^2u_\mu|0\rrang\bm{e}_3.
 \end{equation}
 Using the fact that 
 \begin{equation}
 \llang 0|\pi^a(x)\partial\pi^b(x)|0\rrang
 =\Big[\partial_y\llang 0|T\pi^a(x)\pi^b(y)|0\rrang\Big]_{y=x},
 \end{equation}

\begin{equation}
\llang\bm{V}_{\mu}\rrang = 8mf^2\alpha\epsilon(\mui)Ju_\mu\bm{e}_3
~=~\llang n_I\rrang u_\mu\bm{e}_3.
\end{equation}

\begin{figure}
\centering
\includegraphics[scale=1]{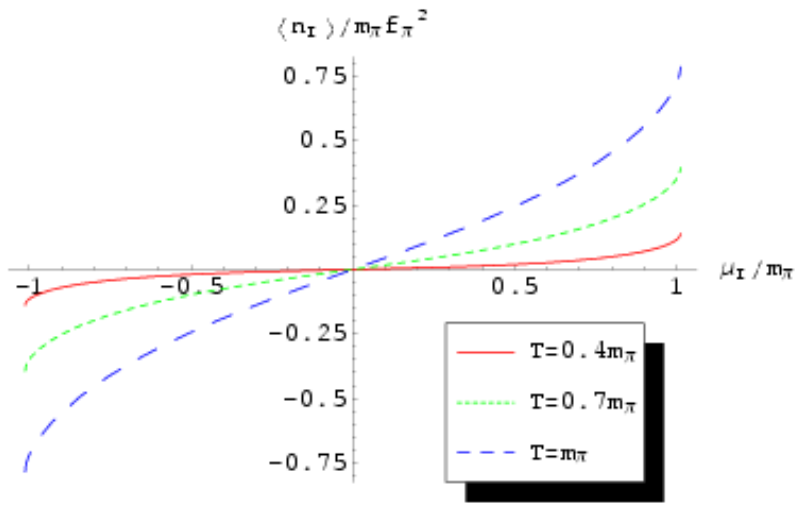}
\caption{\footnotesize Isospin-number density condensate as a function of the 
isospin chemical potential for different values of the temperature.}
 \label{figure.phase1.ni}
\end{figure}

Figure \ref{figure.phase1.ni} shows the isospin-number density expectation value
in the thermal vacuum. 
It gets a non vanishing value only inside the thermal bath.

If we compare the results of the PCAC relation for charged pions in equation   
(\ref{PCAC.pm-mp}), we find that the term proportional to $u_\mu$ is exactly
the isospin-number density condensate
\begin{equation}
\llang 0|A_\mu^\pm|\pi^\mp\rrang = i(p_\mu\pm\mui u_\mu)f_{\pi^\pm}(T,\mui)\mp
iu_\mu\llang n_I\rrang.
\end{equation}

\chapter{Thermal pions in the second phase for $|\mui|\gtrsim m$}\label{musimm}

In chapter \ref{prop}, I presented a systematic expansion method for radiative
calculations in the second phase. For the case of the chemical potential near
the transition point between the two phases, it is possible to expand the
radiative corrections in powers \mbox{of $s^2$}:
\begin{equation}
\Sigma(p;f,m;T,\mui)=\alpha'\sum_{n=0}\sigma_n(\bar p,\bar T)s^{2n}.
\end{equation}
 If we consider the lowest order in this expansion, ${\cal O}(s^0)$, we find
that the vertices and propagators are the same as in the case of the first
phase. The only difference is that the factors $m$ are changed by $|\mui|$. 
I remark that in this phase a negative chemical potential $\mui=-|\mui|$, will
be considered where the negative pion will condense.

\section{Masses}

Proceeding with the relevant vertices and propagators, the loop
corrections to the  pion propagator matrix are shown in Figures
\ref{fig props} and \ref{fig verts}, respectively.

Defining $\underline\lambda' = \lambda -\ln(\mui^2/\Lambda^2)$, the
self-energy is
\begin{eqnarray}
 \Sigma(p)_{\tin{00}}
  &=&\alpha'\Big\{-\mu_\tin{I}^2\big[ \fr{4}{3}\underline\lambda
   +\fr{1}{2}\bar l_3 -2\bar l_4+32\pi^2\epsilon-{ud}^2l_7
   +2I'_0-\fr{4}{3}I'\big]
   \nonumber\\
  && \hspace{6cm}
  +p^2\big[\fr{4}{3}\underline\lambda
   -2\bar l_4+\fr{8}{3}I'\big]\Big\},\\
 \Sigma(p)_{\tin{-+}}
  &=& \alpha'\Big\{ -\mu_\tin{I}^2\big[\fr{1}{2}\bar
l_3+32\pi^2\epsilon_{ud}^2l_7-2I'_0-8J'\big]
   +p^2\big[\fr{4}{3}\underline\lambda
    -2\bar l_4+\fr{4}{3}I'_0+\fr{4}{3}I'\big]
   \nonumber\\
   && \hspace{4.5cm} -2p_0|\mu_\tin{I}|\big[\fr{4}{3}\underline\lambda-2\bar l_4
    +\fr{4}{3}I'_0+\fr{4}{3}I'+4J'\big]\Big\},\\
 \Sigma(p)_{\tin{+-}}
  &=& \Sigma(-p)_{\tin{-+}}.
\end{eqnarray}
These self-energies include also higher correction terms of ${\cal O}(s
^2)$. Now $\alpha'=(\mui/4\pi f)^2$ is the
perturbative term, and the functions $I'$, $J'$, $I'_n$ are defined as follows:
\begin{eqnarray}
I' &\equiv& \int_1^\infty dx\sqrt{x^2-1}
 \Big[ n_B\big(|\mui |(x-1)\big) +n_B\big(|\mui |(x+1)\big)
  \Big],\\
J'&\equiv& \int_1^\infty dxx\sqrt{x^2-1}
 \Big[ n_B\big(|\mui |(x-1)\big)-n_B\big(|\mui |(x+1)\big)
  \Big],\\
I'_n &\equiv&
 \int_1^\infty dx\sqrt{x^2-1}x^{2n}2n_B\big(|\mu_\tin{I}|x\big).
\end{eqnarray}

Note that these definitions are almost the same ones we used in
the first phase. Here, however, the term that
multiplies $x$ in the argument of the Bose-Einstein distribution
is $\mui $ instead of $m$.

Following the prescription, indicated in section \ref{renormalization}, that the
renormalized self-energy does not depend on $p^2$, we have that the
renormalization constants are
\begin{eqnarray}
 Z_0 &=&  1+\alpha' \big[\fr{4}{3}\underline\lambda
   -2\bar l_4+\fr{8}{3}I'\big],\nonumber\\
 Z_\pm &=& 1+\alpha'\big[ \fr{4}{3}\underline\lambda
   -2\bar l_4 +\frac{4}{3}I'_0+\fr{4}{3}I'\big],
\end{eqnarray}
and the renormalized self-energy corrections are then
\begin{eqnarray}
\Sigma_{R\tin{00}} &=& \mui^2\alpha'
 \big[-\fr{1}{2}\bar l_3-32\pi^2\epsilon_{ud}^2l_7-2I'_0+4I'\big],\nonumber\\
\Sigma_{R}(p_0)_\tin{-+} &=& \alpha'\Big\{\mu_\tin{I}^2
 \big[-\fr{1}{2}\bar l_3+2I'_0+8J'\big]
    -8|\mu_\tin{I}|J'p_0\Big\},\nonumber\\
\Sigma_{R}(p_0)_\tin{+-}
  &=& \Sigma_{R} (-p_0)_\tin{-+},
\end{eqnarray}
plus higher corrections of order $\mui^2\alpha' s^2$. 
We will use these results to extract the renormalized temperature- and
chemical potential-dependent masses.

To extract the  $m_{\pi^\tin{0}}$ mass, we do not need any further
effort, since it is already diagonal in the matrix propagator, and
its renormalized self-energy is constant in the external momenta,
i.e., it does not depend on $p_0$:

\begin{equation}
m_{\pi^\tin{0}}=|\mui|\Big\{ 1+\alpha'\big[ -\fr{1}{4}\bar
l_3-16\pi^2\epsilon_{ud}^2l_7-I'_0+2I'\big]\Big\}.
\end{equation}

For the case of $m_{\pi^\tin{\pm}}$, the vanishing determinant of
the charged matrix propagator, provides us with a second-order
equation in powers of $p_0^2$:
\begin{equation}
 p_0^4-p_0^2\Big[ m_+^2 +2\Sigma_{R\tin{-+}}^\tin{0}
+4|\mui|c\Sigma_{R\tin{-+}}^\tin{1}
+(\Sigma_{R\tin{-+}}^\tin{1})^2\Big] +s^2\mui^2\Sigma_{R\tin{-+}}^\tin{0}
+(\Sigma_{R\tin{-+}}^\tin{0})^2=0.
\end{equation}
The solution of this equation gives the $\pi^\pm$ masses,
recognizing them respectively according to the tree-level case.
We can write the expression for the masses in a more compact way, by expanding
around the $m_+$ term that appears in the last equation. 
The $\pi^+$ mass is given by
\begin{equation}
m_{\pi^+} = m_+ +\alpha' |\mui|\big[ -\fr{1}{4}\bar l_3 +I'_0-4J' +{\cal
O}(s^2)\big].
\end{equation}

Some care has to be taken in the extraction of the $\pi^-$ mass. For the case
of $m_{\pi^0}$ and $m_{\pi^+}$, since they have a tree-level mass, it is
possible to consider them as 
\begin{equation}
m_R(T,\mui) =m_t(\mui)\left[1+\alpha'\sigma(T,s)\right],
\end{equation}
and, then, we can expand the corrections in terms of $s$ (or $c$ in the
other limit). 
For the case of the $\pi^-$ mass, if we consider the expansion in
mass corrections explained in chapter \ref{renormalization}, 
calculating the determinant of the propagator matrix, at the lowest order, we
obtain a second-order equation for the $\delta m_-$ mass correction (which is
the
$\pi^-$ mass since it does not have a tree-level mass). The result will be of
the form
\begin{equation}
|\mui|\delta m_- = s^2\Sigma_A+\sqrt{s^2\mui^2\Sigma_B
+(s^2\Sigma_A)^2+\Sigma_C\Sigma_D
+\Sigma_E^2}.
\end{equation}
This mixes terms of order $\alpha'$ and $\sqrt{{\cal O}(\alpha') 
+{\cal O}(\alpha'^2)}$, so all these self-energy terms must be expanded
independently  in powers of $s$ (or $c$). 
With this criterion in mind, the $\pi^-$ mass is then 
\begin{equation}
m_{\pi^-}(T,\mui)=
\theta(T_c)\fr{1}{2}\sqrt{s^2\mui^2\Sigma_{R\tin{-+}}^\tin{0}
+(\Sigma_{R\tin{-+}}^\tin{0})^2 },
\end{equation}
with
\begin{equation}
\Sigma_{R\tin{-+}}^\tin{0}= \alpha'[-\fr{1}{2}\bar
l_3 +2I'_0 +8J'],
\end{equation}
and $T_c$ is the critical temperature, where the $\pi^-$ mass is zero at the
transition point from the first to the second phase, i.e., when $\mui=m$:
\begin{equation}
m_{\pi^-}(T_c,m)=0,\qquad T_c\approx 0.135m_\pi=18.9\mbox{MeV} 
\end{equation}

\begin{figure}
\centering
\includegraphics[scale=1.2]{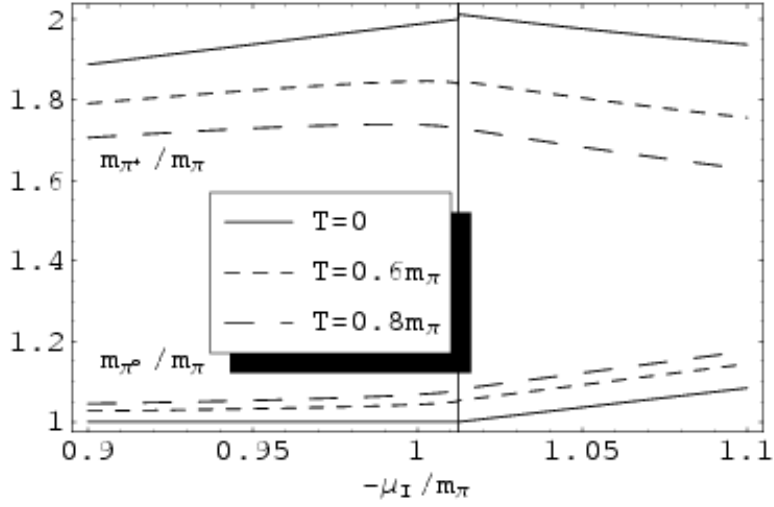}
\caption{\footnotesize $m_{\pi^0}$ and $m_{\pi^+}$ as a function of high values
of the isospin
chemical potential at different values of temperature. All
parameters are scaled with $m_\pi$}
\label{musimm.figure.m_pi0-m_pip}
\end{figure}

To start the discussion, I would like to mention that, near the
transition point, $m_{\pi^+}$ decreases, as in the tree-level
approximation, and this behavior is enforced with temperature. In
this region, $m_{\pi^0}$ grows with both parameters (see Figure
\ref{musimm.figure.m_pi0-m_pip}).

\begin{figure}
\centering
\includegraphics[scale=1]{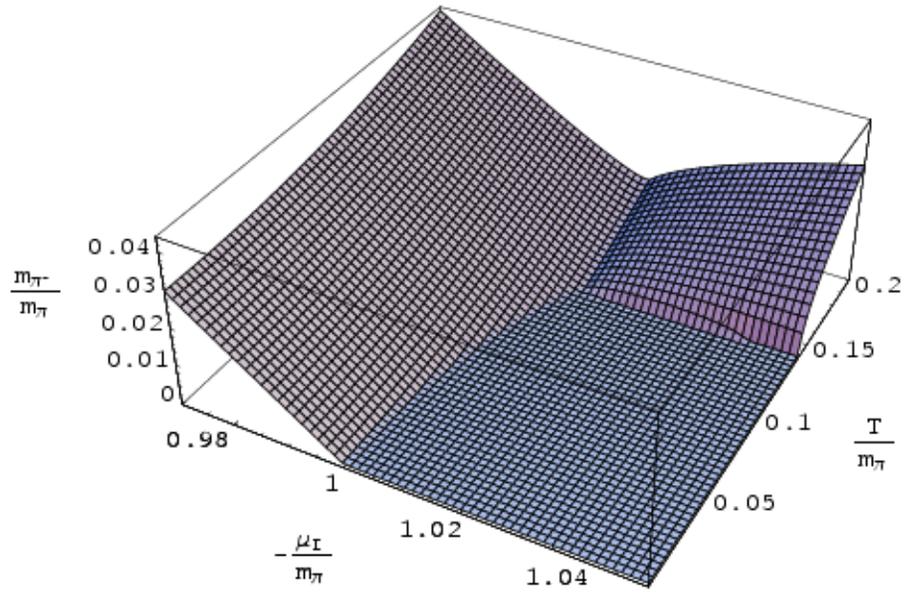}
\caption{\footnotesize $m_{\pi^-}(T,\mui)$. All the parameters are scaled with
$m_\pi$}
\label{musimm.figure.m-s3D}
\end{figure}

\begin{figure}
\centering
\includegraphics[scale=.8]{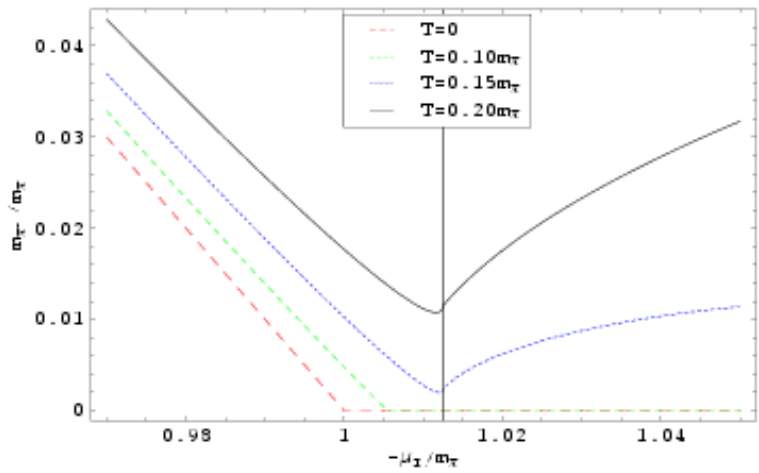}
\caption{\footnotesize $m_{\pi^-}$ plot versus isospin chemical potential for
different temperatures. All values are scaled with $m_\pi$. The
vertical line corresponds to $\mui = m$.}
\label{musimm.figure.m-s}
\end{figure}

\begin{figure}
\centering
\includegraphics[scale=1.2]{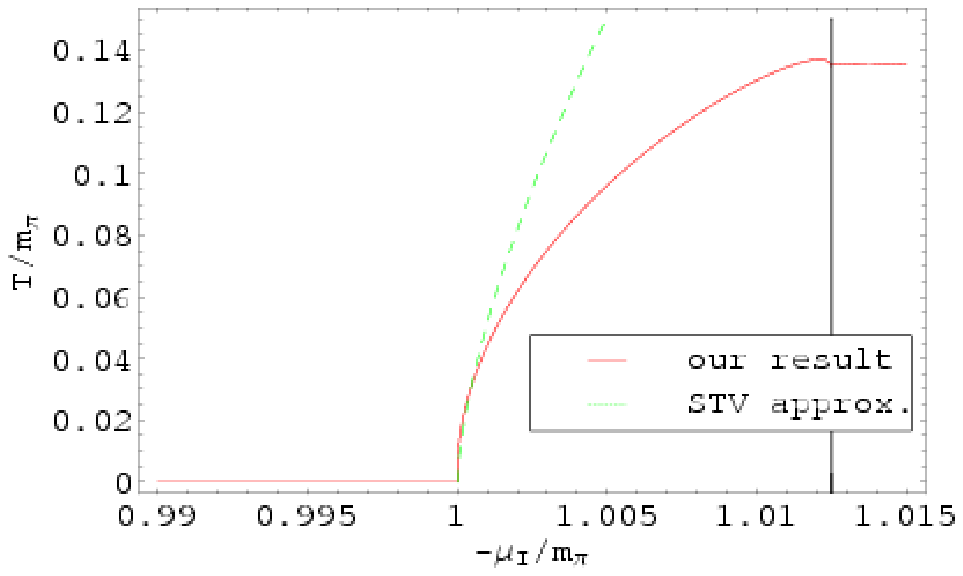}
\caption{\footnotesize Phase diagram of the condensation point of temperature
versus isospin chemical potential. The dashed line corresponds to
the approximation made by Splittorff, Toublan and Veerbarschot
\cite{Splittorff:2002xn}.}
 \label{musimm.figure.phase}
\end{figure}

For $m_{\pi^-}$, according to Figures \ref{musimm.figure.m-s3D} and 
\ref{musimm.figure.m-s},
we see that a certain critical temperature must be reached before
the $\pi^-$ is removed from the condensed state. This is precisely
the condition that determines the phase transition curve (Fig.
\ref{musimm.figure.phase}).

Starting from the first phase, according to the previous chapter, for $T<m_\pi$,
it is possible to find, at first order in
$T$, an expression for the transition line $\mui (T)$ that
coincides with the results given in eq. (8.91) from \cite{Splittorff:2002xn}. In
Fig. \ref{musimm.figure.phase}, the dashed line corresponds to this
approximation, valid at low temperature. 
However, this result gives us the transition line from the viewpoint of the
first phase. 

 The reader could think from Fig. \ref{musimm.figure.phase} that the
 critical temperature remains constant in the second phase.
 However, the temperature actually rises as a function of
 the chemical potential, in the form
 \begin{equation}
 T_{trans}(\mui)=\frac{T_c}{m}|\mui| \approx 0.134 |\mui|.
 \end{equation}
 The slope is very smooth,  and therefore we cannot appreciate
this behavior from the values shown in the figure.
 This growing behavior of the critical temperature is of course consistent with
general statements about phase transitions in the Ginzburg-Landau theory and it
has been actually calculated in the lattice for two or three color QCD.
Nevertheless, if we consider the thermal corrections in the second phase, it
happens that the condensation phenomenon starts to disappear for a certain
value of $T>18.9$MeV near the transition point, leaving a kind of superfluid
state which includes the condensed as well as the normal phase (massive pion
modes).

\section{Condensates.}

As was said at the beginning of this chapter, the different propagators and
vertices are almost the same as in the first phase if we consider the zeroth
order in the radiative corrections (just changing $m$ by $|\mui|$ and $\mui$ by
$-\mui$). 
In this case, the same occurs, as in the expansion of the different currents.

\subsection{Chiral condensate.}
Following the same procedure in the case of the first phase, the non-vanishing
components of the chiral condensate at order $s^0$ are
\begin{eqnarray}
\llang J_{s(1,0)}\rrang&=&-2Bf^2c,\\
\llang J_{s(1,3)}\rrang&=&Bc\llang 0|\pi_0^2+2|\tilde\pi|^2|0\rrang,\\
\llang J_{s(1,0)}\rrang&=&-4B\mui^2c(l_3+l_4+\tilde h_1).
\end{eqnarray}
The term $c$ must be kept because it is a global constant.

The resulting chiral condensate at finite temperature and isospin chemical
potential is then 
\begin{equation}
\llang\bar qq\rrang 
= -2Bf^2c\{1+\alpha'[-\fr{1}{2}\bar l_3+2\bar l_4-2I'_0-4I'\}.
\end{equation}

\begin{figure}
\centering
\includegraphics[scale=1]{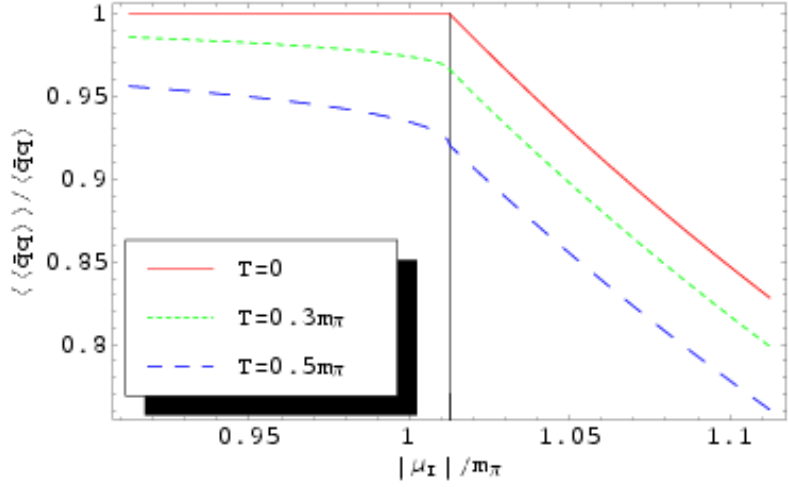}
\caption{\footnotesize Chiral condensate as a function of the isospin chemical
potential for different values of the temperature. The vertical line denotes
the transition point between the two phases.}
\label{figure.qqs}
\end{figure}

In Figure \ref{figure.qqs}, we can see the transition of the chiral condensate
from the first phase to the second phase. Now, the chiral condensate in the
second phase tends to decrease abruptly with the chemical potential. 
This effect is enhanced by  temperature.\footnote{Due to  the arguments
of the previous chapter, we can set the constant $\tilde h_1\sim 0$.}. 

\subsection{Isospin-number density.}

The non-vanishing components of the condensed vector current at order $s^0$ are
\begin{eqnarray}
\llang\bm{V}\rrang_{(1,0)}&=&-f^2|\mui|s^2{\bf e}_3u,\\
\llang\bm{V}\rrang_{(1,2)}&=&\llang 0|
-i(\tilde\pi_1\partial\tilde\pi_2-\tilde\pi_2\partial\tilde\pi_1)
-2|\mui||\tilde\pi|^2u|0\rrang{\bf e}_3.
\end{eqnarray}
Like in the case fot the calculation of $m_{\pi^-}$, we have to keep
the tree-level part
proportional to $s^2$. 

The isospin-number density condensate at finite temperature and chemical
potential is then
\begin{equation}
\llang n_I\rrang=-|\mui|f^2[s^2+8\alpha'J'].
\end{equation}

\begin{figure}
\centering
\includegraphics[scale=1]{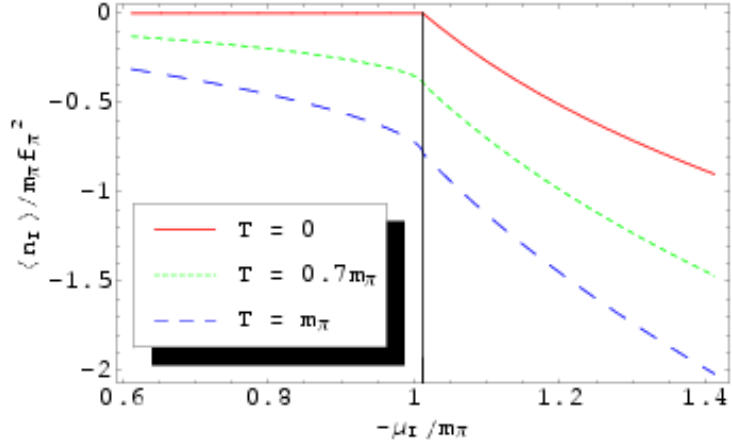}
\caption{Isospin number density as a function of the isospin chemical potential
for different values of the temperature.}
\label{figure.nis}
\end{figure}

In Figure \ref{figure.nis}, we can see the transition of the isospin number
density
from the first phase to the second phase. 
Like the chiral condensate, in
the second phase $\llang n_I\rrang$ tends to decrease abruptly with the chemical
potential. Once again this effect is enhanced by temperature.

\subsection{Pion condensate.}

Now, in the second phase, the pion condensate is finite and give us important 
information about the condensed phase. The pion condensate is
defined as 
\begin{equation}
\llang\pi^a\rrang\equiv \llang 0|\bar qi\gamma_5\tau^aq|0\rrang
=\llang 0|J_p^a|0\rrang,
\end{equation}
and is, like the other condensates, a QCD vacuum-state operator with the
same quantum numbers of the pion field.  
From Appendix \ref{expansion.J_p}, the non-vanishing components of the pion
condensates are
\begin{eqnarray}
\llang\bm{J}_{p(1,0)}\rrang&=&-2Bf^2s\tilde{\bf e}_1\\
\llang\bm{J}_{p(1,3)}\rrang
&=&Bc\llang 0|\pi_0^2+2|\pi|^2|0\rrang\tilde{\bf e}_1\\
\llang\bm{J}_{p(1,0)}\rrang&=&-4B\mui^2c(l_3+l_4)\tilde{\bf e}_1.
\end{eqnarray}
As we did for the chiral condensate, we must keep the term $s$, since it is a
global factor.
We can see that the pion condensate is oriented in the direction ${\bf e}_1$,
i.e., $\llang\bm{\pi}\rrang=\llang\tilde\pi_1\rrang\tilde{\bf e}_1$. 

Proceeding in the same way as we did with the other condensates, the pion
condensate at finite temperature and isospin chemical potential is
\begin{equation}
\llang\tilde\pi_1\rrang =2Bf^2s\{1+\alpha'[-\fr{1}{2}\bar l_3 +2\bar l_4
-2I'_0-4I']\}.
\end{equation}
It is important to recall that the pion condensates, like the chiral condensate,
give an information about the quarks condensed in the vacuum; in this case the
pion
condensate is a mixture of $\bar u\gamma_5d$ and $\bar d\gamma_5
u$.\footnote{Although related, do not confuse the pion condensate with the
number density of condensed pions.}
 This condensate breaks parity.

Note that this result is basically the same as the chiral condensate (except the
global factors $s$ and $-c$) if we neglect the term $\tilde h_1$. Then the
chiral condensate and pion condensate satisfy the relation
\begin{equation}
s\llang\bar qq\rrang +c\llang\tilde\pi_1\rrang=0.
\end{equation}
We will see that this relation is valid in the limit when $|\mui|\gg
m$.

\begin{figure}
\centering
\includegraphics[scale=1]{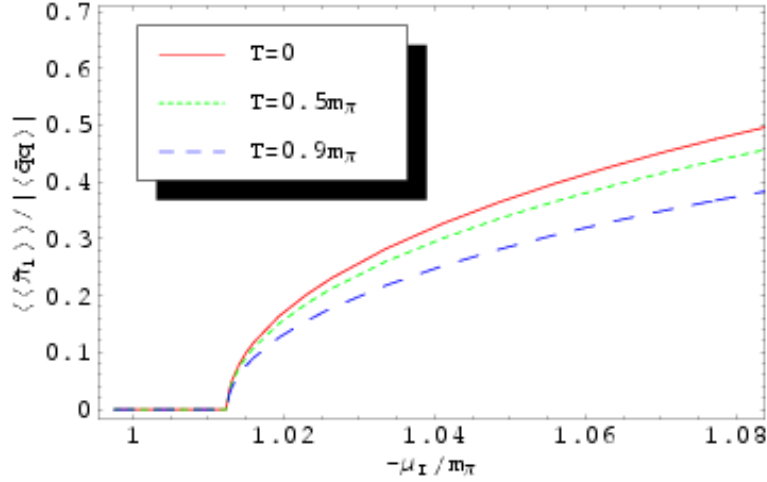}
\caption{\footnotesize Pion condensate as a function of the isospin chemical
potential for different values of the temperature. The vertical line denotes the
transition point between the two phases.}
\label{figure.pis}
\end{figure}

Figure \ref{figure.pis} shows the behavior of the pion condensate at finite
temperature and isospin chemical potential. 
If we compare with Figure \ref{figure.qqs}, we can see that the chiral
condensate
diminishes with the chemical potential in a similar rate as the pion condensate.
This means that the quarks in the chiral condensate mix together to form mixed
$u$-$d$ pseudo-scalar Cooper-pairs state. 
Due to the thermal bath, both chiral and pion condensates decrease, as
expected, due to the increase of the $\pi^-$ mass due to the thermal bath.

\chapter{Thermal pions in the second phase for $|\mui|\gg m$}
\label{muggm}

In this chapter, we will finish the analysis of the in-medium radiative
correction to the masses and condensates in the region of high isospin
chemical potential. 
As in the previous section, we will expand the
different corrections to the propagators and the condensates, but this time in
powers of $c$. 
Nevertheless, we have to include a logarithmic term in
the expansion since loop corrections give rise to terms of the form
$\lambda-\ln(\mui^2/\Lambda^2)$, which cancel with the terms
$\lambda-\ln(m^2/\Lambda^2)$ coming from the $l_i$ coupling
constants. The radiative corrections can be expanded as
\begin{equation}
\Sigma(p;T,\mui;m,f)=\alpha'\sum_{n=0}\left[\sigma_n(\bar p,\bar T)
  +\sigma_n^{\mbox{\tiny log}}(\bar p)\ln c\right]c^n.
\end{equation}

\section{Masses}

As we did for the $|\mui |\gtrsim m$ case, we use the relevant
vertices and propagators to compute the self-energy corrections.
In our approximation (${\cal O}(c^0)$), the corrections to the
$D_{\tin{00}}$ and $D_{\tin{11}}$ are the same. This happens
because the  vertices and propagators used in the loop corrections differ in
quantities of higher
order  than $c$.

\begin{figure}
\begin{minipage}[b]{.5\textwidth}
\centering
\fbox{
\includegraphics[scale=1]{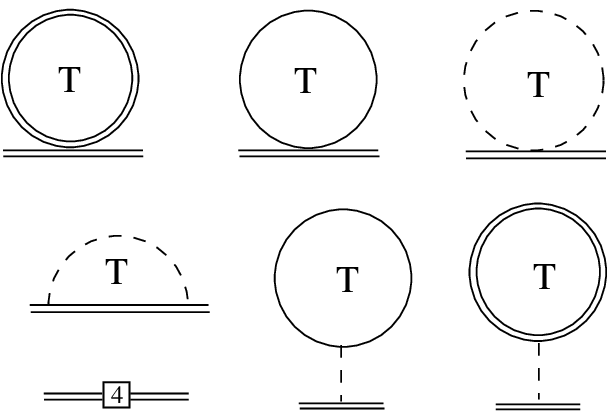}}\\
{\bf a.}
\end{minipage}
\begin{minipage}[b]{.5\textwidth}
\centering
\fbox{
\includegraphics[scale=1]{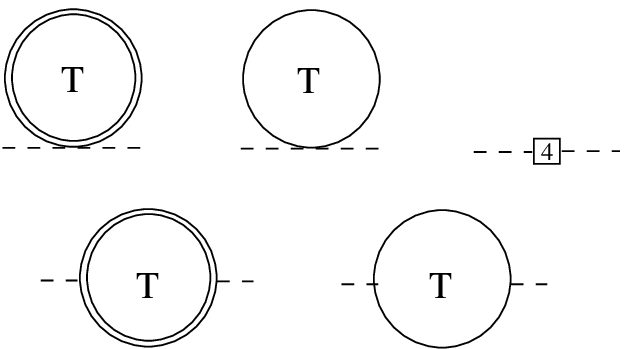}}\\
\vspace{.3cm}
{\bf b.}
\end{minipage}
\caption{\footnotesize Loop contributions to {\bf a.} $\Sigma_{\tin{00}}$ (or
$\Sigma_{\tin{11}}$) and  {\bf b.} $\Sigma_{\tin{22}}$ at order $c^0$}
\label{figure.loops-c}  
\end{figure}

The loop corrections are shown in figure \ref{figure.loops-c}.a, for
$D_{\tin{00}}$ (which are the same as those of $D_{\tin{11}}$,
exchanging the double line for a single line), and figure
\ref{figure.loops-c}.b for $D_{\tin{22}}$.

Note that, in this region, tadpole diagrams appear, which are absent
in the previous case. However, at the order $c^0$, it turns out
that these tadpoles vanish, because the vertex is proportional to
$p_0$, and the tail of the tadpole does not carry momentum. At the
order $c$,  non-vanishing tadpole diagrams will appear (and
also new ${\cal L}_4$ corrections). Nevertheless, for high values
of the chemical potential, the leading behavior will be given by
our approximation.

The self-energy corrections are
\begin{eqnarray}
\Sigma(p_0)_\tin{00} &=& \alpha'\mui^2 \Big\{ (p_0^2-1)\big[
\fr{2}{3}\underline\lambda' +2\ln c -\fr{2}{3}\bar l_1
-\fr{4}{3}\bar l_2 +\fr{4}{3}+\fr{4}{3}I'_0 +\fr{4}{9}\pi^2T^2\big]\nonumber\\
    &&  +\fr{4}{3} -4I'_0 +2A_1(p_0) +16B_0(p_0) +16B_1(p_0) -32\pi
iC_1(p_0)\Big\},\\
\Sigma(p_0)_\tin{11} &=& \Sigma(p_0)_\tin{00},\\
 \Sigma(p_0)_\tin{22}
&=& 2\alpha'\mui^2\Big\{ p_0^2\big[ \fr{2}{3}\underline\lambda' -\bar
l_1 -2\bar l_2 +3\ln c +2
-\fr{4}{3}I'_0\big]\nonumber\\
 && \hspace{1.5cm} +2A_2(p_0) +16B_2(p_0) -32\pi iC_2(p_0)\Big\},
\end{eqnarray}
plus corrections of order $\alpha'\mui^2 c^2$. The functions $A_n$,
$B_n$, $C_n$ are defined as
\begin{eqnarray}
A_1(\bar p_0) &=& \int_0^1dx\big[ 3\bar p_0^2x^2 +(\bar p_0^2-1)x\big]
    \ln\big[\bar p_0^2(x^2-x) +x -i\epsilon\big],\label{muggm.A_1}\\
A_2(\bar p_0) &=& \int_0^1dx\bar p_0^2\ln \big[ \bar p_0^2(x^2-x) +1
-i\epsilon\big],\\
B_0(\bar p_0) &=& \int_0^\infty dxxn_B(|\mui|x)\left[\frac{x^2}{\bar p_0^2
-2\bar p_0x -1 +i\epsilon}+x\rightarrow
    -x\right],\\
B_1(\bar p_0) &=& \int_1^\infty
dx\sqrt{x^2-1}n_B(|\mui|x)\left[\frac{(\bar p_0 -x)^2}{\bar p_0^2-2\bar p_0x +1
+i\epsilon}+x\rightarrow
    -x\right],\\
B_2(\bar p_0) &=& \int_1^\infty dx\sqrt{x^2-1}n_B(|\mui|x)
\left[\frac{\bar p_0^2}{\bar p_0^2-2\bar p_0x+i\epsilon}+x\rightarrow
    -x\right],\\
C_1(\bar p_0) &=& \left| \frac{\bar p_0^2-1}{2\bar p_0}\right| n_B\!\!\left(
\left| \frac{\bar p_0^2-1}{2\bar p_0}\right|\right) n_B\!\!\left( \left|
\frac{\bar p_0^2+1}{2\bar p_0}\right|\right),\\
C_2(\bar p_0) &=& \big[ \theta (\bar p_0-2)+\theta
(-\bar p_0-2)\big]\bar p_0^2\sqrt{(\bar p_0/2)^2-1}n_B(|\bar
p_0/2|)^2.\label{muggm.C_2}
\end{eqnarray}

As we can see from the previous equations, it is highly non-trivial to identify
the renormalized mass, as was the case in the region $|\mui |\gtrsim m$.
Therefore, we need to expand the
propagator, as was explained in section \ref{renormalization}, around the
tree-level masses, identifying the term proportional to $p_0^2$,
or in this case $(m_t+\delta m)^2$.

For the case of $\Sigma_{R\tin{00}}$, it is enough to evaluate
$\Sigma_\tin{00}$ with $p_0=m_{\pi^0}=m_0+\delta m_0$, because
$D_{R\tin{00}}^{-1}$ is diagonal with respect to the charged-pion
propagator. We expand around $m_0$ and renormalize it,
according to section \ref{renormalization}. 
$\Sigma_{R\tin{11}}$ and $\Sigma_{R\tin{22}}$  need to be evaluated 
for two values, $m_{\pi^{\pm}}=m_\pm +\delta m_\pm$.
The expansion in mass corrections of the functions (\ref{muggm.A_1}) -
(\ref{muggm.C_2}) is explained in Appendix \ref{formulas.2DJp}.

The renormalization constants for the different masses are
(evaluated at these masses)
\begin{eqnarray}
Z_{0}[m_0] &=& 1+2\alpha'\Big\{ \fr{1}{3}\big[ \underline\lambda +2\ln
c -\bar l_1 -2\bar l_2\big] -7 +\fr{38}{9}\pi^2\fr{T^2}{\mui^2}\nonumber\\
    &&\hspace{3cm} -\fr{34}{3}I'_0 -2K_{21}
    -16\pi i\fr{T}{|\mui|}n_B(|\mui|)\Big\},\\
Z_1[m_+] &=& Z_0+{\cal O}(\alpha' c^2),\\
Z_1[m_-] &=& 1+\fr{2}{3}\alpha'\Big\{ \underline\lambda +3\ln c -\bar
l_1 -2\bar l_2 +\fr{3}{4} +\fr{2}{3}\pi^2\fr{T^2}{\mui^2}\nonumber\\
    && \hspace{2cm} -\fr{16}{5}\pi^4\fr{T^4}{\mui^4}
-\fr{512}{21}\pi^6\fr{T^6}{\mui^6} -\fr{34}{3}I'_0 -6K_{21}\Big\},\\
Z_2[m_+] &=& 1+2\alpha'\Big\{ \fr{2}{3}\underline\lambda \bar l_1
-2\bar l_2 +3\ln c +\fr{4}{3} -\fr{10\pi}{9\sqrt{3}}
    -8K_{10} -2K_{12}\Big\},\\
Z_2[m_-] &=&  1+2\alpha'\Big\{ \fr{2}{3}\underline\lambda \bar l_1
-2\bar l_2 +3\ln c +2 -\fr{4}{3}I'_0 -8K_{00}\Big\},
\end{eqnarray}
where the mass inside the brackets is the mass around which the
self-energy expansion was computed. The function $K_{nm}$ is
defined as
\begin{equation}
K_{nm}\equiv \int_1^\infty dx\frac{\sqrt{x^2-1}n_B(|\mui|x)}{\big
[x^2-(n/2)^2\big]^{m+1}}.
\end{equation}

The self-energy corrections, associated with the different
tree-level pion masses, as functions of the corresponding
renormalized masses, are
\begin{eqnarray}
\Sigma_R(m_{\pi^0})_\tin{00} &=& 4\alpha'|\mui|\Big\{ \big[ 6
-\fr{14}{3}\pi^2\fr{T^2}{\mui^2} +13I'_0 +2K_{21}\nonumber\\
     &&\hspace{1.5cm}+16\pi i\fr{T}{|\mui|}n_B(|\mui|)\big](m_{\pi^0}-|\mui|)
+3I'_0\Big\},\\
 \Sigma_R(m_{\pi^+})_\tin{11} &=&
\Sigma_R(m_{\pi^+})_\tin{00}
+\mui^2{\cal O}(\alpha' c^2),\\
\Sigma_R(m_{\pi^-})_\tin{11}  &=&  -2\alpha'\mui^2\Big\{
\fr{1}{6} +\fr{512}{63}\pi^6\fr{T^6}{\mui^6} +16I'_0 -8I'_1
-2K_{21}\Big\},\\
 \Sigma_R(m_{\pi^+})_\tin{22} &=& 2\alpha'|\mui|\Big\{
\big[
-\fr{2}{3} -\fr{4\pi}{9\sqrt{3}} +16K_{10} -5K_{11}\big]|\mui|\nonumber\\
    && \hspace{1.5cm} +\big[ -\fr{8}{3} +\fr{32\pi}{9\sqrt{3}} +20K_{11}
    +3K_{12}\big]m_{\pi^+}\Big\},\\
\Sigma_R(m_{\pi^-})_\tin{22} &=& \mui^2{\cal O}(\alpha'c^2).
\end{eqnarray}
Note that the self-energy corrections actually have the form
$\Sigma_R(m_{\pi^i}) =
\Sigma_R^\tin{0}[m_i]+\Sigma_R^\tin{1}[m_i]m_{\pi^i}$
 presented in Section \ref{renormalization}.

Since we have already expanded the self-energy corrections in
powers  of the mass corrections $(\delta m_i)^n$, we can neglect
higher terms in the determinant of the propagator, finding 
the solution for $\delta m_i$ from the pole condition. 
For $m_{\pi^0}$ and $m_{\pi^+}$, the corrections $\delta m_0$ and $\delta
m_+$ are of order $\mui\alpha'$ (neglecting terms of order $\alpha' c$).
For $m_{\pi^-}$, because there is no finite tree-level mass, we
need to keep terms of the order $\alpha'^2$ in the expansion; i.e. we
need to consider, in principle, $(\delta m_-)^2 \sim \mui\alpha'^{1/2}$.
Remembering that $\Sigma_R^\tin{0}$ is of order $\mui^2\alpha'$, and
$\Sigma_R^\tin{1}$ of order $\mui\alpha'$, we can expand the
propagator, neglecting higher order terms.

The resulting expressions for the renormalized  masses are
surprisingly simple. Many terms vanish, including also some
complex contributions.

\begin{eqnarray}
m_{\pi^0} &=& |\mui|\big[ 1+6\alpha' I'_0\big]\\
m_{\pi^+} &=& |\mui|\big[ \sqrt{1+3c^2} +6\alpha' I'_0\big]\\
m_{\pi^-} &=& \mui{\cal O}(\alpha' c^2).
\end{eqnarray}

Note that $m_{\pi^-}$ vanishes once again in this region, i.e.,
the pion condenses again, in spite of the fact that the thermal
corrected mass started to grow near the phase transition point.
This behavior, however,  could be just a fictitious result from
our expansion up to order $c^0$.

\begin{figure}
\centering
\includegraphics[scale=1.3]{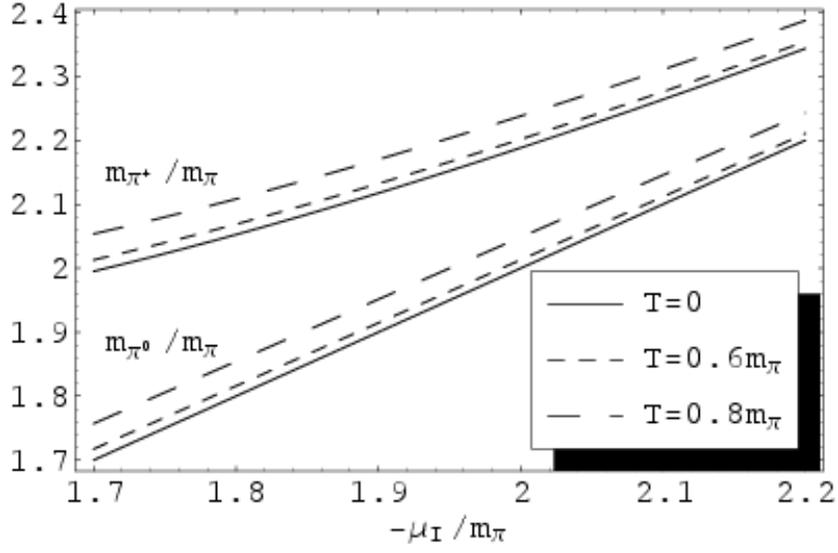}
\caption{\footnotesize $m_{\pi^0}$ and $m_{\pi^+}$ as a function of high values
of the isospin
chemical potential at different values of temperature.}
\label{figure.m0mpc}
\end{figure}

In this region, $m_{\pi^0}$ 
increases monotonically both with temperature and chemical
potential. 
$m_{\pi^+}$, as the temperature and the chemical
potential rise, becomes asymptotically  close to $m_{\pi^0}$, as
was expected (see figure. \ref{figure.m0mpc}). 
In contrast with the first region, the $\pi^+$ mass grows with temperature, and
a crossover occurs somewhere in the intermediate region of the chemical
potential for the temperature dependence, since near the phase
transition point we have $m_{\pi^+}(T_1,\mui)>m_{\pi^+}(T_2,\mui)$,
and this behavior changes in the high chemical potential region in
such a way that $m_{\pi^+}(T_1,\mui)<m_{\pi^+}(T_2,\mui)$.

\section{Condensates}

\subsection{Chiral condensate}

The non-vanishing components of the chiral condensate, neglecting higher
corrections of order $c^2$, are
\begin{eqnarray}
\llang J_s\rrang_{(1,0)}&=&-2Bf^2c,\\
\llang J_s\rrang_{(1,2)}&=&Bc\llang
0|\pi_0^2+\tilde\pi_1^2+\tilde\pi_2^2|0\rrang,\\
\llang J_s\rrang_{(3,0)}&=&-2Bf^2\mui^2c(l_4+2\tilde h_1),
\end{eqnarray}
keeping the global term $c$. Following the same procedure as in the previous
chapters, the chiral condensate at finite temperature and isospin chemical
potential is
\begin{equation}
\llang\bar qq\rrang=-2Bf^2c\left\{1+\alpha'\left[\bar l_4+32\pi^2\tilde h_1
+2\ln c-4I'_0-\frac{2}{3}\left(\frac{\pi T}{\mui}\right)^2\right]\right\}.
\label{equation.qqc}
\end{equation}

\begin{figure}
  \centering
\includegraphics[scale=1]{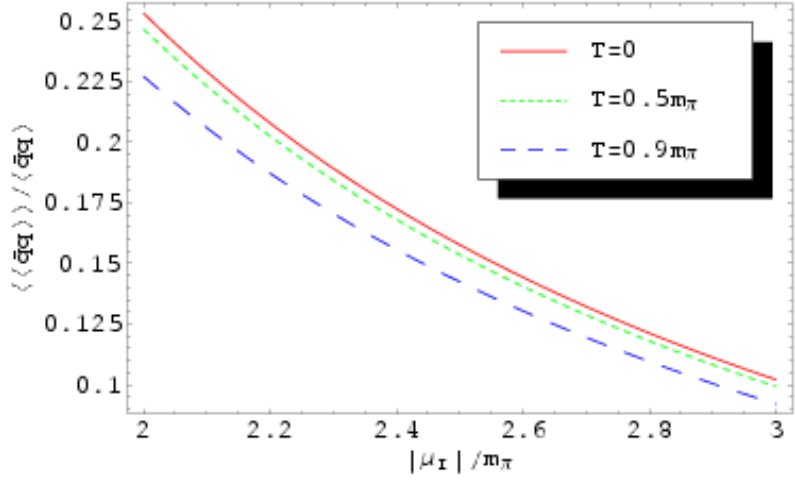}
\caption{\footnotesize Chiral condensate as a function of the isospin chemical
potential for different values of the temperature for high values of the
chemical potential.}
\label{figure.qqc}
\end{figure}

Figure \ref{figure.qqc} shows the chiral condensate behavior at finite
temperature for high values of the chemical potential. It continues decreasing
with both parameters.

\subsection{Isospin-number density.}

The non-vanishing components of the vector current, within the thermal
vacuum, neglecting higher corrections of order $c^2$, are

\begin{eqnarray}
\llang\bm{V}_\mu\rrang_{(1,0)}&=&-|\mui|f^2s^2u_\mu{\bf e}_3,\\
\llang\bm{V}_\mu\rrang_{(1,2)}&=&|\mui|s^2\llang
0|\pi_0^2+\tilde\pi_1^2|0\rrang u_\mu{\bf e}_3,\\
\llang\bm{V}_\mu\rrang_{(3,0)}&=&-4|\mui|^3s^2(l_1+l_2) u_\mu{\bf e}_3.
\end{eqnarray}
The isospin-number density is then
\begin{equation}
\llang n_I\rrang=-|\mui|f^2s^2\left\{1
+2\alpha'\left[\fr{1}{3}\bar l_1+\fr{2}{3}\bar l_2+2\ln c-4I'_0\right]\right\}.
\end{equation}

\begin{figure}
\centering
\includegraphics[scale=1]{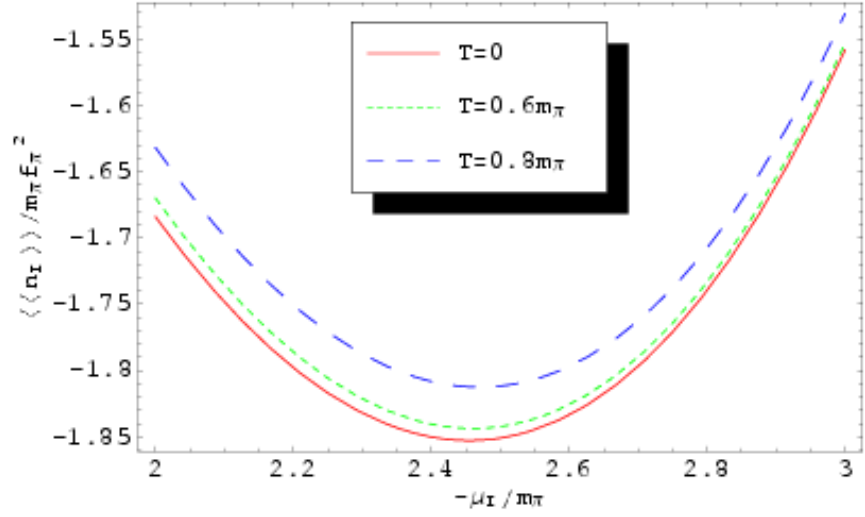}
\caption{\footnotesize Isospin-number density as a function of the isospin
chemical potential at different values of the temperature in the high chemical
potential limit.}
\label{figure.nic}
\end{figure}

Figure \ref{figure.nic} shows the isospin-number density for different
temperatures and high values of the isospin chemical potential. 
Due to the logarithmic term, it starts to grow again with the chemical
potential. 
As in the case of the $\pi^+$ mass in this high chemical potential limit, it
grows with temperature, in contrast to the $|\mui|\sim m$ case. 
A crossover of the different temperature lines must occur somewhere in the
intermediate region of the chemical potential.

\subsection{Pion condensate.}

The non-vanishing components of the pseudo-scalar current, within the thermal
vacuum and neglecting higher corrections of order $c^2$, are
\begin{eqnarray}
\llang\bm{J}_{p(1,0)}\rrang&=& 2Bf^2s\tilde{\bf e}_1,\\
\llang\bm{J}_{p(1,3)}\rrang&=& -Bs\llang
0|\pi_0^2+\tilde\pi_1^2+\tilde\pi_2^2\rrang\tilde{\bf e}_1,\\
\llang\bm{J}_{p(1,3)}\rrang&=& 2\mui^2Bsl_4.
\end{eqnarray}
The pion condensate is then
\begin{equation}
\llang\bm{\pi}\rrang=2Bf^2s\left\{1+\alpha'\left[\bar l_4
+2\ln c-4I'_0-\left(\frac{\pi T}{\mui}\right)^2\right]\right\}.
\end{equation}

\begin{figure}
\centering
\includegraphics[scale=1]{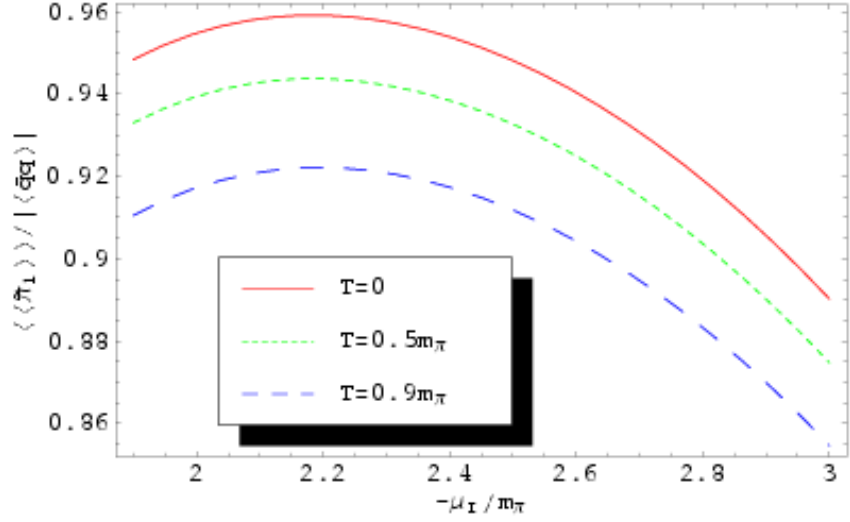}
\caption{\footnotesize Pion condensate as a function of the isospin chemical
potential for different temperatures in the limit of high chemical potential.}
\label{figure.pic}
\end{figure}

Figure \ref{figure.pic} shows the pion condensate at finite temperature and
high values of the chemical potential. Like the isospin chemical potential, it
has a deflection and starts to decrease, due to the logarithmic terms. 
Comparing with equation (\ref{equation.qqc}), if we neglect the term $\tilde
h_1$, the pion condensate and the chiral condensate
 follow the relation $s\llang\bar qq\rrang
+c\llang\tilde\pi_1\rrang=0$, as was the case in the region $|\mui|\gtrsim m$.

\chapter{Conclusions}\label{conc}

In this thesis, different aspects of pion dynamics within a thermal bath at
finite isospin chemical potential were developed in the frame of Chiral
Perturbation Theory and Thermo-field Dynamics. The extraction of the pion
masses, pion decay constants, and different condensates were analyzed
under these conditions in order  to get a physical insight about the pion
properties in a hot and dense pion gas. 

First, an expansion criterion was developed to calculate
radiative corrections when the chemical potential becomes bigger than the pion
mass (second phase), separating the analysis for the two cases, $|\mui|\gtrsim
m$ and $|\mui|\gg m$. 
Nevertheless, in our analysis we restrict ourselves to
$|\mui|\le m_\eta$ in order to avoid strangeness effects. 
We also considered the so called first phase
$|\mui|<m$, finding also corrections to masses and decay constants.

Then we calculated the three pion masses at finite
temperature and isospin chemical potential. 
Each of the three masses behaves differently under the introduction of the
isospin chemical potential. 
For values of the chemical potential near the tree-level pion mass, one of the
charged pions will condense (in this case, the $\pi^-$ when $\mui$ becomes
negative).
The condensation produces a symmetry breaking in the flavor group
SU(2)$\rightarrow$U(1), inducing a two-phase structure.
However, the thermal bath shifts the condensation point with respect to the
tree-level case. It was also possible to extract the decay constants(which now
are different for the $\pi^0$ and the $\pi^\pm$), the chiral condensate and the
isospin number density. 
In the case of the masses, the decay constants and the chiral condensate, it was
possible to re define and generalize the G-MOR relation for finite isospin
chemical potential.

The fact that the charged pions acquire a shifted mass at zero temperature is a
special property of the pions. 
Except for the charged pion masses, the isospin chemical potential effects
appear only for finite temperature; this is also known as the ``isospin
chemical potential silver blaze problem"\footnote{From Arthur Conan Doyle's
story, where Sherlock Holmes knew the identity of the thief because a dog doing
nothing.} \cite{Cohen:2004qp} which indicates that at zero temperature nothing
happens, unless the isospin chemical potential becomes bigger than the pion
mass.

Then, we proceed with the consideration of pion masses and condensates for
isospin chemical potential values in the vicinity of the transition point.
It was possible to find the transition where the $\pi^-$ mass vanishes. 
For a certain temperature $T_c$, which depends on $\mui$, a normal phase and a
condensed phase, as in a superfluid, coexist. 
With increasing temperature, the thermal $\pi^-$ mass grows and thus 
the pion condensate and the chiral condensate start to decrease. 
As the temperature and chemical potential increase, the isospin number
density (absolute value) grows more rapidly than in the first phase.

Finally, in chapter \ref{muggm}, the pion behavior for the high isospin chemical
potential values was studied, obtaining the pion masses and 
condensates. According to our results, the $\pi^-$ normal phase dissappears
once again, but this is an artifact of the approximation involved.


A further analysis should include aditional corrections due to higher order
diagrams in our expansion parameter. The extension of the present discussion to
the strange meson sector could also be interesting to pursue.

\appendix

\chapter{Expansion in Terms of Pion Fields.}\label{expansion}

This appendix summarizes the main steps in the construction of the moment
expansion of the
chiral
Lagrangian and currents in both phases. Here we will use the vacuum expectation
value in the second phase 
\begin{equation}
\bar U = e^{i\tilde\tau_1\varphi} 
= \cos\varphi +i\tilde\tau_1\sin\varphi
\equiv c+i\tilde\tau_1s,
\end{equation}
where for any vector, the 
$\tilde{\mbox{v}}_i$ components refers to the components of the vector rotated
in an azimutal
angle $\phi$:
\begin{equation}
 \tilde{\mbox{v}}_1 =\mbox{v}_1\cos\phi +\mbox{v}_2\sin\phi,
 \qquad
 \tilde{\mbox{v}}_2 = -\mbox{v}_1\sin\phi +\mbox{v}_2\cos\phi,
 \qquad
 \tilde{\mbox{v}}_3 = v_3.
\end{equation}
For the case of the first phase we only have to set $\varphi = 0$
(i.e. $c=1$, $s=0$) and 
$\phi =0$ in all the results of this appendix.

The unitary $U$ fields can be
expressed as 
\begin{equation}
U=U_0+i\bm{\tau}\cdot\bm{U} =
U_0+i\tilde{\bm{\tau}}\cdot\tilde{\bm{U}},
\end{equation}
which satisfy the unitarity relation $U_0^2+\tilde{\bm{U}}^2=1$.
On the other side, the field $U$ is defined as 
$U=\bar U^\tin{1/2}e^{i\xi}\bar U^\tin{1/2}$,  with
\begin{equation}
\xi = \bm{\tau}\cdot\bm{\xi}\equiv\bm{\tau}\cdot\frac{\bm{\pi}}{f},
\qquad \mbox{and} \qquad
\bar U^\tin{1/2} = e^{i\tilde\tau_1\varphi/2} = \cos\fr{\varphi}{2}
+i\tilde\tau_1\sin\fr{\varphi}{2}.
\end{equation}

The exponential can be expanded in terms of the $\xi$ fields. Using the fact
that $\xi^2=\bm{\xi}^2$ we have that
\begin{equation}
e^{i\xi} = \sum_{n=0}\frac{(-\bm{\xi}^2)^n}{(2n)!} 
 +i\xi \sum_{n=0}\frac{(-\bm{\xi}^2)^n}{(2n+1)!}.
\end{equation}
Now putting the previous equation between the vacuum field $\bar
U^\tin{1/2}$
\begin{equation}
U = \big(c+i\tilde\tau_1s\big)\sum_{n=0}\frac{(-\bm{\xi}^2)^n}{(2n)!} 
 +i\big(\xi +is\tilde\xi_1 +(c-1)\tilde\tau_1\tilde\xi_1\big) 
 \sum_{n=0}\frac{(-\bm{\xi}^2)^n}{(2n+1)!}
\end{equation}
then
\begin{eqnarray}
U_0 &=& c\sum_{n=0}\frac{(-\bm{\xi}^2)^n}{(2n)!}
 -s\tilde\xi_1\sum_{n=0}\frac{(-\bm{\xi}^2)^n}{(2n+1)!}\\
\tilde{\bm{U}} &=& s\tilde{\bm{e_1}}\sum_{n=0}\frac{(-\bm{\xi}^2)n}{(2n)!}
 +\big(\bm{\xi}+(c-1)\tilde\xi_1\tilde{\bm{e_1}}\big)
 \sum_{n=0}\frac{(-\bm{\xi}^2)n}{(2n+1)!}.
\end{eqnarray}
As we need the terms only up to ${\cal O}(\xi^4)$, the previous expression
reduces to\\

\fbox{
\begin{Beqnarray}
U_0 &=& c-s\tilde\xi_1 -\fr{1}{2}c\bm{\xi}^2 +\fr{1}{6}s\tilde\xi_1\bm{\xi}^2
 +\fr{1}{24}c\bm{\xi}^4\label{U_0}\\
\tilde{\bm{U}} &=& \big(s +c\tilde\xi_1 -\fr{1}{2}s\bm{\xi}^2
-\fr{1}{6}c\tilde\xi_1\bm{\xi}^2 +\fr{1}{24}s\bm{\xi}^4\big)\tilde{\bm{e_1}}
 +\big(\tilde\xi_2\tilde{\bm{e_2}}
+\xi_3\bm{e_3}\big)\big(1-\fr{1}{6}\bm{\xi}^2\big)\label{U_i}
\end{Beqnarray}}\\
$+$ terms of ${\cal O}(\xi^5)$.\\

Another important thing to keep in mind is the fact
that the terms $L_{\mu\nu}$ and $R_{\mu\nu}$ vanish for the values of the
external fields $v=\fr{1}{2}\mui\tau_3u$ and $a=0$.

\section{Expansion of the Chiral Lagrangian}\label{expansion.lagrangian}

The lagrangian with massive quarks at finite isospin chemical potential is
\begin{eqnarray}
{\cal L}_2(M,0,\fr{1}{2}\mui\tau_3 u,0)  &=& \frac{f^2}{4}Tr\left[(D_\mu U)^\dag
 D^\mu U+2BM(U^\dag +U)\right]\\
{\cal L}_4(M,0,\fr{1}{2}\mui\tau_3 u,0) &=&  \fr{1}{4}l_1\left( Tr\left[(D_\mu
 U)^\dag D^\mu U\right]\right)^2  \nonumber\\
 && +  \fr{1}{4}l_2Tr\left[(D_\mu U)^\dag D_\nu
           U\right]Tr\left[(D^\mu U)^\dag D^\nu U\right]\nonumber\\ 
 &&+  \fr{1}{4}(l_3+l_4)\left(Tr\left[BM( U^\dag
           +U)\right]\right)^2\label{expansion.L_4}\\ 
 &&+  \fr{1}{4}l_4Tr\left[(D_\mu U)^\dag D^\mu U\right]Tr\left[BM(
           U^\dag + U)\right]\nonumber\\
 && -\fr{1}{4}l_7\left(Tr\left[BM(
           U^\dag-U)\right]\right)^2\nonumber\\ 
 &&+  \mbox{constants}\nonumber
\end{eqnarray} 
with the covariant derivative
\begin{equation}
DU = \partial U -i\mui u[\fr{1}{2}\tau_3,U]
 = \partial U_0 +i\bm{\tau}\cdot (\partial\bm{U} 
  -\mui u\bm{U}\times\bm{e_3})
\label{DU}
\end{equation}

First of all, we need to compute the traces. It is enough to calculate the
traces of the main structures of the Lagrangian:  
$Tr[(D_\mu U)^\dag D_\nu U]$, $Tr[M(U^\dag +U)]$, and 
$Tr[M(U^\dag +U)]^2$.

Using the shape of the covariat derivative in Eq. (\ref{DU})
\begin{eqnarray}
Tr\left[(D_\mu U)^\dag D_\nu U\right] &=& Tr\left[\left\{\partial_\mu U_0
 -i\bm{\tau}\cdot (\partial_\mu\bm{U} -\mui
 u_\mu\tilde{\bm{U}}\times\bm{e_3})\right\}\Big\{h.c.,\quad
\mu\leftrightarrow\nu\Big\}\right]\nonumber\\
&=& 2\big[\partial_\mu U_0\partial_\nu U_0 +(\partial_\mu
 \bm{U} -\mui u_\mu\bm{U}\times\bm{e_3})\cdot (\mu\rightarrow\nu
 )\big]\nonumber\\
&=& 2\big[\partial_\mu U_0\partial_\nu U_0 +\partial_\mu
 \bm{U}\cdot \partial_\mu\bm{U}+\mui^2(\tilde U_1^2 +\tilde U_2^2)u_\mu
 u_\nu\nonumber\\
 && \hspace{2cm} +\mui (\{\tilde U_1\partial_\nu\tilde U_2 -\tilde
U_2\partial_\nu\tilde U_1\}u_\mu +\mu\leftrightarrow\nu )\big].
\label{TrD_muUD_nuU}
\end{eqnarray}

Now, using the fact that 
\begin{equation}
BM = B\big[\fr{1}{2}(m_u+m_d)+\fr{1}{2}(m_u-m_d)\tau_3\big] =
\fr{1}{2}m^2(1+\epsilon_{ud}\tau_3),
\end{equation}
the other terms we want to calculate are
\begin{eqnarray}
Tr\left[BM( U^\dag +U)\right] &=& Tr\left[2BMU_0\right] 
= 2m^2U_0\label{TrM(U+U)}\\
Tr\left[BM( U^\dag -U)\right]^2 &=& Tr\left[-2iBM\bm{\tau}\cdot\bm{U}\right]^2
= -4m^4\epsilon_{ud}^2U_3^2.\label{TrM(U-U)^2}
\end{eqnarray}

\subsection{${\cal L}_2$ up to order $\pi^4$}

Now we are going to expand the the structures that appear in 
${\cal L}_2$, $Tr[(D_\mu U)^\dag D^\mu U]$ and $Tr[M( U^\dag +U)]$, neglecting
higher terms of ${\cal O}(\xi^5)$.
From Eq. (\ref{TrD_muUD_nuU})
\begin{equation}
Tr\left[(D_\mu U)^\dag D^\mu U\right] = 2\big[(\partial U_0)^2
 +(\partial\bm{U})^2 +2\mui (\tilde U_1\partial_0\tilde U_2 
 -\tilde U_2\partial_0\tilde U_1) +\mui^2(\tilde U_1^2 
 +\tilde U_2^2)\big].
 \label{TrDUDU}
\end{equation}
Using the results of Eqs. (\ref{U_0}) and (\ref{U_i})
\begin{eqnarray}
(\partial U_0)^2 +(\partial \tilde{\bm{U}})^2 &=& (\partial\bm{\xi})^2
 -\fr{1}{3}(\partial\bm{\xi})^2\bm{\xi}^2 
 +\fr{1}{3}(\bm{\xi}\cdot\partial\bm{\xi})^2\\
\tilde U_1\partial_0\tilde U_2 -\tilde U_2\partial_0\tilde U_1 &=&
 s\partial_0\tilde\xi_2 +c(\tilde\xi_1\partial_0\tilde\xi_2
   -\tilde\xi_2\partial_0\tilde\xi_1)
  -\fr{2}{3}s(\partial_0\tilde\xi_2\bm{\xi}^2 
   -\tilde\xi_2\bm{\xi}\cdot\partial_0\bm{\xi})\nonumber\\
 && -\fr{1}{3}c(\tilde\xi_1\partial_0\tilde\xi_2
  -\tilde\xi_2\partial_0\tilde\xi_1)\bm{\xi}^2\\
\tilde U_1^2 +\tilde U_2^2 &=& s^2 +2cs\tilde\xi_1 
  +c^2\tilde\xi_1^2 +\tilde\xi_2^2 -s^2\bm{\xi}^2\nonumber\\ 
 && -\fr{4}{3}cs\tilde\xi_1\bm{\xi}^2
 -\fr{1}{3}(c^2\tilde\xi_1^2 +\tilde\xi_2^2
-s^2\bm{\xi}^2)\bm{\xi}^2.
\label{TrDUDU-3}
\end{eqnarray}
Even though we neglect constants and total derivative terms in the
lagrangian, we will keep them in the previous equation for further
considerations in  ${\cal L}_4$.

For the case of $Tr[M( U^\dag +U)]$, from Eq. (\ref{TrM(U+U)}), we
just need to replace the results of Eq. (\ref{U_0}).

\subsection{${\cal L}_4$ up to order $\pi^2$}

In the next part, the structures of the ${\cal L}_4$ Lagrangian will be
expanded in order of apparition from Eq. (\ref{expansion.L_4}), neglecting
higher terms of ${\cal O}(\xi^3)$.
In this case the resulting constant terms and total derivative terms will be
neglected, since they do not contribute. It will be used also the fact that
$\xone\partial_0\xtwo = -\xtwo\partial_0\xone $ + a total derivative.

\medskip

Using the results of Eqs. (\ref{TrDUDU})-(\ref{TrDUDU-3}) up to order $\xi^2$
\begin{eqnarray}
Tr\left[(D_\mu U)^\dag D^\mu U\right]^2 &=& 4\big[(\partial\bm{\xi})^2
  -\fr{1}{3}(\partial\bm{\xi})^2 +2\mui\{s\partial_0\tilde\xi_2
   +c(\tilde\xi_1\partial_0\tilde\xi_2
   -\tilde\xi_2\partial_0\tilde\xi_1)\}\nonumber\\
  && +\mui^2[s^2 +2cs\tilde\xi_1 -s^2\bm{\xi}^2
   +c^2\tilde\xi_1^2]\big]^2\nonumber\\
 &=& 8\mui^2s^2\big[2\mui^2cs\tilde\xi_1 +(\partial\bm{\xi})^2
   +2(\partial_0\tilde\xi_2)^2 
   +8\mui c\tilde\xi_1\partial_0\tilde\xi_2\nonumber\\
   && \qquad -\mui^2s^2\bm{\xi}^2 +3\mui^2c^2\tilde\xi_1^2
+\mui^2\tilde\xi_2^2\big]\\
&&  +{\cal O}(\xi^3)+\mbox{tot. deriv.} +\mbox{const.}\nonumber
\end{eqnarray}

\medskip

Using Eq. (\ref{TrD_muUD_nuU}) and inserting Eqs. (\ref{U_0}) and (\ref{U_i}) up
to order $\xi^2$
\begin{eqnarray}
Tr\left[(D_\mu U)^\dag D_\nu U\right]^2 &=& 
 4\big[\partial_\mu\bm{\xi}\cdot\partial_\nu\bm{\xi} 
 +\mui^2\{s^2 +2cs\tilde\xi_1 +c^2\tilde\xi_1^2 +\tilde\xi_2^2
  -s^2\bm{\xi}^2\}u_\mu u_\nu\nonumber\\ 
 && +\mui\{[s\partial_\mu\tilde\xi_2 
   +c(\tilde\xi_1\partial_\mu\tilde\xi_2
    -\tilde\xi_2\partial_\mu\tilde\xi_1)]u_\nu 
  +\mu\leftrightarrow\nu\}\big]^2\nonumber\\
 &=& 8\mui^2s^2\big[2\mui^2cs\tilde\xi_1 +(\partial_0\bm{\xi})^2
 +(\partial\tilde\xi_2)^2 +(\partial_0\tilde\xi_2)^2\nonumber\\ 
 && \qquad +8\mui c\tilde\xi_1\partial_0\tilde\xi_2
 +\mui^2(3c^2\tilde\xi_1^2 +\tilde\xi_2^2 -s^2\bm{\xi}^2)\big]\\
 && +{\cal O}(\xi^3) +\mbox{tot. deriv.} +\mbox{const.}\nonumber
\end{eqnarray}

\medskip

From Eq. (\ref{TrM(U+U)}) and Eq. (\ref{U_0}) up to order $\xi^2$
\begin{eqnarray}
Tr\left[BM(U^\dag +U)\right] &=& 4m^4\left[c-s\tilde\xi_1
 -\fr{1}{2}c\bm{\xi}^2\right]^2 \nonumber\\
 &=& -4m^4\left[2cs\xone +c^2\bm{\xi}^2
-s^2\xone^2\right] +{\cal O}(\xi^3)+\mbox{const.}
\end{eqnarray}

\medskip

From Eqs. (\ref{TrDUDU})-(\ref{TrDUDU-3}), Eq. (\ref{TrM(U+U)}) and Eq.
(\ref{U_0}) up to order $\xi^2$
\begin{eqnarray}
&& Tr\left[(D_\mu U)^\dag D^\mu U\right] Tr\left[BM(U^\dag +U)\right]\nonumber\\
&& \qquad =4m^2[(\partial\bm{\xi})^2 +2\mui\{s\partial_0\xtwo
+c(\xone\partial_0\xtwo -\xtwo\partial_0\xone)\}\nonumber\\
&& \qquad\qquad\quad +\mui^2\{s^2 +2cs\xone +c^2\xone^2 +\xtwo^2
-s^2\bm{\xi}^2\}]
 [c -s\xone -\fr{1}{2}c\bm{\xi}^2]\nonumber\\
&& \qquad = -4m^2[\mui^2s(1 -3c^2)\xone +2\mui(1-3c^2\xone\partial_0\xtwo
-c(\partial\bm{\xi})^2\nonumber\\
&& \qquad\qquad\qquad +\mui^2c\{\fr{3}{2}s^2\bm{\xi}^2 -(1 -3s^2)\xone^2
+\xtwo^2\}]\\
&& \qquad\quad +{\cal O}(\xi^3) +\mbox{tot. deriv.} + \mbox{const.} \nonumber
\end{eqnarray}

\medskip

From Eqs. (\ref{TrM(U-U)^2}) and (\ref{U_i}) up to order $\xi^2$
\begin{equation}
Tr\left[BM(U^\dag -U)\right]^2 = -4M^4\epsilon_{ud}^2\xi_3^2 +{\cal O}(\xi^3)
\end{equation}

\section{Expansion of currents}\label{expansion.currents}

In this section the currents that will
be used in this thesis to obtain
information on condensates and on the decay
constant will be expand in terms of pion fields.

The currents are sorted in powers of momentum
\begin{equation}
J=J_{(1)}+J_{(3)}+\cdots 
\end{equation}

For the case of the PCAC relation, we need to expand the axial current up to
order $\pi^3$ in $A_{(1)}$ and order $\pi$ in $A_{(3)}$. As this relation
will be used only in the first phase, it will be calculated only with the values
$c=1$ and $s=0$. So, the effective axial current will be of the form
\begin{equation}
A_{\mbox{\small eff}}=A_{(1,1)}+A_{(1,3)}+A_{(3,1)}
\end{equation}

For the case of currents that will be used in the
calculation of condensates, an expansion will be developed in the
second phase as I did the previous section,. In the case of the first phase, we
just need to take the terms $c=1$ and
$s=0$.

For one loop corrections to the condensates, we need to expand up
to order $\pi^2$ in $J_{(1)}$ and order zero in $J_{(3)}$ according to the
power counting in chapter \ref{intro.power-counting}. The effective
current in this case will be
\begin{equation}
J_{\mbox{\small eff}}=J_{(1,0)}+J_{(1,1)}+J_{(1,2)}+J_{(3,0)}, 
\end{equation}
and then, the respective condensate is
\begin{equation}
\llang J\rrang =\llang 0|J_{(1,0)}|0\rrang
+\llang 0|J_{(1,2)}|0\rrang +\llang 0|J_{(3,0)}|0\rrang.
\end{equation}

\subsection{Axial current.}\label{expansion.axial}

As was said before, we need the axial current in the first phase, to get, via
the PCAC relation, the
pion decay constant. Since it will be saturated with a single pion, we need to
expand it up to order
$(\pi)^3$. The axial current for massive quarks at finite chemical potential
is
\begin{equation}
A_\mu^a
=\frac{\delta S_\chi}{\delta a_\mu^a}(M,0,\fr{1}{2}\mui u\tau_3,0).
\end{equation}
The expressions for $A_{(1)}$ and $A_{(3)}$ with finite quark masses and
finite chemical potential are
\begin{eqnarray}
A_{(1)\mu}^a&=&\frac{f^2}{4}
Tr\left[i\big\{\fr{1}{2}\tau^a,U^\dag\big\}D^\mu U~+h.c.\right] \\
A_{(3)\mu}^a&=&\fr{1}{2}l_1
Tr\left[i\big\{\fr{1}{2}\tau^a,U^\dag\big\}D^\mu U~+h.c.\right]
Tr\left[(D_\alpha U)^\dag D^\alpha U\right]\nonumber\\
&&+\fr{1}{2}l_2
Tr\left[i\big\{\fr{1}{2}\tau^a,U^\dag\big\}D_\beta U \delta^\mu_\alpha
~+h.c.,~\alpha\leftrightarrow\beta\right]
Tr\left[(D^\alpha U)^\dag D^\beta U\right]\nonumber\\
&&+\fr{1}{4}l_4
Tr\left[i\big\{\fr{1}{2}\tau^a,U^\dag\big\}D^\mu U~+h.c.\right]
Tr\left[BM(U^\dag +U)\right].
\end{eqnarray}
First we need to calculate the trace of the following term
\begin{eqnarray}
Tr\left[i\big\{\fr{1}{2}\tau^a,U^\dag\big\}DU\right]
&=& Tr\left[\big\{i\tau^aU_0+U^a\big\}
\big\{\partial U_0+i\bm{\tau}\cdot[\partial\bm{U}
-\mui u\bm{U}\times{\bf e}_3]\big\}\right]\nonumber\\
&=&2\left[\bm{U}\partial U_0-U_0\partial\bm{U}
+\mui uU_0\bm{U}\times{\bf e}_3\right]^a.
\end{eqnarray}
Using the expansion of the fields in equations (\ref{U_0}) and (\ref{U_i}) up
to order $\xi^3$ with $c=1$ and $s=0$ we get
\begin{equation}
Tr\left[i\big\{\fr{1}{2}\bm{\tau},U^\dag\big\}DU\right]
=-2\left[\partial\bm{\xi}-\mui
u\bm{\xi}\times{\bf e}_3\right]\left[1-\fr{2}{3}\bm{\xi}^2\right]
-\fr{4}{3}\bm{\xi}(\bm{\xi}\cdot\partial\bm{\xi}).
\end{equation}
This expression starts with terms of order $\xi$. In the case of $A_{(3)}$, we
need just expressions up to order $\xi$, but from equation (\ref{TrD_muUD_nuU})
we can see that the term $Tr\big[(D_\mu U)^\dag D_\nu U\big]$ is of order
$\xi^2$. Then we can neglect the terms proportional to $l_1$ and $l_2$, and
using eq. (\ref{TrM(U+U)}) we obtain
\begin{eqnarray}
\bm{A}_{(1)}&=&-f^2\left\{\left[\partial\bm{\xi}-\mui
u\bm{\xi}\times{\bf e}_3\right]\left[1-\fr{2}{3}\bm{\xi}^2\right]
+\fr{2}{3}\bm{\xi}(\bm{\xi}\cdot\partial\bm{\xi})\right\}\\
\bm{A}_{(3)}&=&-2l_4m^2\left[\partial\bm{\xi}-\mui
u\bm{\xi}\times{\bf e}_3\right]
\end{eqnarray}

\subsection{Scalar current.}

We need the scalar current to calculate radiative corrections to the quark
condensate (or chiral condensate) defined as
\begin{equation}
\llang\bar qq\rrang \equiv\llang 0|\bar qq|0\rrang =\llang
0|J_{s}|0\rrang  
\end{equation}
 
The scalar current  for massive
quarks and finite isospin
chemical potential is
\begin{equation}
J_s=-\frac{\delta S\chi}{\delta s^0}(M,0,\fr{1}{2}\mui
u\tau_3,0).
\end{equation}
The expressions for $J_{s(1)}$ and $J_{s(3)}$ with finite quark masses and
finite chemical potential are
\begin{eqnarray}
J_{s(1)}&=&-\fr{1}{2}f^2BTr\big[U^\dag+U\big]\\
J_{s(3)}&=&-\fr{1}{2}(l_3+l_4)BTr\big[U^\dag+U\big]
Tr\big[BM( U^\dag+U)\big] \nonumber\\
&&-\fr{1}{4}l_4BTr\big[(D_\mu U)^\dag D^\mu U\big] 
Tr\big[U^\dag+U\big] \\
&&-4\tilde h_1BTr\big[BM\big]\nonumber
\end{eqnarray}
For $J_{s(2)}$, expanded up to oreder $(\pi)^2$, we use the results of equation
(\ref{U_0})
\begin{equation}
J_{s(1)}=-2f^2BU_0
=-2f^2B\left(c-s\tilde\xi_1-\fr{1}{2}c\bm{\xi}^2\right)
\end{equation}

$J_{s(3)}$ at order $(\pi)^0$ is just setting the fields $U=\bar U$. Then
using the equation (\ref{TrDUDU}), with $\bar U= c+i\tilde\tau_1s$  
\begin{eqnarray}
J_{s(3,0)}&=&-4(l_3+l_4)Bm^2\bar U_0^2 
-2l_4B\mui^2(\tilde{\bar U}_1^2+\tilde{\bar U}_2^2)\bar U_0 
-4\tilde h_1Bm^2\nonumber\\
&=&-2B\big[2(l_3+l_4)m^2c^2+l_4\mui^2s^2c+2\tilde h_1m^2\big].
\end{eqnarray}

\subsection{Vector current.}\label{expansion.vector}

The vector current will be used to calculate the isospin number density
condensate, where
\begin{equation}
\llang n_I\rrang \equiv\llang 0|\fr{1}{2}q^\dag\tau_3q|0\rrang =\llang
0|V_{0}^3|0\rrang.
\end{equation}
The pseudo-scalar current for massive quarks at finite isospin chemical
potential is
\begin{equation}
V_{\mu}^a=\frac{\delta S_\chi}{\delta v_\mu^a}(M,0,\fr{1}{2}\mui u\tau_3,0) 
\end{equation}

The expressions for $V_{(1)\mu}^a$ and $V_{(3)_\mu}^a$ are
\begin{eqnarray}
V_{(1)\mu}^a&=&\frac{f^2}{4}
Tr\left[-i\big[\fr{1}{2}\tau^a,U^\dag\big]D^\mu U~+h.c.\right] \\
V_{(3)\mu}^a&=&\fr{1}{2}l_1
Tr\left[-i\big[\fr{1}{2}\tau^a,U^\dag\big]D^\mu U~+h.c.\right]
Tr\left[(D_\alpha U)^\dag D^\alpha U\right]\nonumber\\
&&+\fr{1}{2}l_2
Tr\left[-i\big[\fr{1}{2}\tau^a,U^\dag\big]D_\beta U \delta^\mu_\alpha
~+h.c.,~\alpha\leftrightarrow\beta\right]
Tr\left[(D^\alpha U)^\dag D^\beta U\right]\nonumber\\
&&+\fr{1}{4}l_4
Tr\left[-i\big[\fr{1}{2}\tau^a,U^\dag\big]D^\mu U~+h.c.\right]
Tr\left[BM(U^\dag +U)\right]
\end{eqnarray}

First we need to calculate the trace of the following term
\begin{eqnarray}
Tr\left[-i\big[\fr{1}{2}\tau^2,U^\dag\big]DU\right]
&=& Tr\left[-i\big(\bm{U}\times\bm{\tau}\big)^a \big\{\partial U_0
+i\bm{\tau}\cdot[\partial\bm{U}
-\mui u\bm{U}\times{\bf e}_3]\big\}\right]\nonumber\\
&=&2\left[\bm{U}\times\partial\bm{U}
-\mui u\bm{U}\times\big(\bm{U}\times{\bf e}_3\big)\right]^a
\end{eqnarray}

Expanding up to order $\xi^2$
\begin{eqnarray}
&&Tr\left[-i\big[\fr{1}{2}\bm{\tau},U^\dag\big]DU\right]\nonumber\\
&&\quad=2\left[-\mui s\xi_3u+\xtwo\partial\xi_3-\xi_3\partial\xtwo -\mui
c\xone\xi_3\right]\tilde{\bf e}_1\nonumber\\
&&\qquad+2\left[-s\partial\xtwo
+c\xi_3\partial\xone-c\xone\partial\xi_3
-\mui\xtwo\xi_3u\right]\tilde{\bf e}_2\nonumber\\
&&\qquad+2\Big[\mui s^2u+2\mui cs\xone u+s\partial\xtwo +c\xone\partial\xtwo
-c\xtwo\partial\xone\\
&&\hspace{5cm}-\mui\left(s^2\bm{\xi}^2-c^2\xone^2-\xtwo^2\right)u\Big]{\bf e}_3,
\nonumber
\end{eqnarray}
we see that we only need to replace it to obtain $V_{(1)\mu}^a$ up to
order $\xi^2$. For
the case of $V_{(3,0)\mu}^a$ we only need to set $\xi=0$. Using equation
(\ref{TrM(U+U)}) and (\ref{TrD_muUD_nuU}), we get
\begin{equation}
\bm{V}_{(3,0)\mu}=2s^2\left[2\mui^2s^2(l_1+l_2)+m^2l_4\right]{\bf e}_3u_\mu.
\end{equation}

\subsection{Pseudo-scalar current.}\label{expansion.J_p}
The pseudo-scalar current will be used to calculate the pion condensate, where
\begin{equation}
\llang\pi^a\rrang \equiv\llang 0|\fr{1}{2}i\bar q\gamma_5\tau^aq|0\rrang=\llang
0|J_{p}^a|0\rrang.
\end{equation}
The pseudo-scalar current for massive quarks at finite isospin chemical
potential is
\begin{equation}
J_{p}^a=\frac{\delta S_\chi}{\delta p^a}(M,0,\fr{1}{2}\mui u\tau_3,0) 
\end{equation}

The expressions for $J_{p(1)}^a$ and $J_{p(3)}^a$ are
\begin{eqnarray}
J_{p(1)}^a&=&\fr{1}{2}f^2BTr\left[i\tau^a(U^\dag-U)\right]\\
J_{p(3)}^a&=&\fr{1}{2}(l_3+l_4)BTr\left[i\tau^a(U^\dag-U)\right]
Tr\left[BM(U^\dag+U)\right]\nonumber\\
&&+\fr{1}{4}l_4BTr\left[i\tau^a(U^\dag-U)\right]
Tr\left[(D_\mu U)^\dag D^\mu U\right].
\end{eqnarray}

Considering
\begin{equation}
Tr\left[i\tau^a(U^\dag-U)\right]=2U^a,
\end{equation}
then the expansion of $J_{p(1)}^a$ up to order $\xi^2$ corresponds just to re
placethe
expansion of $U^a$ (eq. (\ref{U_i}) and $J_{p(3,0)}^a$ replacing the previous
equation, equation (\ref{TrM(U+U)}) and (\ref{TrD_muUD_nuU}) with $\xi=0$
\begin{eqnarray}
\bm{J}_{p(1)}&=&2Bf^2\left[\left(s+c\xone -\fr{1}{2}s\bm{\xi}^2\right)
 \tilde{\bf e}_1
+\xtwo\tilde{\bf e}_2+\xi_3{\bf e}_3 +{\cal O}(\xi^3)\right]\\
\bm{J}_{p(3,0)}&=& 2Bs\left[2m^2c(l_3+l_4)+\mui^2s^2l_4\right]\tilde{\bf e}_1
\end{eqnarray}

\chapter{Some useful formulas, identities and constants}

\section{$l_i$ constants}\label{l-constants}

The  $\bar l_i$ coupling constants are defined as the value of the
corresponding couplings at the scale $\Lambda =m$, subtracting the divergent
part proportional to
$\lambda$:
\begin{equation}
l_i(\Lambda)=\frac{\gamma_i}{32\pi^2}\left[ \bar l_i
  +\ln\frac{m^2}{\Lambda^2} -\lambda\right],
\end{equation}
with
\begin{equation}
\lambda=\frac{2}{4-d}+\ln 4\pi +\Gamma^\prime (1)+1
\end{equation}
for $i=1,\dots ,6$.
The table \ref{l-table} was obtained from \cite{Scherer:2002tk};
it shows the different
$l_i(\Lambda)$ and $l_7$
constants evaluated at the scale $\Lambda = m$, the corresponding $\gamma_i$
factors and indicates also from which processes they where obtained. The
other constants, $\tilde
h_i$, that appear in the chiral ${\cal L}_4$ Lagrangian, are model dependent
and do not involve divergences.

\begin{table}
\begin{tabular}{|c|c|c|c|}
\hline
$\bar l_i$ & Value & Obtained from & $\gamma_i$ \\
\hline
 & $-2.23\pm 3.7$ \cite{Gasser:1983yg} & $\pi\pi$ D-wave scattering lengths
  ${\cal O}(P^4)$ & \\
 & $-1.7\pm 1.0$ \cite{Bijnens:1994ie} & $\pi\pi$ and $K_{l4}$ & \\
 $\bar l_1$ & $-1.5$ \cite{Bijnens:1997vq} & $\pi\pi$ D-wave scattering lengths
  ${\cal O}(P^6)$ & $1/3$ \\
 & $-1.8$ \cite{Colangelo:2001df} & $\pi\pi$ scattering ${\cal O}(P^4)$ + Roy
  equations & \\
 & $-0.4\pm 0.6$ \cite{Colangelo:2001df} & $\pi\pi$ scattering ${\cal O}(P^6)$
  + Roy equations & \\
\hline
  & $6.03\pm 1.3$ \cite{Gasser:1983yg} & $\pi\pi$ D-wave scattering lengths
${\cal O}(P^4)$ & \\
  & $6.1\pm 0.5$ \cite{Bijnens:1994ie} & $\pi\pi$ and $K_{l4}$ & \\
 $\bar l_2$ & $4.5$ \cite{Bijnens:1997vq} & $\pi\pi$ D-wave scattering lengths
${\cal O}(P^6)$ & $2/3$ \\
  & $5.4$ \cite{Colangelo:2001df} & $\pi\pi$ scattering ${\cal O}(P^4)$ + Roy
equations & \\
 & $4.3\pm 0.1$ \cite{Colangelo:2001df} & $\pi\pi$ scattering ${\cal O}(P^6)$
+ Roy equations & \\
\hline
 & $2.9\pm 2.4$ \cite{Gasser:1983yg} & SU(3) mass formulae & \\
 $\bar l_3$ & $|\bar l_3|<16$ \cite{Colangelo:2001df} & $K_{l4}$ decay & 
  $-1/2$ \\
\hline
 & $4.3\pm 0.9$ \cite{Gasser:1983yg} & $F_K/F_\pi$ & \\
 $\bar l_4$ & $4.4\pm 0.3$ \cite{Bijnens:1998fm} & scalar form factor ${\cal
  O}(P^6)$ & 2 \\
 & $4.4\pm 0.2$ \cite{Colangelo:2001df} & $\pi\pi$ scattering ${\cal O}(P^6)$
  + Roy equations & \\
\hline
 & $13.9\pm 1.3$ \cite{Gasser:1983yg} & $\pi\rightarrow e\nu\gamma$
  $  {\cal O}(P^4)$ & \\
 $\bar l_5$ & $13.0\pm 0.9$ \cite{Bijnens:1998fm} & $\pi\rightarrow e\nu\gamma$
   ${\cal O}(P^6)$ & $-1/6$\\
\hline
 & $16.5\pm 1.1$ \cite{Gasser:1983yg} & $\langle r^2\rangle_\pi$ ${\cal
O}(P^4)$ & \\
 $\bar l_6$ & $16.0\pm 0.5\pm 0.7$ \cite{Bijnens:1998fm} & vector form factor
  ${\cal O}(P^6)$ & $-1/3$ \\
\hline
 $l_7$ & $\sim 5\times 10^{-3}$ \cite{Gasser:1983yg} & $\pi^0-\eta$ mixing &
0
  \\
\hline
\end{tabular}
\caption{\footnotesize $l_i$ constants values.}
\end{table}\label{l-table}

\section{$f$, $B$ and $\epsilon_{ud}$.}\label{f-B-e}
The one-loop corrections for the pion mass and the
pion decay constant are
\begin{eqnarray}
 m_\pi &=& m[1-\alpha\bar l_3/4] ~~\approx~~ 139.57~\mbox{MeV}\\
f_\pi &=& f[1+\alpha\bar l_4] ~~\approx~~ 92.42~\mbox{MeV},
\end{eqnarray}
with $\alpha =(m/4\pi f)^2$. From these definitions and with the value of $\bar
l_3$ and $\bar l_4$ it is easy to derive the value of the tree-level pion mass
and decay constant.
\begin{eqnarray}
\alpha &\approx& 0.017\\
m &\approx& 141~\mbox{MeV}\\
f &\approx& 86~\mbox{MeV}.
\end{eqnarray}

From the definition of the tree-level mass, $m^2 = B(m_u+m_d)$, using the
current-quark masses values we can extract the value of $B$
\begin{equation}
m_u+m_d=5~\mbox{to}~11~\mbox{MeV}\qquad\Longrightarrow\qquad B\approx
1.8~\mbox{to}~4~\mbox{GeV}
\end{equation}

From the value of the light quark mass ratios, we can evaluate the
$\epsilon_{ud}$ constant
\begin{equation}
\frac{m_u}{m_d}=0.2~\mbox{to}~ 0.7\qquad\Longrightarrow\qquad
\epsilon_{ud}=\frac{m_u-m_d}{m_u+m_d}=-0.2~\mbox{to}~-0.7 
\end{equation}

\section{Functions and definitions}\label{formulas.functions}

The functions involved in the radiative corrections in the first phase are
\begin{eqnarray}
\alpha &=& (m/4\pi f)^2\\
\underline\lambda &=& \lambda -\ln (m^2/\Lambda^2)\\
I   &=& \int_1^\infty dx\sqrt{x^2-1}
        [n_B(mx-|\mui|)+n_B(mx+|\mui|)]\\
J   &=& \int_1^\infty dxx\sqrt{x^2-1}
        [n_B(mx-|\mui|)-n_B(mx+|\mui|)]\\
I_n &=&  \int_1^\infty dxx^{2n}\sqrt{x^2-1}~2n_B(mx)
\end{eqnarray}
being $\alpha$ the perturbative parameter. These integrals do not depend on
the chemical potential sign, and grow with both, temperature and chemical
potential

In the case of the second phase, we
denote the same functions with a prime: $\alpha'$, $I'$, $J'$, $I'_n$ which are
the same functions, but with $|\mui$ instead of $m$.
\begin{eqnarray}
\alpha' &=& (\mui/4\pi f)^2\\
\underline\lambda' &=& \lambda -\ln (\mui^2/\Lambda^2)\\
I'   &=& \int_1^\infty dx\sqrt{x^2-1}
        \big[n_B\big(|\mui|(x-1)\big)+n_B\big(|\mui|(x+1)\big)\big]\\
J'   &=& \int_1^\infty dxx\sqrt{x^2-1}
        \big[n_B\big(|\mui|(x-1)\big)-n_B\big(|\mui|(x+1)\big)\big]\\
I'_n &=&  \int_1^\infty dxx^{2n}\sqrt{x^2-1}~2n_B(|\mui|x).
\end{eqnarray}
These integrals are also growing functions of the temperature but
decrease with the chemical potential.

Other intermediate function that appears in the self-energy corrections for
$|\mui|\gg m$ is
\begin{equation}
K_{mn}
=\int_1^\infty dx\frac{\sqrt{x^2-1}~n_B(|\mui|x)}{\big[x^2-(n/2)^2\big]^{m+ 1}}
\end{equation}

\section{Dimensional Regularization}\label{dim_reg}

In the case of a loop formed by one
free-propagator and without thermal insertions: 
\includegraphics[scale=.3]{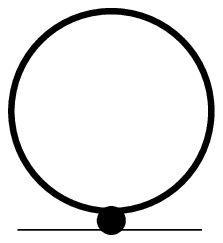}, the different integrals that
appear are 
\begin{eqnarray}
\Lambda^{4-d}\int\frac{d^dk}{(2\pi)^d}\frac{i}{k^2-\Delta} &=&
\frac{\Delta}{(4\pi)^2}\left[\ln\frac{\Delta}{\Lambda^2}
  -\lambda\right]\\
\Lambda^{4-d}\int\frac{d^dk}{(2\pi)^d}\frac{ik^2}{k^2-\Delta} &=&
\frac{\Delta^2}{(4\pi)^2}\left[\ln\frac{\Delta}{\Lambda^2}
  -\lambda\right]\\
\Lambda^{4-d}\int\frac{d^dk}{(2\pi)^d}\frac{ik^\mu  k^\nu}{k^2-\Delta} &=&
\frac{g^{\mu\nu}\Delta^2}{4(4\pi)^2}\left[\ln\frac{\Delta}{\Lambda^2}
 -\frac{1}{2} -\lambda\right] 
\end{eqnarray}

In the case of a loop formed with two free-propagators and without thermal
insertions: \includegraphics[scale=.3]{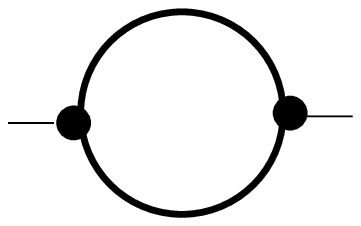}, first we combine
the denominators of the propagators using Feynman's identity
\begin{equation}
\frac{1}{AB}=\int_0^1 dx\frac{1}{[xA+(1-x)B]^2}, 
\end{equation}
then, the different integrals that appear are
\begin{eqnarray}
\Lambda^{4-d}\int\frac{d^dk}{(2\pi)^d}\frac{i}{[k^2-\Delta]^2} &=&
\frac{1}{(4\pi)^2}\left[\ln\frac{\Delta}{\Lambda^2}
  +1-\lambda\right]\\
\Lambda^{4-d}\int\frac{d^dk}{(2\pi)^d}\frac{ik^2}{[k^2-\Delta]^2} &=&
\frac{2\Delta}{(4\pi)^2}\left[\ln\frac{\Delta}{\Lambda^2}
  +\frac{1}{2}-\lambda\right]\\
\Lambda^{4-d}\int\frac{d^dk}{(2\pi)^d}\frac{ik^\mu  k^\nu}{[k^2-\Delta]^2} &=&
\frac{g^{\mu\nu}\Delta}{2(4\pi)^2}\left[\ln\frac{\Delta}{\Lambda^2}
 -\lambda\right]. 
\end{eqnarray}

Integrals proportional to $k^\mu$ are equal to zero. The diagrams
with thermal insertions do not need to be regularized.

\section{Diagrams with one thermal insertions.}\label{formulas.thermal_ints}

In the case of a loop formed by one thermal insertion
(and without free-propagator):
\includegraphics[scale=.3]{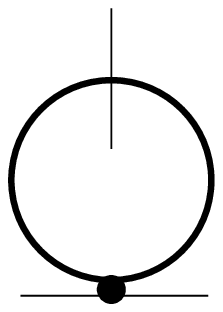}, the general integral that
appear is 
\begin{eqnarray}
F_n(a, b)
&\equiv&\int\frac{d^4k}{(2\pi)^3}k_0^nn_B(|k_0-b|)\delta(k^2-a^2)\nonumber\\
&=&\frac{4a^{2+n}}{(4\pi)^{2} }
\int_1^\infty dxx^n\sqrt{x^2-1}\big[n_B(|ax-b|)+(-)^nn_B(|ax+b|)\big].
\end{eqnarray}
Table \ref{table.1T-ins} shows the function $F_n$ with the
parameters $a$ and $b$  evaluated at values that appear in
the self-energy corrections.
\begin{table}
$$
\begin{array}{|rcl|rcl|}
\hline
\mbox{First phase}& & &\mbox{Second phase}&&\\
\hline &&&&&\\
F_0(m, \pm\mui) &=& \frac{4m^2}{(4\pi)^2}I & 
F_0(|\mui|, \pm|\mui|) &=& \frac{4\mui^2}{(4\pi)^2}I'\\ 
&&&&&\\ \hline &&&&&\\
F_1(m, \pm\mui) &=& \pm\epsilon(\mui)\frac{4m^3}{(4\pi)^2}J & 
F_1(|\mui|, \pm|\mui|) &=& \pm\frac{4|\mui|^3}{(4\pi)^2}J'\\
&&&&&\\\hline &&&&&\\
F_0(m, 0) &=&  \frac{4m^2}{(4\pi)^2}I_0 & 
F_{2m}(|\mui|, 0) &=& \frac{4\mui^{2}}{(4\pi)^2}\mui^{2m}I'_m\\
&&&&&\\ \hline &&&&&\\
&&& F_{2m}(0,0)&=&\frac{8(2m)!}{(4\pi)^2}\zeta(2m+1)T^{2m+2}\\
&&&&&\\\hline
\end{array}
$$
\caption{\footnotesize $F_n$ evaluated with the different values that appear in
the self-energy corrections.}
\label{table.1T-ins}
\end{table}
$F_{2m+1}(a,0)$ and integrals proportional to $k_i$ are equal to zero.

\section{Diagrams with two Dolan-Jackiw
propagators.}\label{formulas.2DJp}

The apparition of vertex with three legs in the second phase, in the high
isospin chemical potential limit, produces loops with two DJp.

\begin{eqnarray}
i~\includegraphics[scale=.6]{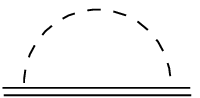}&=&\alpha'\mui^2
\left[(\bar p_0^2+1)\underline\lambda'
+\fr{4}{3}\bar p_0^2+2A_1(\bar p_0)\right]\\
i~\includegraphics[scale=.6]{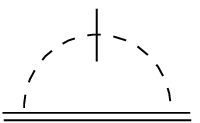}
&=&16\alpha'\mui^2B_0(\bar p_0)\\
i~\includegraphics[scale=.6]{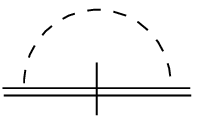}
&=&16\alpha'\mui^2B_1(\bar p_0)\\
i~\includegraphics[scale=.6]{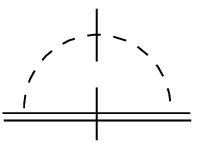}
&=&-32\pi i\alpha'\mui^2C_1(\bar p_0)
\end{eqnarray}
\begin{eqnarray}
i~\includegraphics[scale=.6]{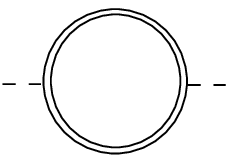}
~=~i~\includegraphics[scale=.6]{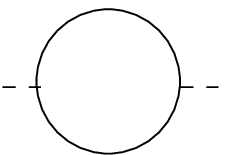}
&=& 2\alpha'\mui^2\left[\bar p_0^2(1-\underline\lambda')+A_2(\bar p_0)\right]\\
i~\includegraphics[scale=.6]{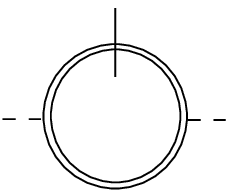}
~=~i~\includegraphics[scale=.6]{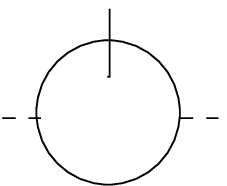}
&=& 8\alpha'\mui^2B_2(\bar p_0)\\
i~\includegraphics[scale=.6]{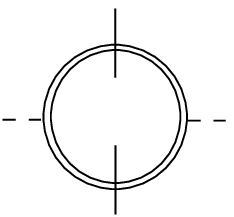}
~=~i~\includegraphics[scale=.6]{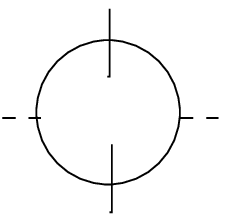}
&=& -32\pi i\alpha'\mui^2C_2(\bar p_0)
\end{eqnarray}
with
\begin{eqnarray}
A_1(\bar p_0) &=& \int_0^1dx\big[ 3\bar p_0^2x^2 +(\bar p_0^2-1)x\big]
    \ln\big[\bar p_0^2(x^2-x) +x -i\epsilon\big]\\
A_2(\bar p_0) &=& \int_0^1dx\bar p_0^2\ln \big[ \bar p_0^2(x^2-x) +1
-i\epsilon\big]\\
B_0(\bar p_0) &=& \int_0^\infty dxxn_B(|\mui|x)\left[\frac{x^2}{\bar p_0^2
-2\bar p_0x -1 +i\epsilon}+x\rightarrow
    -x\right]\\
B_1(\bar p_0) &=& \int_1^\infty
dx\sqrt{x^2-1}n_B(|\mui|x)\left[\frac{(\bar p_0 -x)^2}{\bar p_0^2-2\bar p_0x +1
+i\epsilon}+x\rightarrow
    -x\right]\\
B_2(\bar p_0) &=& \int_1^\infty dx\sqrt{x^2-1}n_B(|\mui|x)
\left[\frac{\bar p_0^2}{\bar p_0^2-2\bar p_0x+i\epsilon}+x\rightarrow
    -x\right]\\
C_1(\bar p_0) &=& \left| \frac{\bar p_0^2-1}{2\bar p_0}\right| n_B\!\!\left(
\left| \frac{\bar p_0^2-1}{2\bar p_0}\right|\right) n_B\!\!\left( \left|
\frac{\bar p_0^2+1}{2\bar p_0}\right|\right)\\
C_2(\bar p_0) &=& \big[ \theta (\bar p_0-2)+\theta
(-\bar p_0-2)\big]\bar p_0^2\sqrt{(\bar p_0/2)^2-1}n_B(|\bar p_0/2|)^2,
\end{eqnarray}
where $\bar p_0\equiv p_0/\mui$. 

To extract the masses, we need to expand the self energy in terms of the
mass corrections around the tree-level masses. Then, we need to know the
different values of the functions described in the last equations up to the
second derivative in the energy, as was described in chapter
\ref{renormalization}.

\begin{table}
$$
\begin{array}{|l|l|l|}
\hline
\qquad G_n(1) & \hspace{1.5cm} G'_n(1) & \hspace{2cm} G''_n(1)\\
\hline &&\\
A_1(1)=-\frac{2}{3} & A_1'(1)=-\frac{10}{3} & A''_1(1)=-\frac{46}{3}\\
&&\\\hline&&\\
A_2(1)=-2 +\frac{\pi}{\sqrt{3}} & A_2'(1)=-2+\frac{2\pi}{3\sqrt{3}} &
A''_2(1)=-\frac{2}{3}-\frac{10\pi}{9\sqrt{3}}\\
&&\\\hline&&\\
B_0(1)=0 & B'_0(1)=-\frac{1}{6}\left(\frac{\pi T}{\mui}\right)^2 &
B''_0(1)=\frac{1}{2}\left(\frac{\pi T}{\mui}\right)^2\\
&&\\\hline&&\\
B_1(1)=I'_0 & B'_1(1)=I'_0 & B''_1=-3I'_0-\frac{1}{2}K_{21}\\
&&\\\hline&&\\
B_2(1)=\frac{1}{4}K_{10} & B'_2(1)=-K_{10}-\frac{1}{16}K_{12} &
B''_2(1)=-K_{10}-\frac{5}{4}K_{11}-\frac{1}{4}K_{12}\\
&&\\\hline&&\\
C_1(1)=0 & C'_1(1)=0 & C''_1(1)=2\frac{T}{|\mui|}n_B(|\mui|)\\
&&\\\hline&&\\
C_2(1)=0 & C'_2(1)=0 & C''_2(1)=0\\
&&\\\hline
\end{array}
$$
\caption{\footnotesize Zero, first and second derivative of the functions $A_n$,
$B_n$, $C_n$, evaluated with $\bar p_0=1$.}
\label{table.G_n(1)}
\end{table}

\begin{table}
$$
\begin{array}{|l|l|l|}
\hline
\qquad G_n(0) & \quad G'_n(0) & \hspace{2cm} G''_n(0)\\
\hline &&\\
A_1(0)=\frac{1}{4}  & A_1'(0)=0 & A''_1(1)=-\frac{5}{6}\\
&&\\\hline&&\\
A_2(0)=0  & A_2'(0)=0 & A''_2(0)=0\\
&&\\\hline&&\\
B_0(0)=\frac{2}{15}\left(\frac{\pi T}{\mui}\right)^4 & B'_0(0)=0 &
B''_0(0)=-\frac{4}{15}\left(\frac{\pi T}{\mui}\right)^4
-\frac{128}{63}\left(\frac{\pi T}{\mui}\right)^6\\
&&\\\hline&&\\
B_1(0)=I'_1 & B'_1(0)=0 & B''_1(0)=2I'_0-10I'_1+8I'_2\\
&&\\\hline&&\\
B_2(0)=0 & B'_2(0)=0 & B''_2(0)=-K_{00}\\
&&\\\hline&&\\
C_1(0)=0 & C'_1(0)=0 & C''_1(0)=0\\
&&\\\hline&&\\
C_2(0)=0 & C'_2(0)=0 & C''_2(0)=0\\
&&\\\hline
\end{array}
$$
\caption{\footnotesize 
Zero, first and second derivative of the functions $A_n$,
$B_n$, $C_n$, evaluated with $\bar p_0=0$.}
\label{table.G_n(0)}
\end{table}

\begin{equation}
G_n(\bar m_R)=G_n(\bar m_t)+G'_n(\bar m_t)\delta m 
+\fr{1}{2}G''_n(\bar m_t){\delta m}^2+{\cal O}({\delta m}^3)
\end{equation}
with $G_n=A_n,B_n ~\mbox{or}~ C_n$. and $\bar m_t$ can be
\begin{equation}
\bar m_0 =1, \qquad \bar m_+=\sqrt{1+3c^2},\qquad m_-=0.
\end{equation}
As we are working in the expansion at order $c^0$, then
$G_n^{(k)}(\bar m_+)=G_n^{(k)}(1)+{\cal O}(c^2)$. So we need to evaluate these
expressions at $\bar p_0 =0 ~\mbox{and}~1$. Tables \ref{table.G_n(1)} and
\ref{table.G_n(0)} show the different functions and their derivatives evaluated
in $\bar p_0=1$ and $\bar p_0=0$, respectively.

\section{LSZ reduction formula for the PCAC relation.}\label{LSZ}

Due to the fact that in the first phase the physical pions are identified as
$\pi^0$, $\pi^\pm$ and the equations of motion do not mix these fields, the
generalization of the LSZ reduction formula is 
\begin{eqnarray}
&&\llang 0|
\pi^{a_1}(x_1)\pi^{a_2}(x_2)\cdots\pi^{a_n}(x_n)|\pi^b(p)\rrang\nonumber\\
&&\qquad =i\int d^4xe^{ipx}\llang 0| {\cal
D}_x^bT[\pi^{a_1}(x_1)\pi^{a_2}(x_2)\cdots\pi^{a_n}(x_n)\pi^b(x)]|0\rrang,
\end{eqnarray}
with $a_i, b= 0,+,-$ and where ${\cal D}$ is the Euler-Lagrange operator
\begin{eqnarray}
{\cal D}_x^0 &=&-\partial^2-m^2\\
{\cal D}_x^\pm &=&-\partial^2\mp 2i\mui\partial_0+\mui^2-m^2.
\end{eqnarray}
Then in the PCAC relation, the reductions made when we saturate the axial
current with a single pion are
\begin{eqnarray}
&& \llang 0|\pi^a(x)|\pi^b(p)\rrang = (\delta_{b0}\delta_{a0}
 +\delta_{b\pm}\delta_{a\mp})e^{-ipx}\\ 
&&\llang 0|\pi^{a_1}(x_1)\pi^{a_2}(x_2)\pi^{a_3}(x_3)|\pi^b(p)\rrang \nonumber\\
&& \qquad = (\delta_{b0}\delta_{a_10}
 +\delta_{b\pm}\delta_{a_1\mp})e^{-ipx_1}\llang
  0|T\pi^{a_2}(x_2)\pi^{a_3}(x_3)|0\rrang\nonumber\\  
  && \quad\qquad +(\delta_{b0}\delta_{a_20}
 +\delta_{b\pm}\delta_{a_2\mp})e^{-ipx_2}\llang
   0|T\pi^{a_1}(x_1)\pi^{a_3}(x_3)|0\rrang\\
  && \quad\qquad+(\delta_{b0}\delta_{a_30}
 +\delta_{b\pm}\delta_{a_3\mp})e^{-ipx_3}\llang
   0|T\pi^{a_1}(x_1)\pi^{a_2}(x_2)|0\rrang\nonumber.
\end{eqnarray}

\addtocontents{toc}{\mbox{}}

\listoffigures
\addcontentsline{toc}{section}{List of figures}

\listoftables
\addcontentsline{toc}{section}{List of tables}

\bibliography{bib}
\addcontentsline{toc}{section}{Bibliography}

\end{document}